\documentclass[12pt]{article}
\pdfoutput=1
\usepackage{amsmath,amssymb,epsfig,amsfonts}
\usepackage{graphicx,subfigure}
\usepackage[usenames, dvipsnames]{color}
\usepackage[backref]{hyperref}
\usepackage{cite}
\usepackage{verbatim} 
\usepackage{float}
\restylefloat{table}
\usepackage[all]{xy}
\usepackage{pdflscape}
\usepackage{tikz}
\usepackage{array}
\usepackage{youngtab}
\usepackage{extarrows}

\usepackage{color}
\usepackage{ulem}

\makeatletter

\newcommand{\be}{\begin{equation}}
\newcommand{\ee}{\end{equation}}
\newcommand{\ba}{\begin{aligned}}
\newcommand{\ea}{\end{aligned}}

\newcommand{\bea}{\begin{eqnarray}}
\newcommand{\eea}{\end{eqnarray}}

\def\Tr{\mathop{\mathrm{Tr}}\nolimits}

\def\unit{{1\kern-.65ex {\rm l}}}
\def\1{{1\kern-.65ex {\rm l}}}

\newcount\hour \newcount\minute
\hour=\time \divide \hour by 60
\minute=\time
\def\now{%
\ifnum \hour<13
  \ifnum \hour=0 \advance \hour by 12 \number\hour:\else \number\hour:\fi%
     \ifnum \minute<10 0\fi%
     \number\minute%
\ A.M.%
\else \advance \hour by -12 \number\hour:%
  \ifnum \minute<10 0\fi%
  \number\minute%
  \ P.M.%
\fi%
}

\makeatother

\begin{document}

\baselineskip=18pt  
\numberwithin{equation}{section}  
\allowdisplaybreaks  



\vspace*{-2cm} 
\begin{flushright}
{\tt KCL-MTH-16-01}\\
\end{flushright}

\vspace*{0.8cm} 
\begin{center}
 {\LARGE  F-theory and 2d $(0,2)$ Theories 
}
 
 \vspace*{1.8cm}
{Sakura Sch\"afer-Nameki$\,^1$\ and Timo Weigand$\,^2$}\\

 \vspace*{1.0cm} 
{\it $^1$ Department of Mathematics, King's College London \\
  The Strand, London WC2R 2LS, UK}\\
  {\tt {gmail:$\,$sakura.schafer.nameki}}\\
\smallskip
{\it $^2$ Institut f\"ur Theoretische Physik, Ruprecht-Karls-Universit\"at,\\
 Philosophenweg 19, 69120 Heidelberg, Germany }\\
  {\tt {email:$\,$ t.weigand\phantom{@}thphys.uni-heidelberg.de}}

\vspace*{0.8cm}
\end{center}
\vspace*{.5cm}

\noindent
F-theory compactified on singular, elliptically fibered Calabi-Yau five-folds gives rise to  two-dimensional gauge theories preserving $N=(0,2)$ supersymmetry. In this paper we initiate the study of such compactifications and determine the dictionary between the geometric data of the elliptic fibration and the 2d gauge theory such as the matter content in terms of $(0,2)$ superfields and their supersymmetric couplings.
We study this setup both from a gauge-theoretic point of view, in terms of the partially twisted 7-brane theory, and provide a global geometric description based on the structure of the elliptic fibration  and its singularities. 
Global consistency conditions are determined and checked against the dual M-theory compactification to one dimension. This includes a discussion of gauge anomalies, the structure of the Green-Schwarz terms and the Chern-Simons couplings in the dual M-theory supersymmetric quantum mechanics. 
Furthermore, 
by interpreting the resulting 2d $(0,2)$ theories as heterotic worldsheet theories,
we propose a correspondence between the geometric data of elliptically fibered Calabi-Yau five-folds and the target space of a heterotic gauged linear sigma-model (GLSM).
In particular the correspondence between the Landau-Ginsburg and sigma-model phase of a 2d $(0,2)$ GLSM is realized 
via different T-branes or gluing data in F-theory.

\newpage

\tableofcontents

\section{Introduction}

Two-dimensional $N=(0,2)$ supersymmetric gauge theories occupy a sweetspot in field theory and string theory. 
Their relation to superconformal field theories in higher dimensions is in part responsible for the recently revived interest in their dynamics. At the same time,  in combination with conformal invariance, two-dimensional field theories with $(0,2)$ supersymmetry lie at the very heart of string theory since they describe the worldsheet of the heterotic theories. Following  the seminal paper \cite{Witten:1993yc} much interest was sparked also in non-conformal $(0,2)$ gauge theories which flow  to a $(0,2)$ superconformal field theory (SCFT) in the infrared.
Recent years have seen intensified efforts to understand the properties of 2d $(0,2)$ theories from first principles, as well as through the connection with higher-dimensional theories. For instance, defects of supersymmetric three-dimensional gauge theories are described in terms of 2d $(0,2)$ theories \cite{Gadde:2013wq}. Another avenue is to consider the dimensional reduction of supersymmetric gauge theories, such as 4d $N=1$ theories \cite{Kutasov:2013ffl,Kutasov:2014hha} or twisted reductions of 4d $N=4$ theories  \cite{Bershadsky:1995vm,Benini:2012cz,Benini:2013cda}. Among the most intriguing connections is the relation to the enigmatic 6d $(0,2)$ theory which captures the effective theory of M5-branes. Dimensionally reducing the 6d $(0,2)$ theory to 2d on a four-manifold (embedded as a co-associate cycle in a $G_2$ manifold) results in a $(0,2)$ supersymmetric gauge theory \cite{Gadde:2013sca, Assel:2016lad}, whose characteristics are encoded in the geometry of the four-manifold. Much progress has been made in uncovering the properties of such theories. 

An alternative way to obtain large, and at times comprehensive, classes of gauge theories is to geometrically engineer these within string theory. Geometric engineering of 2d $N=(0,2)$ gauge theories has thus far been somewhat  confined to a sparce set of examples. Compatifications of Type II and heterotic supergravity to two dimensions, mostly with focus on models with four supercharges, have been analyzed e.g. in \cite{Dasgupta:1996yh,Forste:1997bd,Gukov:1999ya,Gates:2000fj,Haack:2000di,Font:2004et,Greiner:2015mdm}, and  \cite{Mohri:1997ef,GarciaCompean:1998kh,Franco:2015tna,Franco:2015tya} have obtained 
$(0,2)$ models from D1-branes at  local singularities. 
Here, our goal is to {develop} a geometric engineering framework for 2d $N=(0,2)$ theories which generates both a large class of examples and potentially even a classification by means of constraining the gauge theory from the geometry of the compactification spaces.

In the 20 years after its uncovering, F-theory \cite{Vafa:1996xn,Morrison:1996na,Morrison:1996pp} has established itself as a powerful framework for geometric engineering of gauge theories in even dimensions, specifically 8d, 6d, and 4d. Recent work has exemplified the strength of this approach, which resulted in a classification of 6d $N=(0,1)$ SCFTs  \cite{Heckman:2013pva}.  
Thus far, entirely unexplored are compactifications of F-theory to two dimensions, whose analysis we initiate in this paper by constructing 2d $N=(0,2)$ theories from F-theory on  elliptically fibered Calabi-Yau five-folds $Y_5$.

  As for any geometric engineering framework, we first have to develop the precise correspondence between the gauge theoretic ingredients in 2d and the intricate structures of the underlying five-fold geometry. {Going beyond the geometric realization of gauge theories,} this approach even offers the prospect of interpreting the 2d $(0,2)$ theory obtained by F-theory compactification on $Y_5$  as a (heterotic) worldsheet theory in its own right, thereby establishing a new correspondence between the original compactification space $Y_5$ and the target space associated with the {resulting} 2d heterotic worldsheet theory. 
To pursue this program, much of our interest will be focused on the elliptic fiber of the Calabi-Yau variety $Y_5$, as this will govern the gauge degrees of freedom, matter and supersymmetric couplings in 2d and geometrically encode the 7-brane degrees of freedom in F-theory.

In carrying out this program we {benefit}  from the considerable progress that has been achieved in the study of {lower}-dimensional elliptic Calabi-Yau varieties in analysing 6d and 4d vacua with $N=1$ supersymmetry. The latter case was partially motivated by the construction of phenomenologically relevant string vacua \cite{Beasley:2008dc, Beasley:2008kw, Donagi:2008ca} (for recent reviews of F-theory see e.g. \cite{Denef:2008wq,Weigand:2010wm,Maharana:2012tu}).
The advances made in this  {active} field of studying F-theory on Calabi-Yau three- and four-folds will provide an ideal setting to venture into the study of elliptic Calabi-Yau five-folds. The geometric lessons learned on lower-dimensional compactification spaces will serve as crucial input into our analysis. But various higher-dimensional intricacies will be encountered along the way, making five-folds a much richer class of Calabi-Yau varieties than the ones thus far studied. This is mirrored in the more complex structure of the 2d $N=(0,2)$ landscape of gauge theories. 
In particular, the theories we set out to study {seem to be}  genuine $(0,2)$ models insofar as they are not in any way closely related to $N=(2,2)$ theories, which for many constructions in the past have been the starting point in the construction of $(0,2)$ theories. 

There are various approaches to studying the 2d theories that emerge from F-theory on Calabi-Yau five-folds. 
A gauge theory with gauge group $G$ arises as the world-volume theory of 7-branes wrapping a complex three-fold $M_G$ (i.e., counting real dimensions, a six-cycle) in the {complex dimension four} base $B_4$  of the elliptic fibration. From this point of view,  the 2d theory is described as a partially topologically twisted 8d supersymmetric Yang-Mills (SYM) theory, where the twist is along the compact directions. The supersymmetric vacua of this gauge theory are characterized in terms of {generalized} Hitchin equations on $M_G$  for a Higgs bundle $(A, \varphi)$.  
Such a gauge theoretic point of view, which also formed the basis of the work \cite{Beasley:2008dc, Beasley:2008kw, Donagi:2008ca} on four-dimensional F-theory compactifications, is particularly useful in determining the precise correspondence between the geometric data of $M_G$ and the 2d spectrum. We therefore begin our analysis by studying the dimensional reduction of the partially topologically twisted 8d SYM theory to 2d. 

Much of the properties of 7-branes in the F-theory compactification {are} encoded in the geometry of the elliptic fiber, in particular its singularities above $M_G$. Aspects of the base of the fibration will for this paper not play a central role, but are key to the study of superconformal points \cite{TBA1}. 
{Due to the absence of} a first-principle formulation of F-theory, dualities are of {particular} importance in {identifying}  the compactification data. The most important of these is the duality with M-theory, compactified on the Calabi-Yau five-fold $Y_5$ to one dimension, which yields an $N=2$ supersymmetric quantum mechanics. The super-mechanics obtained from M-theory on smooth, not necessarily elliptically fibered Calabi-Yau five-folds has been studied in \cite{Haupt:2008nu}. As we will discuss, in the presence of a fibration structure this super-mechanics theory lifts to a 2d $N=(0,2)$ theory in the F-theory limit of vanishing fiber volume. Amongst other things, this approach will turn out to be useful in studying the global consistency conditions of the compactification, the rich structure of gauge anomalies in {chiral gauge theories} and the inclusion of gauge backgrounds in form of M-theory fluxes. 
The perturbative limit of the F-theory construction is described by a Type IIB orientifold on a Calabi-Yau four-fold. This point of view provides us with invaluable intuition in particular in studying the sector of D3-branes, whose dynamics in the dual M-theory compactification, where they correspond to M2-branes, is considerably more elusive. Another useful approach in studying F-theory compactifications is to consider heterotic/F-theory duality, which is applicable when the base $B_4$ of the five-fold is a $\mathbb P^1$-fibration over $B_3$.  The 2d $(0,2)$ F-theory vacuum is then mapped to the theory obtained from compactification of the heterotic string over an elliptic fibration over $B_3$. {The Higgs bundles and their spectral covers that we discuss for the 7-branes in Calabi-Yau five-folds should then have a counterpart in terms of spectral covers for the heterotic duals.}
The exploration of this duality is left for future work.{\footnote{Note that the correspondence with heterotic GLSMs which we will discuss in this paper is of a different nature than this more canonical heterotic/F-theory duality. }}

The theories we obtain from F-theory by combining these various angles have the following structure.
There are two sources for the vector {multiplets:} from the gauge fields on the 7-branes as well as from extra D3-branes wrapping holomorphic curves {inside} $B_4$. 
Charged massless matter arises by dimensional reduction of the {\it bulk} modes, by which we mean the gauge degrees of freedom along the worldvolume of the 7-branes (and in principle also the D3-branes),  from $[p,q]$-string excitations localised at the intersection of two 7-branes over a complex surface (which will be referred to as {\it surface matter}), and  from $[p,q]$-strings at the intersection points between the 7- and the D3-branes, respectively.
This matter organizes into 2d $(0,2)$ chiral and Fermi multiplets, which are counted by certain cohomology groups {that we determine}.

The matter interacts via  non-derivative couplings allowed by the $(0,2)$ structure of the effective theory which can be computed by evaluating the overlap of the internal zero-mode wavefunctions. Apart from pure 7-brane bulk and bulk-surface matter couplings, such interactions localize at the intersection of matter surfaces. Holomorphic couplings charcterized in terms of the fields $E$ and $J$ arise from both codimension three and four loci in the base, which give rise to cubic and quartic couplings, respectively. The pure surface-matter couplings arise from the wavefunction overlap at distinguished curves in the base over which the singularity structure of the fiber enhances further. Generically, at such codimension three loci several types of gauge invariant interactions coalesce due to the strong fiber enhancement. The interactions have contributions at leading order from the point of view of the 7-brane theory, which give rise to so-called $E$- and $J$-type couplings in the 2d theory. These are always cubic in nature. More general interactions arise by integrating out massive fields. We indicate this latter point in an example which realizes the quintic hypersurface sigma-model.  At points in the base of the fibration, i.e. over codimension four loci in the five-fold, additional quartic interactions arise.

  The specific multiplicities of massless charged matter depends, apart from the topology of the wrapped cycles, on the gauge background, which translates, via M/F-theory duality,  into 3-form gauge data.
 Even in the absence of gauge fluxes, chirality of the theory requires the cancellation of gauge anomalies. In particular, the 3-form tadpole cancellation condition from M-theory determines the total class of curves wrapped by the D3/M2-branes in such a way that the complete matter from both the 7-branes and the D3-branes is anomaly-free. The structure of anomaly cancellation for abelian gauge symmetries is considerably enriched  due to a wealth of Green-Schwarz terms, which we discuss from the IIB and the M-theory perspective. 
Finally  we find a powerful check of our expressions derived for the chiral index of massless matter by analyzing the Chern-Simons terms in the M-theory super-mechanics and comparing it with the 1-loop generated Chern-Simons obtained from F-theory. This is the 1d/2d analogue of the higher-dimensional correspondence of
\cite{Witten:1996qb,Intriligator:1997pq,Aharony:1997bx,Grimm:2011fx,Bonetti:2011mw,Bonetti:2013ela,Bonetti:2013cza,Grimm:2013oga,Cvetic:2012xn}.

We close this paper with an outlook towards superconformal theories and the relation to gauged linear sigma-models (GLSMs), which have been central in the understanding of the moduli space of 2d $(0,2)$ theories \cite{Witten:1993yc}. Some evidence will be given 
in support of a new correspondence between F-theory compactifications on elliptic Calabi-Yau five-folds $Y_5$ with $G_4$-flux and heterotic compactifications on {three-folds} with vector bundles.  
The idea is here to interpret the 2d $(0,2)$ theory obtained by F-theory compactifiation on $Y_5$ as the GLSM which flows in the infra-red to the non-linear sigma-model describing the propagation of the heterotic string on a Calabi-Yau target space.
The simplest such models correspond to  heterotic sigma-models on toric hypersurfaces. From the F-theory point of view, the underlying GLSMs are somewhat complementary to the ones discussed in the {main part} of the paper, as there is no non-abelian gauge group. The only gauge degrees of freedom are from $U(1)$s, which are realized in terms of rational sections of elliptic fibrations. 
In addition, GLSMs with non-abelian gauge groups do correspond to interesting heterotic theories, e.g. on hypersurfaces of Grassmannians \cite{Witten:1993yc} or even more general varieties (see e.g. \cite{Donagi:2007hi,Jockers:2012zr,Halverson:2013eua} and references therein), and it will be an interesting avenue of research to relate these models with the 2d $(0,2)$ F-theory models obtained in this paper.

Irrespective of the gauge group of the GLSM,  the above correspondence suggests that the various phases of the GLSM are realized in terms of different F-theory Higgs bundle configurations $(A, \Phi)$ which were termed gluing branes or T-branes \cite{Cecotti:2010bp, Donagi:2011jy, Donagi:2011dv, Marsano:2012bf, Anderson:2013rka, Collinucci:2014taa}. These are off-diagonal background values for the Higgs field. Schematically, we find the following identification of GLSM phases, focusing here for simplicity on the GLSM associated with a degree $n$ hypersurface in $\mathbb {CP}^{n-1}$ of \cite{Witten:1993yc}: 

\be\label{Duality}
\begin{array}{ccccc}
{\rm NLSM-phase}             &  & {\rm GLSM}                                       &  &  {\rm LG-phase}   \vspace{2mm} \cr 
G = \emptyset                    &  \xleftarrow[{\ \rm data\ } ]{\ \rm gluing\ }  & G= U(1)                                            &  \xrightarrow[{\ \rm data\ }]{\ \rm gluing\ }  & G = \mathbb Z_5 \vspace{2mm}  \cr
(\tilde{A}, \tilde{\Phi})                           &   &  (A, \Phi)               &   &   (\hat{A}, \hat{\Phi})\vspace{2mm}  \cr
\end{array}
\ee
Here, the GLSM with $U(1)$ gauge group arises from a compactification with rank one Mordell-Weil group (MW), and trivial Tate-Shafarevich (TS) group. The special phases of the GLSM correponding to the non-linear sigma-model (NLSM)
 as well as the Landau-Ginzburg (LG) phase are reached by turning on gluing data on the 7-brane theory in the Calabi-Yau five-fold, which are non-diagonalizable vevs for the Higgs field. While developing such a correspondence in greater depth will be the subject of future work \cite{TBA1}, we shall provide more  details on this idea already, in section \ref{sec:NLSM}. 


{The paper is} organized as follows: After setting the stage in section \ref{sec:GeneralStuff} with a reminder on F-theory as well as 2d $(0,2)$ theories,  we begin our analysis in sections \ref{sec:8dSYM} and \ref{sec:Matter} by first analyzing the compactification of the partially twisted 7-brane theory. Here we characterize the dimensional reduction to a 2d $N=(0,2)$ supersymmetric theory with gauge and matter degrees of freedom in terms of geometric data on the 7-brane compactification cycle.  Some of the details of the computations are relegated to appendix \ref{app:Gamm}. The sector of D3-branes wrapping curves in the {compactification} space is the subject of section \ref{sec:D3Sector}. In section \ref{sec:Singularfibers}, we describe these theories from the point of view of the elliptic Calabi-Yau five-fold underlying the F-theory compactification and identify the gauge theoretic data with the geometric properties of the elliptic fiber. Fluxes, global consistency conditions  such as anomalies and tadpoles and the Chern-Simons couplings are discussed in sections \ref{sec_G4fluxes}, \ref{sec:GlobalConsistency} and  \ref{sec:CScouplings}. A large set of examples can be found in sections \ref{sec:Examples} and \ref{sec:ExamplesGlobal}, {with some of the technical details} provided in appendix \ref{app:Examples}. 
In section \ref{sec:SCFTGLSM} we give a brief outline of the relation of {this new class of} 2d $(0,2)$ theories with 2d SCFTs in the infrared \cite{TBA1}, as well as a more detailed exposition of the correspondence addressed in (\ref{Duality}).
The weakly coupled description of the F-theory compactification in terms of Type IIB orientifolds can be found in appendix \ref{app_IIB}.
We conclude in section \ref{sec_Concl} with a list of future research directions originating from the present paper.

{\it Note added}: 
After this article appeared on the arxiv, \cite{Apruzzi:2016iac} was submitted, which has some overlap with the results presented here.


\section{F-theory, Five-folds and $(0,2)$ Models}
\label{sec:GeneralStuff}

The purpose of this paper is to study the effective theory of F-theory compactified on an elliptically fibered Calabi-Yau five-fold to $\mathbb{R}^{1,1}$. 
The low energy effective theory in 2d is a supersymmetric gauge theory which preserves two chiral supercharges. The dictionary 
between geometric properties of the Calabi-Yau and the gauge theoretic data, which will be estabilished in the course of the next sections, will allow us to construct a rich class of $(0,2)$ supersymmetric gauge theories. 
This section will serve as an overview of the general setup underlying these constructions, as well as a summary of the methods, such as dualities to M-theory, which will be instrumental in the following. We will also give a brief review of 2d gauge theories with $(0,2)$ supersymmetry. 


\subsection{F-theory on Calabi-Yau five-folds}
\label{sec:FCY5}

We construct two-dimensional F-theory \cite{Vafa:1996xn,Morrison:1996na,Morrison:1996pp}, i.e. non-perturbative Type IIB, vacua by dimensional reduction on  elliptically fibered Calabi-Yau varieties $Y_5$ of complex dimension five. {Schematically, such varieties $Y_5$ are of the form
\be
\ba
\pi :\quad \mathbb{E}_\tau \ \rightarrow & \  \ Y_5 \cr 
& \ \ \downarrow \cr 
& \ \  B_4 
\ea
\ee
where $\mathbb{E}_\tau$ is the elliptic fiber.
{We consider non-trivial fibrations, whereby the base $B_4$ is a complex four-dimensional K\"ahler cycle, with non-trivial canonical class}. We shall assume that the fibration has a {zero-section}, corresponding a map $\sigma_0$ from the base to the fiber. This in particular implies the existence of a Weierstrass form for $Y_5$\footnote{Projectivizing this in $\mathbb{P}^{123}[z, x, y]$ realizes the zero-section as $z=0$, also sometimes referred to as $w=0$ in the literature. As the F-theory aficionado will appreciate, {the present notation} was reached 
in a diplomatic {settlement, whereby the authors agreed} to denote the zero-section by $z=0$, whereas the exceptional sections of the resolutions will be referred to as $\zeta_i$.  }
\be\label{Weier}
y^2 = x^3 + f \, x \, z^4 + g \, z^6 \,,
\ee
with $f, g$ sections of suitable powers of the anti-canonical bundle of the base, $K_{B_4}^{-1}$. The zero-section is then realized by $z=0$.\footnote{All that follows can be generalized to settings without a zero-section, so-called genus-one fibrations, along the lines of \cite{Braun:2014oya,Morrison:2014era,Anderson:2014yva,Klevers:2014bqa,Garcia-Etxebarria:2014qua,Mayrhofer:2014haa,Mayrhofer:2014laa,Cvetic:2015moa,Lin:2015qsa,Kimura:2015qpz}. Genus-one fibrations give rise to F-theory models with discrete gauge groups, which will become of some importance for us in section \ref{sec:NLSM}. } The identification of the complex structure of the elliptic fiber with the axio-dilaton $\tau$ of type IIB implies that non-trivial fibrations correspond to vacua with varying string coupling, resulting in not necessarily perturbative vacua. The natural action of $SL(2,\mathbb{Z})$ on the complex structure of elliptic curves geometrizes thereby the S-duality of type IIB string theory. 

\begin{table}\begin{center}
\begin{tabular}{|c|c|}\hline
Singularities above {codim} & 2d $N=(0,2)$ Gauge Theory \cr\hline
1 & Gauge group $G$ \cr 
2  & Matter (chiral and Fermi) in ${\bf R} \oplus \bar{\bf R}$\cr 
& Bulk-surface matter couplings: $E$ and $J$ \cr 
3 & Holomorphic matter couplings:  $E$ and $J$  \cr 
4 & Holomorphic matter couplings: $E$ and $J$   \cr\hline
\end{tabular}
\caption{Identification of singularities in the elliptic fibrations above codimension $d$ loci in the base $B_4$ of the elliptic Calabi-Yau five-fold with 2d gauge theoretic data. \label{tab:GeoGauge}}
\end{center}
\end{table}

Singularities of the elliptic fiber  correspond to divergences in the axio-dilaton  sourced by the presence of  7-branes.
More precisely, the 7-branes correspond to logarithmic singularities creating branch-cuts in the the transverse directions to the branes, and the axio-dilaton undergoes an $SL(2,\mathbb{Z})$ monodromy. Singularities over complex codimension one in $B_4$ thus correpond to 7-branes wrapped on complex three-cycles $M_G$ times $\mathbb{R}^{1,1}$ and give rise to the gauge degrees of freedom in the two-dimensional theory.  
The singularities are characterized in terms of the vanishing of the discriminant of the Weierstrass equation 
\be
\Delta = 4f^3 +27g^2 \,.
\ee
The gauge algebra $\mathfrak{g}$ is encoded in the type of singularity above $M_G$, which can be determined from the vanishing orders of $(f, g,\Delta)$ along these loci.\footnote{{For the present purposes it will not be necessary to distinguish between} the gauge algebra $\mathfrak{g}$ and the gauge group $G$. See e.g. \cite{Mayrhofer:2014opa}  for how this distinction arises in F-theory.}
 We will show as a very first step that {the world-volume theory of the 7-branes, i.e. 8d SYM}, compactified on a complex three-cycle indeed gives rise to a 2d $(0,2)$ supersymmetric theory, whose supersymmetric vacua have a characterization in terms of a Hitchin-type equation. Singularities appearing in codimension two in the base will be shown to correspond to additional matter sectors -- which can be thought of as arising from intersecting 7-branes. 
So far the dictionary is very much alike to the compactification on Calabi-Yau three- and four-folds. The distinction to these earlier cases manifests itself in higher codimension. Like the four-fold case, where codimension three points in the three-dimensional base give rise to cubic Yukawa couplings, here we will find that the cubic holomorphic interactions are generated in codimension three  -- this time over curves in the base. Over point in codimension four, the only additional couplings are quartic. This is summarized in table \ref{tab:GeoGauge}.

In the absence of a first principle definition of F-theory, much of the analysis relies either on inferring properties from the effective 7-brane theory, as will be studied in section \ref{sec:8dSYM},  relations to perturbative string theories, or dualities. 
Surprisingly few backgrounds of this type have been studied in the past. Related perturbative constructions have appeared in \cite{Gates:2000fj} in type IIA and IIB on Calabi-Yau four-folds, which preserve $N=(2,2)$ and $N=(0,4)$, respectively, and torus orbifolds in \cite{Font:2004et}. Compactifications on Calabi-Yau five-folds first appeared, in a rather different context, in \cite{Curio:1998bv}.

Of particular relevance to understanding the low energy effective theory is the duality to M-theory compactified on elliptic Calabi-Yau five-folds. M/F-duality corresponds to taking the volume of the elliptic fiber {in the M-theory compactification} to zero, {which results in a non-perturbative IIB background in 10d}: 
\be\label{MFDuality}
\begin{array}{ccc}
\hbox{M-theory on $Y_5$}\qquad \qquad 
& \xrightarrow[]{\ {\rm Vol} (\mathbb{E}_{\tau}) \rightarrow 0\ } 
&\qquad \hbox{F-theory on $Y_5$}
\cr \cr 
\downarrow\qquad \qquad 
& 
&\qquad \downarrow\cr \cr 
\hbox{1d Super-Mechanics}\qquad \qquad 
& \xrightarrow[]{\ R_A \sim {1\over R_B} \,\rightarrow \, 0 \ } 
&\qquad \hbox{2d $(0,2)$ Gauge Theory}
\end{array}
\ee
{Here the F-theory limit of taking the volume of the elliptic fiber to zero corresponds in the M-theory/IIA language to the zero radius limit $R_A\rightarrow 0$, or equivalently, after T-duality, to the decompactification limit in IIB, which lifts the supersymmetric Quantum Mechanics to a 2d $N=(0,2)$ gauge theory.}
Compactifications of M-theory on smooth (not necessarily elliptically fibered) Calabi-Yau five-folds to supersymmetric quantum mechanics were studied in \cite{Haupt:2008nu}. Applied to elliptic five-folds, these quantum mechanical models are related by M/F-theory duality to the 2d $(0,2)$ theories studied in this paper. For our purposes, this duality plays a crucial role in identifying D3-brane contributions, which in M-theory correspond to M2-branes, Chern-Simons couplings in section \ref{sec:CScouplings} as well as fluxes and tadpole cancellation conditions in section \ref{sec:GlobalConsistency}.



\subsection{Two-dimensional $N=(0,2)$ Theories}
\label{sec:Review02}

In this final overview part, we summarize some properties of 2d $(0,2)$ theories, mostly for future reference and to setup our nomenclature. The conventions followed throughout are those in \cite{Witten:1993yc}.
We consider $\mathbb{R}^{1,1}$ with coordinates $(y^0, y^1)$ or $y^\pm = y^0 \pm y^1$ and derivatives $\partial_{\pm} = \partial_0 \pm \partial_1$ and denote by $SO(1,1)_L$ the two-dimensional Lorentz group.
An $N=(0,2)$  supersymmetric theory in two dimensions has negative chirality supersymmetry variation parameters 
 $\epsilon_-$ and $\bar \epsilon_-$, and corresponding supercharges of positive chirality. 
 There are three multiplets in an $N=(0,2)$ theory: the vector multiplet, the chiral multiplet with components $(\varphi, \chi_+)$ and the Fermi multiplet with leading fermionic component $\rho_-$. The fermions in the chiral multiplet (as well as its complex conjugate) have positive 2d chirality, whereas they have negative chirality in the Fermi multiplet. 

The $(0,2)$ superspace coordinates have positive chirality and will be denoted by $\theta^+$ and $\bar\theta^+$.  
The 2d $N=(0,2)$ supersymmetry variations with respect to $(\epsilon_-, \bar\epsilon_-)$ are 
\be\label{GeneralSUSY}
\ba
\delta \varphi &\ =\ - \sqrt{2} \, \epsilon_- \chi_+ \cr 
\delta \chi_+ &\ = \ i \sqrt{2 }(D_0 + D_1) \varphi\,  \bar{\epsilon}_-  \cr 
\delta \rho_- &\ =\ \sqrt{2} \epsilon_- \, G - i \bar{\epsilon}_- \, E
\ea
\qquad 
\ba
\delta \bar\varphi &\ = \  + \sqrt{2} \bar{\epsilon}_- \, \bar\chi_+  \cr
\delta \bar{\chi}_+ &\ = \ - i \sqrt{2} (D_0 + D_1) \bar\varphi\,  \epsilon_- \cr 
{\delta \bar\rho_- }& \ = \  \sqrt{2 }\bar{\epsilon}_- \bar{G }+ i \epsilon_- \bar{E} \,.
\ea
\ee
Here $D_0 + D_1$ denotes the gauge covariantisation of $\partial_0 + \partial_1$.
The expansion of the vector superfield (in a Wess-Zumino type gauge) is
\be \label{defVfield}
V =  (v_0 - v_1) - 2 i \theta^+ \bar\eta_- - 2 i \bar\theta^+ {\eta_-} + 2 \theta^+\bar{\theta}^+ {\mathfrak D} \,.
\ee
We will occasionally also make use of the superfield
\be \label{defVplusfield}
V_+ = \theta^+ \bar\theta^+ (v_0 + v_1) \,,
\ee 
as well as the field strength 
\be
\Upsilon = - 2 \left(\eta_- - i \theta^+({\mathfrak D} - i F_{01}) - i \theta^+ \bar \theta^+ \partial_+ \eta_- \right) \,.
\ee
The chiral and conjugate-chiral superfields enjoy the expansion
\be\ba
\Phi &= \varphi + \sqrt{2} \theta^+ \chi_+ - i \theta^+ \bar\theta^+(D_0 + D_1)\varphi \cr 
\bar \Phi &= \bar \varphi - \sqrt{2} \bar \theta^+ \bar \chi_+ + i \theta^+ \bar\theta^+(D_0 + D_1)\bar \varphi \,,
\ea\ee
and a Fermi superfield and its conjugate take the form
\be\ba
P &= \rho_- - \sqrt{2} \theta^+ G- i \theta^+ \bar\theta^+(D_0 + D_1)\rho_-  - \sqrt{2} \bar\theta^+ E \cr 
\bar P &= \bar \rho_- - \sqrt{2} \bar \theta^+ \bar G +i  \theta^+ \bar\theta^+(D_0 + D_1)\bar \rho_-  - \sqrt{2} \theta^+ \bar E \,.
\ea\ee
Here  $E$ is a holomorphic function of the chiral superfields, which, like ${\mathfrak{D}}$ and $G$, is an auxiliary field. 
The kinetic term of a chiral multiplet $\Phi_i$, taken for simplicity to be charged under a $U(1)$ gauge group with charge $Q_i$, is
\be\ba
L_{\Phi} &= -{i\over 2} \int d^2y d^2\theta\,  \bar\Phi_i \left(\partial_0 - \partial_1 + i Q_i V \right) \Phi_i \cr 
& = \int d^2y \left(- |D_\mu\varphi_i|^2 +i \bar{\chi}_{+,i}  D_- \chi_{+, i} 
- i Q_i \sqrt{2} \bar\varphi_{i} \eta_{-} \chi_{+, i} + i Q_i \sqrt{2}\varphi_i \bar\eta_- \bar\chi_{+, i} + Q_i \varphi_i\bar\varphi_i {\mathfrak D}   \right)\,.
\ea\ee
A general $(0,2)$ theory with Fermi multiplets $P_{a}$ and chiral multiplets $\Phi_i$ can exhibit non-trivial superpotential couplings, also sometimes referred to as $J$-term couplings. These take the form
\be \label{LJsuperspace}
L^J= - \frac{1}{\sqrt{2}} \int d^2 y \, d \theta^+  \left. P_a J^a(\Phi_i)\right|_{\bar \theta^+ = 0} - {\rm c.c.} \,,
\ee
which in components reads
\be \label{LJint}
L^J = - \int d^2y \left(G_a J^a + \rho_{-,a} \chi_{+,i} \frac{\partial J^a}{\partial \varphi_i}\right) - {\rm c.c.} \,.
\ee
The superpotential  $J^a(\Phi_i)$ is a holomorphic function of the chiral superfields and is subject to the constraint 
\be \label{JEzero}
\Tr J^a(\Phi) E_a(\Phi)  = 0 \,,
\ee
 where $E_a$ is the holomorphic combination of chiral superfields appearing in the definition of the Fermi superfields. Together with ${\cal D}_+ P_a = \sqrt{2} E_a$ with  ${\cal D}_+$ the gauge covariant derivative in superspace \cite{Witten:1993yc}
 this constraint ensures that (\ref{LJsuperspace}) represents a supersymmetric interaction.
The kinetic term and some of the interactions for the Fermi multiplet arise from
\be
L^{F}= -{1\over 2} \int d^2 y d^2 \theta \, P \bar P \,.
\ee
The  induced interaction terms can be summarized  as 
\be\label{LFint}
L^{F, {\rm int}}= -\int  d^2 y\,  \left(\bar\rho_- {\partial E \over \partial \varphi_i} \chi_{+, i}  + {\partial \bar{E}\over \partial \bar{\varphi}_i} \bar\chi_{+, i} \rho_- \right)\,.
\ee
Note that in addition to these standard couplings, the following type of  interactions 
\be \label{derivativecoupling1}
\int d^2\theta P \bar{P} \bar\Phi \supset \rho_- \, \bar{\rho}_-\,  (D_0 + D_1)\,  \bar\varphi + \ldots 
\ee
induce derivative couplings, which do not affect the scalar potential.

Let us also indicate the kinetic term for the gauge field strength, for simplicity written only for an abelian gauge field,
\bea
L_{\Upsilon} = - \frac{1}{8 e^2} \int d^2y  \, d^2\theta\,\bar\Upsilon \Upsilon = \frac{1}{e^2}  \int d^2 y \left( \frac{1}{2} F^2_{01} + 
 i \bar \eta_- \partial_+ \eta_-   + \frac{1}{2} {\mathfrak D} ^2 \right)\,.
\eea
Of special importance for us is the Fayet-Iliopoulos (FI) term for an abelian gauge field
\bea \label{FIterm-general}
\frac{1}{4} \int d \theta^+\,  \left( t \, \Upsilon|_{\bar \theta^+ =0} + c.c.\right) = - r {\mathfrak D}  + \frac{\theta}{2\pi} F_{01}, \qquad t= \frac{\theta}{2\pi} + i r.
\eea
In supergravity the constant FI parameter $t$ will be promoted to a chiral superfield. 

The superpotential, the Fermi interactions and the FI term then result in a scalar potential
\bea \label{scalar1}
V = \frac{1}{2 e^2} {\mathfrak D} ^2 + \sum_a \left( |J^a|^2 + |E_a|^2 \right)\,,
\eea
where the $G_a$ auxiliary fields have been integrated out and the $U(1)$ D-term is
\bea\label{scalar2}
{\mathfrak D}  = e^2 \left(\sum_i Q_i \varphi_i \bar \varphi_i  - r\right) \,.
\eea
With the FI parameter $t$ replaced by a chiral superfield, this induces a scalar potential for its imaginary part.
In the following, we will identify how each of these fields arises from the 7-brane theory reduced on a three-cycle in a Calabi-Yau five-fold, and determine the geometric origin of the couplings $J$ as well as $E$. 


\section{Partially Twisted 8d Super-Yang-Mills Theory}
\label{sec:8dSYM}

We begin our exploration of 2d $(0,2)$ theories from F-theory by considering the gauge theory approximation, where the degrees of freedom are only those realized on 7-branes. The 8d supersymmetric Yang-Mills theory (SYM) with gauge group $G$ on the world-volume of a stack of 7-branes will be dimensionally reduced on a complex three-cycle $M_G$ in the Calabi-Yau five-fold $Y_5$. To preserve supersymmetry in the transverse $\mathbb{R}^{1,1}$ one has to perform a partial topological twist. This means that the R-symmetry of the 8d SYM is combined with a subgroup of the holonomy group  acting on the tangent bundle of $M_G$ in such a way that some of the  supercharges become scalars under this new, twisted symmetry and are thus globally well-defined. This process was studied  for 7-branes wrapped on four-cycles in Calabi-Yau four-folds in \cite{Donagi:2008ca, Beasley:2008dc, Beasley:2008kw}. We will find that the vacua of this partially twisted SYM theory are characterized in terms of {generalized} Hitchin equations on $M_G$. Furthermore, we determine the spectrum of the theory and formulate it in terms of 2d $(0,2)$ supermultiplets. 


\subsection{Scalar Supercharges}
\label{sec:Twist}

The effective theory on a stack of 7-branes wrapping a K\"ahler three-cycle $M_G$ is a partially twisted 8d $N=1$ SYM theory with gauge group $G$. It can be obtained from compactification of 10d SYM by decomposing the 10d gauge potential and the gaugino field as
\be
\ba
SO(1,9)_L \ &\rightarrow \ SO(1,7)_L \times U(1)_R \cr 
A_\mu: \qquad {\bf 10} \ & \rightarrow \  {\bf 8^v}_0 \oplus {\bf 1}_{+2} \oplus {\bf 1}_{-2} \cr 
\Psi :\qquad {\bf 16} \ & \rightarrow \ {\bf 8^c}_{+1} \oplus {\bf 8^s}_{-1} \,,
\ea
\ee
where ${\bf 1}_{\pm 2 } =  \Phi_{\pm}$ are the two scalars in 8d. 
Upon {dimensional reduction}  on a compact six-manifold, the Lorentz group is further reduced as follows
\be
\ba
SO(1,7)_L   \ &\rightarrow \ SO(1,1)_L  \times SO(6)_L \cr 
{\bf 8^v} \ & \rightarrow \ {\bf 1}_{+2} \oplus {\bf 1}_{-2} \oplus {\bf 6}_0 \cr
{\bf 8^c} \ &\rightarrow\ {\bf 4}_{+1} \oplus \overline{\bf 4}_{-1} \cr 
{\bf 8^s} \ &\rightarrow\ {\bf 4}_{-1} \oplus \overline{\bf 4}_{+1} \,.
\ea
\ee
Since in the present case the six-cycle is in fact a K\"ahler three-cycle, the holonomy is reduced further to $U(3)$, resulting in
\be
\ba
SO(6)_L\  & \rightarrow \ SU(3)_L \times U(1)_L \cr 
{\bf 4} \ & \rightarrow  \ {\bf 1}_{+3} \oplus {\bf 3}_{-1} \cr 
{\bf 6} \ & \rightarrow \ {\bf 3}_{+2} \oplus \overline{\bf 3}_{-2} \,.
\ea
\ee
Putting it all together the spinors decompose as 
\be \label{decomp81}
\ba
SO(1,7)_L  \times U(1)_R \ &\rightarrow \ SU(3)_L \times SO(1,1)_L  \times (U(1)_L \times U(1)_R) \cr 
{\bf 8^c}_{+1}  \ & \rightarrow \ {\bf 1}_{1; 3, 1} \oplus   {\bf 1}_{-1; -3, 1} \oplus {\bf 3}_{1; -1, 1}  \oplus \overline{\bf 3}_{-1; 1, 1}  \cr 
{\bf 8^s}_{-1} \ & \rightarrow \ {\bf 1}_{-1; 3, -1} \oplus   {\bf 1}_{1; -3, -1} \oplus {\bf 3}_{-1; -1, -1}  \oplus \overline{\bf 3}_{1; 1, -1}  \,.
\ea
\ee
To find a singlet supercharge we need to twist $U(1)_L$ with the $U(1)$ R-symmetry, which leaves us with the two possible choices $J_{twist} = {1\over 2}\left( J_{L} \pm 3 J_R\right)$.
 We fix conventions by defining the twisted $U(1)$ generator  as 
\be
J_{twist} = {1\over 2}\left( J_{L} + 3 J_R\right) \,,
\ee
where the generator was normalized such that it act as $\mp1$ on the (anti-)holomorphic  cotangent bundle of $M_G$. 
This twist gives rise to two supersymmetry parameters $\epsilon_-$ and $\bar \epsilon_-$
 of the same (negative) chirality in 2d,
\be \label{epsidbulk}
\bar\epsilon_- = {\bf 1}_{-1;-3,1;0_{twist}} \,, \qquad   \epsilon_- =  {\bf 1}_{-1;3,-1;0_{twist}} \,.
\ee
Correspondingly, the supercharges are right chiral, and form the foundation for the $(0,2)$ supersymmetry of the theory in two dimensions. Note that $\bar\epsilon_-$ originates from ${\bf \bar 4}_{-1}$ contained in ${\bf 8^c}$ while $\epsilon_-$ originates from ${\bf 4}_{-1}$ contained in ${\bf 8^s}$.
Our conventions here follow \cite{Witten:1993yc} in that the supersymmetry parameters generating the $(0,2)$ SUSY transformations have negative 2d chirality.


\subsection{Field Content and Supersymmetry}

The dimensionally reduced partially twisted 8d SYM theory has the following spectrum 
\be\label{TwistedBulkMatter}
\ba
SO(1,7)_L  \times U(1)_R \ &\rightarrow \ SU(3)_L \times SO(1,1)_L  \times U(1)_{twist} \cr 
{\bf 8^v}_0 \qquad & \rightarrow \quad  {\bf 1}_{2; 0} \oplus {\bf 1}_{-2; 0} \oplus ({\bf 3}_{0; 1} \oplus \overline{\bf 3}_{0; -1 })  \equiv ( v_0, v_1,  {a}, \bar{a}) \cr 
\Phi_{\pm}= {\bf 1}_{2} \oplus   {\bf 1}_{-2} \quad & \rightarrow \quad {\bf 1}_{0; + 3 } \oplus   {\bf 1}_{0; - 3 }    \equiv (
\Phi_+={\bar\varphi},  \Phi_-=\varphi) \cr
{\bf 8^c}_{+1} \qquad & \rightarrow \quad {\bf 1}_{-1;0} \oplus {\bf 1}_{1;3} \oplus \overline{\bf 3}_{-1; 2}  \oplus {{\bf 3}}_{1; 1} \equiv (\bar\eta_-,  \bar\chi_+, \bar \rho_-,  \psi_+) \cr 
{\bf 8^s}_{-1} \qquad & \rightarrow \quad {\bf 1}_{-1;0} \oplus  {\bf 1}_{1; -3} \oplus {\bf 3}_{-1; -2}  \oplus \overline{\bf 3}_{1; -1} \equiv (\eta_-, \chi_+, \rho_-, \bar\psi_+)
 \,.
\ea
\ee
These fields give rise to the  {\it bulk matter}\footnote{Here the term bulk refers to the theory on the entire complex three-cycle $M_G$, and not to the gravitational theory on the ambient Calabi-Yau into which $M_G$ is embedded.}.
Interpreting the charge under  $U(1)_{twist}$ as minus the degree of the form, i.e.  charge $n\leq0 $ corresponds to $\Omega^{(n, 0)}(M_G) $ and $n\geq 0$ to  $\Omega^{(0,n)}(M_G)$, the spectrum of the twisted theory is counted by the following cohomology groups on $M_G$:

\be \label{table-matter1}
\begin{array}{c|c|c|c}
\hbox{Cohomology} & \hbox{Bosons} & \hbox{Fermions} & \hbox{Multiplet} \cr\hline
H^{(0,0)}& v_\mu, \mu=0,1  & \eta_-, \bar{\eta}_-  & \hbox{Vector}\cr
H^{(1,0)} \oplus H^{(0,1)} &  \bar{a}_{{m}}, {a}_{\bar m}  & \bar  {\psi}_{+ {m}} ,\psi_{+ \bar m}, & \hbox{Conjugate-chiral + Chiral  (Wilson lines)}\cr 
H^{(2,0)} \oplus H^{(0,2)} &  -&  {\rho}_{- {m}{n}} ,\bar\rho_{- \bar{m}\bar{n}}& \hbox{Fermi + Conjugate-Fermi}\cr 
H^{(3,0)} \oplus H^{(0,3)} & \varphi_{{k}{m}{n}}, {\bar\varphi}_{{\bar k}{\bar m}{\bar n}},  & 
{\chi}_{+ {k}{m}{n}} , \bar\chi_{+ \bar{k} \bar{m} \bar{n}}& \hbox{Chiral + Conjugate-chiral (deformations of $M_G$)}  \cr 
\end{array}
\ee

The subscripts $\pm$ denote the 2d chirality of the fermions. In the fourth column we have indicated how these degrees of freedom organize into $(0,2)$ multiplets according to the conventions set out in section \ref{sec:Review02}.
These assignments follow from the supersymmetry variations of the fields which will be presented in section \ref{BulkSUSY}.
In particular we are finding two types of chiral superfields in the present case given by
\be\ba
\Phi &= \varphi + \sqrt{2} \theta^+ \chi_+ - i \theta^+ \bar\theta^+(D_0 + D_1)\varphi \cr 
A &= a + \sqrt{2} \theta^+ \psi_+ - i \theta^+ \bar\theta^+(D_0 + D_1)a \,,
\ea
\ee
where $a$ corresponds to the internal components of the gauge field. 


\subsection{Massless Spectrum}

With no gauge field backgrounds turned on, all bulk multiplets transform in the adjoint representation of the 7-brane gauge group $G$. The spectrum  (\ref{table-matter1}) counts both all massless particles in the adjoint and their complex conjugate states in the same representation. The latter can be viewed as the associated anti-particles.
The independent massless states are counted by the cohomology groups 
\bea
H^{(0,p)}(M_G) =  H^0(M_G, \bar \Omega^p_{M_G}) = H^p_{\bar \partial}(M_G) \,.
\eea 
Let us introduce the notation 
\be \label{zeromodedecomp}
(\bar\varphi_{\bar k \bar m \bar n}) |_{\rm zero-mode} = \sum_{\kappa}  \bar \varphi^\kappa \otimes  \hat{\overline{\varphi}}_{\bar k \bar m \bar n, \kappa} \,,
\ee
with $ \bar \varphi^\kappa$ the 2d field associated with one of the ${\rm dim} H_{\bar\partial}^{3}(M_G)$ zero modes and   $\hat{\overline{\varphi}}_{\bar k \bar m \bar n, \kappa} $ the associated internal wavefunction.
A similar notation will be used for the other fields. We will suppress the `flavor index' $\kappa$ unless it is explicitly required.

The complex conjugate zero-mode multiplets are counted by the cohomology groups
\be
H^{(p,0)}(M_G) = \overline{H^0_{\bar \partial}(M_G, \bar \Omega^p)}   \ \equiv \  H^p_{\bar \partial}(M_G)^* \,, 
\ee 
which are the  complex conjugate of the cohomology groups $H^p_{\bar \partial}(M_G)$.

More generally, we can consider configurations with a non-trivial gauge background turned on along $M_G$. These configurations are described by a non-trivial principal gauge bundle $L$. Such gauge flux breaks the original gauge group $G$ into a product of  residual gauge groups $H_m$.
Correspondingly, the spectrum decomposes into irreducible representations ${\bf R}$ of unbroken gauge groups,
\be \label{adjointbreaking}
{ \rm \bf Adj}(G) \ \rightarrow \  \bigoplus_{\bf R} {\bf R}\,.    
\ee
These representations include the adjoint representation  ${ \rm \bf Adj}({H_m})$ of each remnant gauge group factor $H_m$. Reality of $ {\rm \bf Adj}(G)$ implies that  in (\ref{adjointbreaking}) every  complex representation ${\bf R} \neq {\bf \bar R}$ is accompanied by its conjugate representation ${\bf \bar R}$, and in this case the matter in ${\bf R}$ and ${\bf \bar R}$ is independent.
The independent massless matter states in representation ${\bf R}$ are counted by  the cohomology groups 
\be
H^{(0,p)}(M_G, L_{\bf R}) =H^p_{\bar \partial}(M_G,L_{\bf R}) \,,
\ee
for some vector bundle $L_{\bf R}$ which descends from the principal gauge bundle $L$. Their anti-particles are counted by the complex conjugate groups.
For  ${\bf R} \neq {\bf \bar R}$, there are independent matter states in the representation ${\bf \bar R}$ 
from the appearance of ${\bf \bar R}$ in (\ref{adjointbreaking}). Since $L_{\bf \bar R } = L_{\bf R }^* $ the latter are counted by 
\be \label{Rbarstates}
 H^{(0,p)}(M_G, L_{\bf R}^*) =H^p_{\bar \partial}(M_G,L_{\bf R}^*) \,,
 \ee
and their anti-particles  in representation ${\bf  R}$ are counted by the complex conjugate groups.
The massless fermionic bulk particles in representation ${\bf R}$ and their anti-particles are in summary accounted for by the following cohomology groups:
\be
\begin{array}{c|c} \label{cohomologies-bulk}
\hbox{Cohomology}  & \hbox{Fermions } \oplus    \hbox{Anti-Fermions }\cr\hline\vspace{-4mm} \cr 
H_{\bar \partial}^{0}(M_G, L_{\bf R}) \oplus H_{\bar \partial}^{0}(M_G, L_{\bf R})^*  & \bar \eta^{\bf R}_- \oplus \eta^{\bf \bar R}_- \\
H_{\bar \partial}^{1}(M_G, L_{\bf R}) \oplus H_{\bar \partial}^{1}(M_G, L_{\bf R})^*  & \psi^{\bf R}_+ \oplus \bar\psi^{\bf \bar R}_+   \\
H_{\bar \partial}^{2}(M_G, L_{\bf R}) \oplus H_{\bar \partial}^{2}(M_G, L_{\bf R})^*   &  \bar\rho^{\bf R}_-    \oplus \rho^{\bf \bar R} _-\\
H_{\bar \partial}^{3}(M_G, L_{\bf R}) \oplus H_{\bar \partial}^{3}(M_G, L_{\bf R})^*  & \bar\chi^{\bf R}_+  \oplus \chi^{\bf \bar R}_+  
\end{array}
\ee
Note again that e.g. the particles  $\psi^{\bf R}_+$ and $\bar\psi^{\bf \bar R}_+$ are just complex conjugate to each other.
For ${\bf R} \neq {\bf \bar R}$ there is an analogous table with $L_{\bf R}$ replaced by $L_{\bf R}^*$
for the states in representation ${\bf \bar R}$ and their anti-particles in representation ${\bf R}$.

According to the Hirzebruch-Riemann-Roch theorem the index  $\chi(M_G, L_{\bf R})$  takes the form
\be \label{chibulk1}
\ba
\chi(M_G,L_{\bf R}) &= h^0_{\bar \partial}(M_G, L_{\bf R}) - h^1_{\bar \partial}(M_G, L_{\bf R}) + h^2_{\bar \partial}(M_G, L_{\bf R})  -  h^3_{\bar \partial}(M_G, L_{\bf R})= \int_{M_G} {\rm ch}(L_{\bf R}) {\rm Td}(M_G) \cr
&=\frac{1}{24} \, {\rm rk}(L_{\bf R})\,  \int_{M_G} c_1(M_G)\,  c_2(M_G) +  \frac{1}{12}  \int_{M_G} c_1(L_{\bf R}) \left(c_1^2(M_G) + c_2(M_G)\right) \cr 
& \ +  \frac{1}{2}  \int_{M_G} {\rm ch}_2(L_{\bf R}) \, c_1(M_G) + \int_{M_G}{\rm ch}_3(L_{\bf R}) \,.
\ea\ee
Similarly, again for ${\bf R} \neq {\bf \bar R}$, 
 \be\ba
 \chi(M_G, L_{\bf R}^*) &= \frac{1}{24} \, {\rm rk}(L_{\bf R})\,  \int_{M_G} c_1(M_G)\,  c_2(M_G) -  \frac{1}{12}  \int_{M_G} c_1(L_{\bf R}) \left(c_1^2(M_G) + c_2(M_G)\right) \cr 
&   \qquad +\frac{1}{2}  \int_{M_G} {\rm ch}_2(L_{\bf R}) \, c_1(M_G) - \int_{M_G}{\rm ch}_3(L_{\bf R})\,,
\ea\ee
where we have used that  ${\rm ch}_k(L_{\bf R}^*) = (-1)^k \, {\rm ch}_k(L_{\bf R})$.


\subsection{Supersymmetry Variations and Hitchin Equations} \label{BulkSUSY}

The supersymmetry variations of the dimensionally reduced and partially topologically twisted 8d SYM theory are derived in appendix \ref{app:SUSYDetails}. We start with the 10d SYM {Lagrangian} 
\be \label{10dSYMaction}
{\cal L}_{10d}=  -{1\over 4 g^2}  \hbox{Tr} \left( F_{MN}F^{MN}\right) - {i \over 2 g^2} \hbox{Tr} \left(  \overline{\Psi} \Gamma^M D_M \Psi\right)\,,
\ee
whose associated action is invariant under the supersymmetry variations 
\be\ba
\delta A_M &=-i \bar\epsilon \Gamma_M \Psi \cr
\delta \Psi & = {1\over 2} F_{MN} \Gamma^{MN} \epsilon \,,
\ea\ee
and apply the dimensional reduction and twist as explained in section \ref{sec:Twist}. 
In terms of the twisted fields, the supersymmetry variations of the bosonic fields follow as 
\be
\ba
\delta \varphi_{kmn} &= - \sqrt{2} \epsilon_- \chi_{+\, kmn} \cr 
\delta a_{\bar{m}} &= - \sqrt{2 }\epsilon_- \psi_{+\, \bar{m}} \cr 
\delta (v_0 - v_1) & =  2i \epsilon_{-} \bar\eta_--2i \bar\epsilon_- \eta_- \,.
\ea
\ee
For the fermionic fields we find the variations
\be\label{BulkSusy}
\ba
\delta \bar\chi_{+\bar{k}\bar{m}\bar{n}} &= -i \sqrt{2} \epsilon_-D_+ \bar\varphi_{\bar{k}\bar{m}\bar{n}} \cr 
\delta \psi_{+\bar{m}} & = i \sqrt{2} \bar{\epsilon}_- ( D_+  a_{\bar{m}}   { - (\bar\partial_a)_{\bar m} v_+)  } \cr
&= i \sqrt{2} {\bar{\epsilon}_- }F_{\mu\bar{m}}\cr 
\delta \eta_- &= \epsilon_- (F_{01} + i \mathfrak{D}) \cr 
\delta \rho_{-mn} & = \epsilon_- \bar{F}_{mn} 
{-} \bar{\epsilon}_- ({\partial^\dagger_{a}  \varphi})_{mn} \cr 
\ea
\qquad \quad
\ba
\delta \chi_{+kmn} &=  i \sqrt{2}\bar\epsilon_- D_+ \varphi_{kmn} \cr 
\delta \bar\psi_{+{m}} & = - i \sqrt{2} {\epsilon}_-  (  D_+ \bar{a}_{m}  {- (\partial_{\bar a})_m v_+)} \cr
&=- i \sqrt{2} \epsilon_- \bar{F}_{\mu m}\cr
\delta \bar\eta_- &= \bar\epsilon_- (F_{01} -i \mathfrak{D}) \cr 
\delta \bar\rho_{-\bar{m}\bar{n}} & = \bar\epsilon_- {F}_{\bar{m}\bar{n}} 
				{-{\epsilon}_- ({\bar\partial^\dagger_{\bar{a}}  \bar\varphi})_{\bar{m}\bar{n}}}\,.
				\ea
\ee
Here we have defined the derivative $D_\pm= D_0\pm D_1$  as well as the $D$-term 
\be\label{DTerm}
\mathfrak{D} =-(F_{23} +F_{45}+ F_{67} - F_{89} ) \,.
\ee
Supersymmetric vacua are characterized in terms of the vanishing  of the fermions as well as their supersymmetry variations.
These BPS equations constrain both the internal profile of the fields and the field components in 2d. 
From $\delta_{\epsilon_-} \rho_-$ and $\delta_{\bar\epsilon_-} \bar{\rho}_-$ we obtain the condition that the field strength $F$ along the compact directions along $M_G$ must have no $(0,2)$ and $(2,0)$ components ${\bar{F}}_{mn} = {{F}}_{\bar{m}\bar{n}}=0$, i.e. the vacuum expectation values satisfy
\be \label{20Fterm}
 F^{(2,0)}= F^{(0,2)} =0 \,.
\ee
Similarly, the vacuum configuration $\varphi_{m n k}$ on $M_G$ is subject to the constraint
{$({\partial^\dagger_{a} \varphi})_{mn} 
 = ({\bar\partial^\dagger_{{\bar{a}}}  {\overline{\varphi}}})_{\bar{m}\bar{n}} =0$}, which is equivalent to
\be \label{daggerphi}
 {\bar\partial_{a} \varphi} 
 =0\,,\qquad  {\partial_{{\bar{a}}}  {\overline{\varphi}}}  =0 \,.
\ee
Note that our identification of $\rho_{-mn}$ as a Fermi as opposed to a conjugate Fermi field is a matter of convention and corresponds to identifying the holomorphic expression  $({\partial^\dagger_{a} \varphi})_{mn} $ as the Fermi $E$-auxiliary field in the off-shell formulation in agreement with (\ref{GeneralSUSY}). Alternatively one can view the expression $ {F}_{\bar{m}\bar{n}}$, which is holomorphic in the chiral superfields, as the $E$-field, thereby exchanging the role of $\rho$ and $\bar \rho$.

The variations of $\psi$ and $\chi$ result in the BPS equations 
\be \label{BPSvplus}
D_+ \varphi = D_+ \bar\varphi =0 \,,\qquad 
D_+ a {- (\bar\partial_a) v_+} = D_+ \bar{a}{- (\partial_{\bar a}) v_+ }  =0 
\ee
Regarding the D-term, note that 
$F_{8,9} = [\Phi_8, \Phi_9] = {i\over 2} [\varphi, \bar\varphi]$. For the remaining terms, 
let $J$ be the K\"ahler form of the three-fold $M_G$, whereby with our choice of coordinates and metric $J^{m\bar{n}}= i g^{m\bar{n}}$  we can write in holomorphic coordinates $2 z_m = \{ x_2 + i x_3, x_4 + i x_5 , x_6 + i x_7\}$  
\be
-\mathfrak{D} = g^{m \bar{n}}  F_{m \bar{n}}  -  {i\over 2} [\varphi,{\overline{\varphi}}] \,.
\ee
With the help of the identity
\be
i g^{m \bar{n}}  F_{m \bar{n}}   = J \wedge \star  F_{M_G}  = (\star J) \wedge  F_{M_G} = {1\over {(n-1)!}} J^{n-1} \wedge  F_{M_G}\,,
\ee
with $n=3$ for $M_G$, the $D$-term becomes
\be  \label{DtermBPS}
\mathfrak{D} = {i\over 2} \left( J \wedge J \wedge  F_{M_G} + [ \varphi, {\overline{\varphi}}]\right) \,.
\ee
The resulting $D$-term condition for the BPS vacuum is  
\be \label{Dtermzero}
J\wedge J \wedge F_{M_G} + [\varphi, {\overline{\varphi}}] =0 \,,
\ee
and generalizes the Hitchin equation \cite{Hitchin} from compactifications of 4d SYM on a Riemann surface to 8d SYM on a  complex three-dimensional K\"ahler cycle. 

A background satisfying (\ref{20Fterm}), (\ref{daggerphi}), and  (\ref{Dtermzero})   gives rise to a 2d $(0,2)$ supersymmetric gauge theory. 
In this theory, the supersymmetry transformations of the 2d bosonic field {fluctuations around the vacuum values} take the form 
\be\ba
 \delta \varphi   &=   - \sqrt{2} \,   \epsilon_- \chi_+ \cr 
 \delta a &=  -\sqrt{2} \, \epsilon_- \psi_+ \cr 
\delta v_0 = - \delta v_1& =  i\epsilon_- \bar\eta_- -i\bar\epsilon_-  \eta_- 
\ea
\qquad 
\ba
 \delta \bar \varphi     &=   + \sqrt{2} \, \bar\epsilon_-  \bar\chi_+\cr
 \delta \bar{a} &= + \sqrt{ 2} \,  \bar\epsilon_- \bar\psi_+ \cr 
 \phantom{.} & \phantom{.}
\ea
\ee
and those of the fermion variations are
\be
\ba
 \delta \bar\chi_+ &  =   - i \sqrt{2} \epsilon_- (D_0 + D_1) \bar{\varphi} \cr 
     \delta \psi_+ &=  i \sqrt{2}\bar\epsilon_- (D_0 + D_1) \, a \cr 
\delta \eta_- & =  \epsilon_-   F_{01} \cr 
 \delta \rho_{-} &=  0 
\ea 
\qquad 
\ba
 \delta \chi_+   &= i \sqrt{2}\bar\epsilon_- (D_0 + D_1)\varphi \cr 
 \delta \bar\psi_+ &=- i \sqrt{2} \epsilon_- (D_0 + D_1) \bar{a} \cr 
  \delta \bar\eta_-  &=  \bar\epsilon_-  F_{01} \cr 
\delta \bar\rho_{-}&= 0 \,,
\ea
\ee
{where we use that $ (\bar\partial_a) v_+= (\partial_{\bar a}) v_+=0  $ in the transition to the 2d effective action.} The 2d supersymmetry variations are in agreement with the general form  (\ref{GeneralSUSY}) of the supersymmetry variations for the chiral and Fermi multiplets and  justify our identification of the 2d superfields.  
In particular, since we are imposing  (\ref{20Fterm}) and   (\ref{daggerphi}) as part of the defining properties of the vacuum, the auxiliary fields $G^{(\bar\rho_-)}$ and $E^{(\bar{\rho}_-)}$ vanish at this level. 


\subsection{Higgs bundles and Hitchin Systems}

The solutions to the  $F$- and $D$-term equations are generalizations of Hitchin equations for a Higgs bundle 
$(A, \Phi)$ over the complex three-cycle $M_G$
 with the following properties
\be\ba \label{BPS-Hitchin}
F^{(2,0)}=  F^{(0,2)} &=0 \cr 
D_+ \varphi = D_+ \bar\varphi =D_+ a {- (\bar\partial_a) v_+} = D_+ \bar{a}{- (\partial_{\bar a}) v_+ }  & =0  \cr 
 \bar\partial_{a} \varphi =\partial_{{\bar{a}}}  {\overline{\varphi}}  &=0 \cr 
{ J \wedge J \wedge F + [\varphi, {\overline{\varphi}}] } & =0 \,.
\ea
\ee
Put differently, the BPS vacua of the twisted 8d SYM theory can be given an interpretation in terms of a gauge field configuration defined by a bundle with connection $A$ and an adjoint-valued Higgs field $\varphi$. 
These take values in a higher rank gauge algebra $\tilde{\mathfrak{g}}\supset \mathfrak{g}$ which contains the gauge algebra $\mathfrak{g}$ of the 2d gauge theory. The F-term conditions ensure holomorphy of the Higgs bundle, whereas the D-term equations are stability conditions. One important caveat is that this approximation in terms of a gauge theory is exact in the limit when the volume of the three-cycle $M_G$ is large, and the stability condition is expected to receive corrections beyond this. 

The first note-worthy point is that this characterization holds  for 7-branes in {\it any} F-theory compactification\footnote{Whenever a heterotic dual exists, the corresponding spectral cover description of the Higgs bundle maps to the spectral cover of the heterotic vector bundle. But this is in no way a necessary condition for a local spectral cover description to exist. For an in depth discussion of the duality from this point of view see \cite{Hayashi:2008ba}.}. 
The Higgs bundle encodes the local geometry of $M_G$ embedded into the five-fold in terms of a local ALE-fibration over $M_G$: the $(1,1)$-forms in the ALE fiber associate the deformations of  the complex structure $\Omega^{5,0}$ to the Higgs field vevs in the Cartan  subalgebra (CSA) of the gauge algebra
\be
\delta \Omega^{5,0} = \sum_{\rm CSA} \omega^{(1,1)}_i \wedge \varphi_i \,,
\ee
and the gauge field configurations arise from the three-form $C_3$. The simplest class of solutions have $\varphi=0$, resulting in flat gauge fields. The second simplest class {has non-trivial $\varphi$, with} $[\varphi, \bar\varphi]=0$, in which case the vacua can be characterized in terms of the spectral data of the Higgs field. The spectral cover defined as $\det (\lambda {\bf 1} - \varphi) =0$ is an $n$-sheeted covering of $M_G$. Likewise, the gauge bundle can be constructed from line bundles over the spectral cover, and in the case of four-folds has been discussed in much detail e.g. in \cite{Donagi:2009ra, Marsano:2009gv}. The local geometry defined by the Higgs bundle allows in particular now to transition from the gauge theoretic description of the 7-branes to a full geometric construction of the Calabi-Yau five-fold. More specifically, the coefficients in the spectral cover have a close relation to the coefficients in the description of the elliptic fibration in terms of the so-called Tate form. {Developing the spectral covers for these generalized Higgs bundles certainly deserves further consideration in the future. }


\subsection{Supersymmetric Bulk Couplings} \label{sec_BulkCouplings}

The supersymmetric couplings in a general $(0,2)$ theory have been reviewed in section \ref{sec:Review02} and can take the form $L^J$ and $L^F$ summarized in (\ref{LJint}) and (\ref{LFint}), respectively. 
In our context, cubic Yukawa type couplings descend from the second term in the gauge interaction (\ref{10dSYMaction}) of the 10d SYM from which we have obtained the $(0,2)$ 2d theory by reduction and twisting.
This can be seen explicitly by plugging the decomposition of the 10d gaugino and the 10d gauge field into the interaction term  (\ref{10dSYMaction}). 
From the perspective of the theory prior to twisting, the resulting couplings realize the different possibilities of forming a singlet with respect to the structure group $U(3)$ of the K\"ahler three-cycle $M_G$.
Those interaction terms involving the 2d gaugino are part of the 2d SYM interactions. The remaining ones are actual Yukawa couplings.

By decomposition we find two possible types of such Yukawa terms. 
The first type of Yukawas corresponds to the existence of a {$U(3)$-invariant interaction} ${\bf 1} \otimes {\bf 3} \otimes {\bf \bar 3}$.
From the perspective of the twisted theory this translates into the possibility of forming a $(3,3)$ form on $M_G$ from the internal wavefunctions, which can then be integrated to obtain the coupling.
Inspection of the form degrees of the internal wavefunctions reveals that the only possible cubic interaction of this type is of the form  (\ref{LFint}) and given by 
\be \label{Sbulk1}
S^{(F)}_{\rm bulk} = {\bf f}_{\alpha \mu \epsilon}  \int  d^2 y  \,    \,  \bar{\rho}_{-}^{\alpha}  \, \left(  \, \varphi^\mu\,  \psi_{+}^\epsilon  + 
 \,  \chi_+^\mu \, a^\epsilon \, \right) + {\rm c.c.} 
\ee
with couplings
\be \label{Sbulk2}
 {\bf f}_{\alpha \mu \epsilon}    = \int_{M_G}  \hat{\overline{\rho}}_{\bar{k}\bar{m},\alpha} \wedge \left(  \hat \varphi_{kmn,\mu} \wedge  \hat \psi_{ \bar{n},\epsilon} \right)   \,,
 \ee
in an expansion of the form (\ref{zeromodedecomp}). 
We are suppressing gauge indices and a gauge invariant contraction of the involved representations is understood.
Algebraically, this way of taking the overlap of the internal wavefunction corresponds to the canonical map
\be
H_{\bar\partial}^2(M_G) \,  \times \, H_{\bar\partial}^1(M_G) \,\times \, H_{\bar\partial}^0(M_G, K_{M_G})  \,  \longrightarrow  \, H_{\bar\partial}^3(M_G,K_{M_G}) \,  \cong \, \mathbb C \,,
\ee
where the last step uses  the identification $H_{\bar\partial}^3(M_G,K_{M_G}) = H^{3,3}(M_G)$, which can be integrated over $M_G$.
The first two cohomology groups count the zero modes $\hat{\overline{\rho}}_{\bar{k}\bar{m},\alpha}$ and  $\hat \psi_{ \bar{n},\epsilon}$ and the third counts $\hat\varphi_{kmn,\mu} \in H_{\bar\partial}^3(M_G)^* =  H_{\bar\partial}^0(M_G, K_{M_G}) $ (or the respective superpartners), {as summarized in (\ref{cohomologies-bulk})}. 
The interaction being of the form $L^F$ it induces a modification of the Fermi auxiliary $E$-field as
\be \label{Erhobulk}
E^{(\rho^\alpha_-)} = -  {\bf f}_{\alpha \mu \epsilon} \,  \Phi^\mu A^\epsilon \,.
\ee

Interestingly, there exists another type of Yukawa couplings, which group theoretically realizes the existence of the singlet ${\varepsilon_{\alpha \beta \gamma} {\bf 3}^\alpha {\bf 3}^\beta {\bf 3}^\gamma }$ with respect to the structure group $SU(3) \subset SU(3) \times U(1)_{twist}$  acting on the tangent bundle of $M_G$. 
By {dimensional reduction of the 10d SYM interactions, we find} that this corresponds to a superpotential coupling
\be \label{bulk-J-coupling}
S_{\rm bulk}^{(J)} = {\bf g}_{\alpha \beta \gamma} \int  d^2 y \, \,    {\rho}_{-}^{\alpha}  \,   a^{\beta} \,  \psi^{\gamma}_{+}  + \rm{c.c.} 
\ee
with
\be \label{gcoupling}
{\bf g}_{\alpha \beta \gamma} =  \int_{M_G}   \tilde{{\rho}}_{{k}{m}n\bar n, \alpha} \wedge   \hat a_{\bar k,\beta} \wedge  \hat \psi_{ \bar{m},\gamma}  \, .
\ee
Again we are suppressing the suitably contracted gauge indices.
Here
\be
 \tilde{{\rho}}_{{k}{m}n\bar n, \alpha} =  \left( \Omega \cdot \hat \rho_\alpha \right)_{{k}{m}n\bar n}  
\ee
is the element of $H_{\bar\partial}^1(M_G,K_{M_G})$  obtained  from the $(2,0)$ form $\hat{{\rho}}_{{k}{m}, \alpha}$  by contraction with the   $(3,3)$ form $\Omega$ on $M_G$. 
Indeed, by Serre duality
\be
H_{\bar\partial}^{2}(M_G)^* = \Big( H_{\bar\partial}^{1}(M_G, K_{M_G})^\vee\Big)^* \cong H_{\bar\partial}^1(M_G,K_{M_G}) \,.
\ee
In the sequel we will usually omit the tilde when we apply operations of this form.  Such a coupling realizes  the canonical map
 \be
 H_{\bar\partial}^1(M_G) \,  \oplus \, H_{\bar\partial}^1(M_G) \,  \oplus \, H_{\bar\partial}^1(M_G,K_{M_G}) \, \longrightarrow \, H_{\bar\partial}^3(M_G,K_{M_G})  \, \,  {\cong} \, \,  \mathbb C \,.
 \ee
The superpotential associated with (\ref{bulk-J-coupling}) is 
\be\label{BulkJ}
J_{(\rho^\alpha_-)} =  -  {\bf g}_{\alpha \beta \gamma}  \, A^\beta \, A^\gamma\,.
\ee
Note that this coupling is only quadratic in the fields. In $(0,2)$ theories that arise from $(2,2)$ supersymmetric ones by deformation, it is known \cite{Witten:1993yc} that $J= \partial_{\Phi} \mathcal{W}$, where $\mathcal{W}$ is a general gauge invariant holomorphic function of the chiral superfields corresponding to the superpotential of the $(2,2)$ theory. 
In a GLSM interpretation of the $(0,2)$ theory, {the locus} $J=0$ determines the target space of the heterotic string as a hypersurface in an ambient space (as well as part of the gauge bundle data), and the form of $J$ is thus of quite some importance.
In this paper, we started our analysis with the 8d SYM theory, taking only the `renormalizable' couplings with us induced by the gauge kinetic terms in 8d. Including higher order terms obtained by integrating out massive fields, as well as non-perturbative contributions, we expect more general couplings to be generated in the effective theory in 2d. In particular, this should give rise to more general GLSMs with non-trivial target manifolds. This will be discussed in more depth in section \ref{sec:SCFTGLSM}. 

Finally, we should address the supersymmetry condition ${\rm Tr} \,  E\cdot J =0$ (see (\ref{JEzero})). Both $E$- and $J$-couplings arise from the kinetic terms in the 8d SYM action upon dimensional reduction. The off-shell action of the dimensionally reduced 2d theory will be determined in \cite{TBA1}. Supersymmetry of the 2d theory, which follows from the higher dimensional supersymmetry, combined with the twisted reduction implies that the couplings ${\bf f}$ and ${\bf g}$ cannot be independent but have to be such that ${\rm Tr} \,  E\cdot J=0$. 
The condition in terms of component fields reads
\be
\Tr E \cdot J= f^{ijk} f^{ilm} \, {\bf f}_{\alpha \beta\gamma} {\bf g}_{\alpha \delta \epsilon}  \,  \Phi_{j}^{\beta} A_{k}^{\gamma} A_l^\delta A_m^\epsilon =0 \,,
\ee
where we have now made the gauge algebra indices $i,j,\ldots$ of the adjoint valued fields manifest and $f^{ijk}$ are the structure constants. One would indeed expect the geometry to imply the condition $\Tr E\cdot J=0$ automatically, and it would be interesting to find its precise geometric origin.


\section{Matter from the 6d Defect Theory}
\label{sec:Matter}

Additional matter arises from defects in the 8d SYM theory. Such defects correspond to intersections of the 7-brane stack on $M_G$ with flavor 7-branes wrapping different cycles. Two K\"ahler three-cycles inside the base $B_4$ of our F-theory compactification generically intersect over a  K\"ahler surface {$S_{\bf R}\subset M_G$}, along which such matter will therefore be localized. 
The theory living on such a defect is an $N=(1,0)$ 6d SYM theory with an $SU(2)$ R-symmetry. We will couple this theory to the bulk theory by performing a topological twist compatible with $(0,2)$ supersymmetry in two dimensions. 
As in F-theory compactifications to four dimensions \cite{Donagi:2008ca, Beasley:2008dc, Beasley:2008kw}  one can think of this theory as a gauge theory with enhanced gauge symmetry due to the collision of the two 7-brane {stacks}. Extra degrees of freedom due to 
{generically} multi-pronged strings stretched between both branes localize on $S_{\bf R}$ and 
give rise to additional matter charged under the 7-brane gauge group. In terms of the Higgs bundle, the matter surfaces are characterized by the vanishing of sections associated to $\varphi$, i.e. sections of $K_{M_G}$. These are precisely the loci where some of the Higgs field vevs vanish and the gauge algebra is locally enhanced, thus resulting in matter through Higgsing the adjoint of the higher-dimensional gauge algebra to 
$\mathfrak{g}$.
After specifying the topological twist {along $S_{\bf R}$}, we will now determine this charged matter, along with its $E$- and $J$-interactions both with the bulk matter and the interactions of surface matter only.


\subsection{Spectrum of Matter Fields} \label{sec_S2Matter}

We adopt the convention that the supercharges of 6d $N=(1,0)$ supersymmetry transform as a $({\bf 4}, {\bf 2}_R)$ under $SO(1,5)_L \times SU(2)_R$ {(see e.g. \cite{Nishino:1997ff})}.
The associated supersymmetry parameters then transform as a $({\bf \bar 4} ,{\bf 2}_R)$. 
The vector fields of the 6d theory will be identified with the restriction to $S_{\bf R}$ of the vector fields on the two intersecting 7-brane stacks. 
Extra matter states from strings localised on $S_{\bf R}$  organize into a hypermultiplet   in the 6d SYM theory in representation ${\bf R}$ of the gauge group.  
With the above choice of supersymmetry parameters the fermions in the hypermultiplet transform as 
$({\bf 4}, {\bf 1}_R)$ and the scalars as $({\bf 1}, {\bf 2}_R)$. 

In coupling this theory to the 7-brane bulk theory we identify the $R$-symmetry obtained from the latter with a $U(1)_R$ subgroup of $SU(2)_R$. 
Upon compactification on the complex K\"ahler two-cycle $S_{\bf R}$,  $SO(1,5)_L$ decomposes  into  $SO(1,1)_L \times SU(2) \times U(1)_L$, where the naive internal tangent bundle structure group $SO(4)$ is reduced to $U(2) \simeq SU(2) \times U(1)_L$ due to K\"ahlerity of $S_{\bf R}$.
The decomposition of the 6d supersymmetry parameters and of the hypermultiplet then yields the following supersymmetry parameters and matter content in two dimensions:
\be \label{tablelocmatter}
\ba
SU(2)_R \times SO(1,5)_L & \quad \rightarrow \quad U(1)_R \times \left(SU(2)\times U(1)_L \times SO(1,1)_L \right) \cr 
( {\bf 2}, \bar{\bf 4}) & \quad \rightarrow \quad   
 ({\bf 1}_{+1} \oplus {\bf 1}_{-1})\otimes \left(  {\bf 1}_{+1,-1} \oplus {\bf 1}_{-1,-1} \oplus {\bf 2}_{0,+1} \right)  \cr 
({\bf 1}, {\bf 4} )  & \quad \rightarrow \quad  {\bf 1}_{0,+1,+1} \oplus {\bf 1}_{0,-1,+1} \oplus \bar{{\bf 2}}_{0,0,-1} \equiv (\bar\sigma_+, \tau_+, { \bar\mu_-} )\cr 
( {\bf 2}, {\bf 1}) & \quad \rightarrow \quad  {\bf 1}_{-1,0,0} \oplus {\bf 1}_{+1,0,0} \equiv  (\bar S, T) \,.
\ea
\ee
In order for the theory on $S_{\bf R}$ to preserve the same supersymmetries as the twisted bulk theory, it must be topologically twisted in such a way  that two negative chirality scalar supersymmetry parameters  transform as singlets under the twisted $U(1)$.
For the choice  
\be \label{Jtwistdefect}
J_{twist} =    J_{U(1)_L} - J_{U(1)_R}  \,,
\ee
the spinors ${\bf 1}_{+1,+1,-1}\oplus {\bf 1}_{-1,-1,-1}$ from the first line have the desired property. 
Their $R$-charges identify these as the supersymmetry parameters $\bar\epsilon_-$ and $\epsilon_-$ of $R$-charge $+1$ and $-1$ in the 2d $(0,2)$ theory (see (\ref{epsidbulk}))
\be \label{epsid6D} 
\bar\epsilon_- = {\bf 1}_{+1,+1,-1} \,,\qquad \quad  \epsilon_-  = {\bf 1}_{-1,-1,-1} \,.
\ee
The decomposition of the hypermultiplet fermion in (\ref{tablelocmatter}) gives rise to two positive-chirality fermions $\bar\sigma_+$ and $\tau_+$ and one negative-chirality fermion $\bar\mu_-$. From the scalar superpartners we obtain two complex scalars $\bar S$ and  $T$.
As we will see below, the fields $(T,\tau_+)$ and $(\bar S, \bar \sigma_+)$ organize into a chiral superfield ${\cal T}$ and, respectively, a conjugate chiral multiplet $\bar{\cal S}$, { while $\bar \mu_-$ forms the lowest component of a conjugate Fermi multiplet.}

To identify the cohomology groups associated with these multiplets, note first that, as in the bulk theory, a section of  $\Omega^{(0,q)}(S_{\bf R})$ has twist charge $q \geq 0$.
That is, sections of  $\Omega^{(0,1)}$ are being identified with sections of the holomorphic tangent bundle.
If a field transforms as a spinor on $S_{\bf R}$, its twist charge receives an extra contribution of $-1$ from each factor of the spin bundle $K_{S_{\bf R}}^{1/2}$.\footnote{This can be seen by locally decomposing the tangent bundle of the surface $S_{\bf R}$ as $T_{S_{\bf R}}  = T_1 \oplus T_2$ via the splitting principle, see e.g. Appendix A of \cite{Donagi:2008ca}. This corresponds to viewing $S_{\bf R}$ locally as a product of two complex curves.
In one complex dimension, massless Dirac spinors transform as sections of  $K^{1/2} \oplus K^{-1/2}$ with $K^{1/2} = T^{-1/2}$. 
Identifying sections of the tangent bundle $T$  with fields of twist charge $+1$, sections of  $K^{1/2}$ then carry twist charge $- \frac{1}{2}$ in one complex dimension.
Using the splitting principle massless spinors on the surface $S_{\bf R}$ transform as sections of 
$(K_1^{{1}/{2}} \oplus K_1^{-{1}/{2}}) \otimes (K_2^{{1}/{2}} \oplus K_2^{-{1}/{2}})$. The summands $K_1^{-1/2} \otimes K_2^{-1/2}$, $K_1^{-1/2} \otimes K_2^{1/2} \oplus K_1^{1/2} \otimes K_2^{-1/2}$ and $K_1^{1/2} \otimes K_2^{1/2}$ carry twist charge $1$, $0$ and $-1$, respectively.}

It is therefore consistent to interpret, in absence of gauge flux, the fermions  $(\bar\sigma_+  , \tau_+,\bar \mu_-)$ appearing in (\ref{tablelocmatter})  with twist charges ($1$, $-1$, $0$) as elements of $H_{\bar \partial}^2(S_{\bf R},  \sqrt{K_{S_{\bf R}}})$, $H^0_{\bar \partial}(S_{\bf R},  \sqrt{K_{S_{\bf R}}})$ and $H^1_{\bar \partial}(S_{\bf R},\sqrt{K_{S_{\bf R}}})$, respectively. This also fits with the twist charges of the scalar superpartners  $T$ and $\bar S$.
The above assignments  lead to a consistent spectrum and are also in perfect agreement with the embedding of the 6d defect into the 8d bulk theory as {will be} discussed momentarily.

In order for this interpretation to make sense we are assuming that, in absence of gauge flux, the K\"ahler surface $S_{\bf R}$ is spin, $c_1(K_{S_{\bf R}}) \equiv 0 \in H^2(S_{\bf R}, \mathbb{Z}_2)$, such that the spin bundle $\sqrt{K_{S_{\bf R}}}$ is well-defined as an honest line bundle. The requirement of $S_{\bf R}$ being spin is modified in the presence of a non-trivial gauge bundle.
Indeed, suppose the 6d hypermultiplet transforms as a representation ${\bf R}$ of the bulk gauge group. For non-zero gauge flux each field in representation ${\bf R}$ is  valued in a bundle $L_{\bf R}$.
Then 
$H_{\bar \partial}^0(S_{\bf R}, L_{\bf R} \otimes \sqrt{K_{S_{\bf R}}})$ and $H_{\bar \partial}^2(S_{\bf R},  L_{\bf R} \otimes \sqrt{K_{S_{\bf R}}})$ respectively count  chiral multiplets $(T,\tau_{+})_{\bf R}$ and conjugate chiral multiplets $(\bar S, \bar \sigma_{+})_{\bf R}$ in representation ${\bf R}$, while $H_{\bar \partial}^{1}(S_{\bf R}, L_{\bf R} \otimes  \sqrt{K_{S_{\bf R}}})$ counts conjugate Fermi multiplets with lowest component $\bar \mu_{-}$ in representation ${\bf R}$. The complex conjugate cohomology groups can be determined using Serre duality as follows
\be \label{conjugate-dual}
 H_{\bar \partial}^{i}(S_{\bf R},     L_{\bf R} \otimes  \sqrt{K_{S_{\bf R}}} )^*  =  \left(H_{\bar \partial}^{2-i}(S_{\bf R}, L^*_{\bf R} \otimes  \sqrt{K_{S_{\bf R}}} )^\vee\right)^\ast      \, { \cong }  \, H_{\bar \partial}^{2-i}(S_{\bf R}, L^*_{\bf R} \otimes  \sqrt{K_{S_{\bf R}}} )  \,,
\ee
and count the  respective  anti-particles in representation ${\bf \bar R}$. 
The structure of the massless localised  spectrum can then be summarized as follows:
\be \label{table-matterspec} 
\begin{array}{c|c}
\hbox{Cohomology}  & \hbox{Fermions } \oplus    \hbox{Anti-Fermions }\cr \hline \vspace{-4mm} \cr 
H_{\bar \partial}^{0}(S_{\bf R},     L_{\bf R} \otimes  \sqrt{K_{S_{\bf R}}} ) \oplus H_{\bar \partial}^{0}(S_{\bf R},   L_{\bf R} \otimes  \sqrt{K_{S_{\bf R}}} )^*            &  \tau^{\bf R}_+     \oplus   \bar\tau^{\bf \bar R}_+ \cr
H_{\bar \partial}^{1}(S_{\bf R},     L_{\bf R}  \otimes  \sqrt{K_{S_{\bf R}}} ) \oplus H_{\bar \partial}^{1}(S_{\bf R},   L_{\bf R} \otimes  \sqrt{K_{S_{\bf R}}} )^*  &         \bar \mu^{\bf R}_-       \oplus          \mu^{\bf \bar R}_-      \cr
H_{\bar \partial}^{2}(S_{\bf R}, L_{\bf R} \otimes  \sqrt{K_{S_{\bf R}}} ) \oplus H_{\bar \partial}^{2}(S_{\bf R},   L_{\bf R} \otimes  \sqrt{K_{S_{\bf R}}} )^*        &            \bar\sigma^{\bf R}_+      \oplus     \sigma^{\bf \bar R}_+    \cr
\end{array}
\ee
In general only the bundle $L_{\bf R} \otimes \sqrt{K_{S_{\bf R}}}$ must be well-defined as an integer quantized bundle even if both factors individually may not be.
This must be guaranteed in a globally consistent F-theory compactification by the tadpole constraints and the Freed-Witten quantization condition on the gauge fluxes.

For a smooth surface $S_{\bf R}$ the chiral index $\chi(S_{\bf R}, {\bf R})$ is computed via the Hirzebruch-Riemann-Roch theorem as
\be\ba  \label{chiS2}
\chi(S_{\bf R}, {\bf R}) &= h_{\bar \partial}^{0}(S_{\bf R},     L_{\bf R} \otimes  \sqrt{K_{S_{\bf R}}} ) - h_{\bar \partial}^{1}(S_{\bf R},     L_{\bf R} \otimes  \sqrt{K_{S_{\bf R}}} ) + h_{\bar \partial}^{2}(S_{\bf R},     L_{\bf R} \otimes  \sqrt{K_{S_{\bf R}}} ) \cr
&= \int_{S_{\bf R}} \left( \frac{1}{12} \left(c_1(S_{\bf R})^2 + c_2(S_{\bf R})\right) + \frac{1}{2} c_1(S_{\bf R})  \, c_1(L_{\bf R} \otimes {K_{S_{\bf R}}^{1/2}})  + {\rm ch}_2 (S_{\bf R}, L_{\bf R} \otimes {K_{S_{\bf R}}^{1/2}})\right)  \\
                      &= \int_{S_{\bf R}}  \left(c_1^2(S_{\bf R}) \left(\frac{1}{12} - \frac{1}{8} {\rm rk}(L_{\bf R}) \right) + \frac{1}{12} c_2(S_{\bf R}) + \left(\frac{1}{2} c_1^2(L_{\bf R}) - c_2(L_{\bf R})\right) \right).       
\ea\ee
Note that the appearance of only even powers of $c_1(L_{\bf R})$ ensures that $\chi(S_{\bf R}, {\bf R}) = \chi(S_{\bf R},{\bf \bar R})$, where the latter is defined in terms of the conjugate gauge bundle $L_{\bf R}^*$.
This expression, which is valid a priori for smooth matter surfaces, receives corrections in the presence of singularities, as will be discussed in section \ref{sec:chiSing}.

Consistency of this spectrum with the bulk spectrum can be seen as follows. 
From the perspective of the theory on $M_G$, the surface $S_{\bf R}$ can be viewed as a defect, and the surface matter corresponds to zero-modes trapped along this defect.
The defect zero modes are related to the bulk field zero modes (\ref{cohomologies-bulk}) in the same way as described  in  \cite{Beasley:2008dc} for a one-dimensional defect inside a surface wrapped by a 7-brane.
In this correspondence, the fields whose bulk zero modes  transform in $H^1(M_G)$ give rise to defect zero modes transforming as sections of the normal bundle  $N_{S_{\bf R}/M_G}$  of the matter surface in the divisor $M_G$.  As explained at the beginning of this section, the matter surfaces $S_{\bf R}$ are loci characterized by an enhanced gauge group, i.e. vanishing of Higgs vevs $\langle\varphi\rangle$. These are sections of the canonical class $K_{M_G}$ of $M_G$. Thus the normal bundle of $S_{\bf R}$ in $M_G$ is isomorphic to $K_{M_G}$. Together with adjunction 
\be
K_{S_{\bf R}} = K_{M_G}|_{S_{\bf R}} \otimes N_{S_{\bf R}/M_G} \,,
\ee
this yields $N_{S_{\bf R}/M_G} = K_{S_{\bf R}}^{1/2}$ \cite{Beasley:2008dc}.
This results in the `identifications'
\be \label{bulk-surface-identification}
\ba
\psi_+ \in H^1_{\bar\partial} (M_G, L_{\bf R} )  \quad &\rightarrow  \quad  \tau_+ \in H^0_{\bar\partial}  (S_{\bf R}, L_{\bf R} \otimes K_{S_{\bf R}}^{1/2})\cr 
\bar\rho_- \in H^2_{\bar\partial} (M_G, L_{\bf R})   \quad &\rightarrow  \quad  \bar \mu_-\in H^1_{\bar\partial}  (S_{\bf R}, L_{\bf R} \otimes K_{S_{\bf R}}^{1/2})  \cr 
\bar\chi_+ \in H^3_{\bar\partial} (M_G, L_{\bf R})   \quad &\rightarrow  \quad \bar\sigma_+ \in  H^2_{\bar\partial}  (S_{\bf R}, L_{\bf R}\otimes K_{S_{\bf R}}^{1/2})\,,
\ea
\ee
 in agreement with the spectrum (\ref{table-matterspec}) obtained through the twisted defect theory.
 
Finally, note that the specific representation ${\bf R}$ in which the defect matter transforms can be deduced geometrically as described in section \ref{sec:Singularfibers}, but a priori it seems that there is an ambiguity
in assigning matter the representation ${\bf R}$ as opposed to its conjugate ${\bf \bar R}$. 
This ambiguity is merely a matter of convention because changing ${\bf R}$ and ${\bf \bar R}$  exchanges the role of the two independent chiral superfields ${\cal S}$ and ${\cal T}$ as well as of the Fermi field  $\mu_-$ and its conjugate, thereby exchanging the role of the $E$ and $J$-type couplings associated with $\mu$.


\subsection{SUSY variation and BPS equations}

To prove that the fermionic and scalar fields organize into 2d $(0,2)$ superfields as claimed above we must decompose the 6d (1,0) supersymmetry variation taking into account the identification (\ref{epsid6D}). 
The 6d SUSY variation of the hypermultiplet fermions $\Psi$ transforming as $({\bf 4},{\bf 1})$ of $SO(1,5) \times SU(2)_R$ is (see e.g. \cite{Nishino:1997ff})
\be
\delta \Psi =- i\sqrt{2} \,  \bar \epsilon_A   \gamma^\mu  D_\mu \, \Phi_B  \varepsilon^{AB} \,.
\ee
The subscripts $A, B=1,2$ refer to the $SU(2)_R$ symmetry representation of the 6d supersymmetry parameters $\epsilon_A$ transforming in a $({\bf \bar 4},{\bf 2})$ and of the hypermultiplet scalars $\Phi_B$ transforming as the $({\bf 1}, {\bf 2})$, and $\varepsilon^{AB}$ is the anti-symmetric tensor.
 After applying the decomposition (\ref{tablelocmatter}) one finds, much like the analysis in appendix \ref{app:Gamm},
\be
\ba
\delta \tau_+ & =  i \sqrt{2} \, (D_{0} + D_1) T \, \bar{\epsilon}_- \cr 
\delta \bar\sigma_+ &= - i \sqrt{2} \, (D_0 + D_1) \bar{S} \, \epsilon_-   \cr 
\delta \bar\mu_-^{\underline{\dot\alpha}} &=  
\sqrt{2} i  \left(
\bar\epsilon_- \, \epsilon^{\underline{\dot\alpha} \underline{\dot\beta}}  D_{\underline{\dot{\beta}}} T -
\epsilon_- \, \bar{D}^{\underline{\dot\alpha}} \bar{S}\right)  \,.
\ea
\ee
These variations are expressed in terms of the $2+4$-dimensional fields, which {for simplicity} we denote by the same symbol as their 2d components. 
In this spirit the index $\underline{\dot\alpha} =1,2$ refers to the doublet structure of $\bar \mu_-$ under the internal 
$SU(2)$-structure group, {as is clear from} (\ref{tablelocmatter}).
The BPS equations are obtained by separately setting to zero the fermionic variations with respect to $\epsilon_-$ and $\bar \epsilon_-$.
{The vacuum expectation values have to satisfy}
 the BPS equation 
\be \label{BPSeq-surface}
\bar\partial_A T = 0\,,  \qquad \partial_A {\bar{S}} = 0 \,.
\ee
Solutions to these equations describe the string vacuum which gives rise to the  effective $(0,2)$ supersymmetric theory in 2d.
In this theory, the 2-dimensional components of the scalars are furthermore subject to the BPS equations
\be
(D_{0} + D_1) T =0\,, \qquad (D_0 + D_1) \bar{S}=0 \,.
\ee
The supersymmetry variations indeed confirm our assertion that out of a single  6d hypermultiplet one obtains one chiral (conjugate chiral) 2d $(0,2)$ superfield with fermionic component $\tau_+$ ($\bar\sigma_+$)  and scalar component $T$ ($\bar S$), and in addition one 2d  conjugate $(0,2)$ Fermi superfield with lowest component $\bar \mu_-$.
The variation of $\bar \mu_-$ is furthermore in perfect agreement with the form of the variation of $\bar \rho_-$ and the bulk-surface matter correspondence (\ref{bulk-surface-identification}).
In the vacuum defined by  solutions to (\ref{BPSeq-surface}) the auxiliary fields in this conjugate Fermi multiplet vanish at this point of the analysis.


\subsection{Bulk-Surface Matter Interactions} \label{sec_BulkSurfaceCouplings}

The localised matter just described interacts with the bulk matter of table (\ref{table-matter1}). 
At the level of cubic non-derivative couplings, these interactions derive from the bulk couplings (\ref{Sbulk1}) and  (\ref{bulk-J-coupling}) by treating the matter on $S_{\bf R}$ as localised zero-modes originating from the bulk modes as in (\ref{bulk-surface-identification}). In this approach, which has been introduced for F-theory compactifications to 4d in \cite{Beasley:2008dc,Donagi:2008ca}, one views the configuration of 7-branes intersecting over  $S_{\bf R}$ as a Higgs bundle over $M_G$ with spatially varying Higgs field $\varphi$. 
By cataloguing all possible resulting couplings {we find}
\be\ba \label{bulkmatterint}
S_{\rm bulk + matter} &= S^{(F)}_{\rm bulk + matter} + S^{(J)}_{\rm bulk + matter} \cr 
S^{(F)}_{\rm bulk + matter} &=  
{\bf b}_{\alpha \beta \gamma}   \int d^2 y  \,   \bar \rho_-^\alpha  \left(\tau_+^\beta \,  S^\gamma + \sigma_+^\gamma \, T^\beta  \right) + {\rm c.c.}\cr 
&\quad +  {\bf e}_{\delta \gamma\epsilon} \int d^2 y \, \bar{\mu}_-^\delta \left(  S^\gamma \psi_+^\epsilon +  \sigma_+^\gamma a^\epsilon  \right)  + {\rm c.c.} \cr 
S^{(J)}_{\rm bulk + matter} &=  {\bf c}_{\delta \beta \epsilon  }   \int d^2 y  \,   \mu_-^\delta \left(T^\beta \, \psi_{+}^\epsilon + \tau_{+}^\beta \, a^\epsilon\right) + {\rm c.c.}\,.
\ea
\ee
We are employing here a similar decomposition as in (\ref{zeromodedecomp}) such that the superscripts denote the different zero modes (`families') of the respective type as counted by the cohomology groups in tables (\ref{cohomologies-bulk}) and  (\ref{table-matterspec}).
The couplings are gauge invariant due to the existence of a singlet in the tensor product ${\rm \bf Adj} \otimes {\bf R} \otimes {\bf \bar R}$ and {a} gauge invariant contraction is understood. 
The two couplings (\ref{bulkmatterint}) {are induced} from the bulk $E$- and $J$-type interactions by replacing two of the  bulk fields with corresponding surface localised zero modes, whereas the third bulk field is merely restricted to $S_{\bf R}$, where it couples to the localised matter modes.

The coupling constants are computed by taking the overlap of the internal wavefunction associated with each zero mode and integrating over the surface $S_{\bf R}$,
\be
\ba \label{bulkoverlaps}
{\bf b}_{\alpha \beta \gamma}  &= \int_{S_R}  \hat{\overline\rho}_{{ \bar m \bar n},\alpha} \wedge  \left(\hat\tau_{  m n,\beta } \,  \hat S_\gamma + \hat \sigma_\gamma \, \hat T_{m n,\beta }\right) \cr
 {\bf e}_{\delta \gamma\epsilon}  &= \int_{S_R}  \hat{\bar{\mu}}_{\bar m , \delta}  \wedge\left(  \hat{S}_{mn, \gamma} \wedge  \hat{\psi}_{\bar{n},\epsilon} +  \hat\sigma_{mn, \gamma} \wedge \hat a_{\bar{n}, \epsilon}  \right)  \cr 
 {\bf c}_{\delta \beta \epsilon  }    &=  \int_{S_R} \hat{\mu}_{  \bar m, \delta} \wedge \left(\hat T_{m n,\beta} \wedge \hat \psi_{ \bar n, \epsilon} + \hat \tau_{ m n,\beta} \wedge  \hat a_{\bar n, \epsilon}\right) \,.
\ea
\ee
Here we have made the form indices $m,n$ and $\bar m ,\bar n$ on $S_{\bf R}$ explicit for the hatted, internal wavefunctions  (but not the additional spinor indices).
These derive from the degrees of the cohomology groups counting the respective matter states. For instance,  the wavefunction $\hat \tau$ transforms as an element of $H_{\bar \partial}^{0}(S_2,     L_{\bf R} \otimes  \sqrt{K_{S_{\bf R}}} )  \cong H_{\bar \partial}^{2}(S_{\bf R},     L^*_{\bf R} \otimes  \sqrt{K_{S_{\bf R}}} )^*$ (see (\ref{conjugate-dual})). Since elements of $\Omega^{(0,q)}$ have $q$ anti-holomorphic indices, the complex conjugate of the cohomology group $H_{\bar \partial}^{2}(S_{\bf R},     L^*_{\bf R} \otimes  \sqrt{K_{S_{\bf R}}} )$ counts $(2,0)$ forms with values in $(L^*_{\bf R} \otimes  \sqrt{K_{S_{\bf R}}})^\ast$.

Equivalently, the first coupling in (\ref{bulkoverlaps}) realizes the map 
\be \label{Map1}
 H^0_{\bar \partial}(S_{\bf R}, L_{\bf R}  \otimes \sqrt{K_{S_{\bf R}}} ) \,  {\oplus}  \, H^0_{\bar \partial}(S_{\bf R}, L^*_{\bf R} \otimes \sqrt{K_{S_{\bf R}}} ) \, {\oplus} \, H^2_{\bar\partial}(S_{\bf R},{\rm \bf Adj}) \, \longrightarrow  \,H^2_{\bar\partial}(S_{\bf R},K_{S_{\bf R}})   \,  {\cong}   \, \mathbb{C} \,,
\ee
where  
\be\ba
\hat \tau  &\in H^0_{\bar \partial}(S_{\bf R}, L_{\bf R} \otimes \sqrt{K_{S_{\bf R}}})\cr 
  \hat S &\in     H^2_{\bar \partial}(S_{\bf R}, L_{\bf R} \otimes  \sqrt{K_{S_{\bf R}}}  )^*\,  \cong \, H^0_{\bar \partial}(S_{\bf R}, L^*_{\bf R} \otimes  \sqrt{K_{S_{\bf R}}}) \,,
\ea
\ee
and $H^2_{\bar\partial}(S_{\bf R},{\rm \bf Adj}) $ in (\ref{Map1}) appears due to the restriction of $\hat {\bar\rho}  \in  H^2_{\bar\partial}(M_G,{\rm \bf Adj})$ to  $S_{\bf R}$. 
In the last step of (\ref{Map1}) we identify $H^2_{\bar\partial}(S_{\bf R},K_{S_{\bf R}}) =  H^{2,2}(S_{\bf R})$ and integrate  over $S_{\bf R}$.
The third coupling (and similarly the second one) corresponds to  the canonical map
\be
H^1_{\bar \partial}(S_{\bf R}, L_{\bf R}^\ast  \otimes \sqrt{K_{S_{\bf R}}} ) \,  {\oplus}  \, H^0_{\bar \partial}(S_{\bf R}, L_{\bf R} \otimes \sqrt{K_{S_{\bf R}}} )  \, {\oplus} \, H^1_{\bar\partial}(S_{\bf R},{\rm \bf Adj} )   \, \longrightarrow  \,H^2_{\bar\partial}(S_{\bf R},K_{S_{\bf R}})   \, {\cong}    \, \mathbb{C} 
\ee
for the cohomology groups
\be \ba
 \hat{ \mu}   &\in H^1_{\bar \partial}(S_{\bf R}, L_{\bf R}  \otimes \sqrt{K_{S_{\bf R}}} )^\ast  \, {\cong} \, H^1_{\bar \partial}(S_{\bf R}, L_{\bf R}^\ast  \otimes \sqrt{K_{S_{\bf R}}} ) \cr
 \hat T &\in H^0_{\bar \partial}(S_{\bf R}, L_{\bf R} \otimes \sqrt{K_{S_{\bf R}}} ) \,,
\ea \ee
 and with $H^1_{\bar\partial}(S_{\bf R},{\rm \bf Adj})$ arising from the restriction of $\hat \psi \in H^1_{\bar \partial} (M_G, {\rm \bf Adj})$ to $S_{\bf R}$.

 The coupling $S^{(F)}_{\rm bulk + matter}$    derives from an interaction of the form (\ref{LFint})  if we modify the auxiliary field $E^{(\rho_-)}$ 
 associated with the bulk Fermi multiplet  of $\rho_-$  as
\be
\ba
E^{(\rho^\alpha_-)}    =  -  {\bf f}_{\alpha \mu \epsilon} \,  \Phi^\mu A^\epsilon  -  {\bf b}_{\alpha \beta \gamma}   {\cal T}^\beta  {\cal S}^\gamma,   
\ea
\ee
where again suitable contraction of gauge indices is understood. The first term in $E^{(\rho^\alpha_-)}$ reproduces the pure bulk couplings (\ref{Sbulk1}). 
Furthermore we find a contribution
\be
E^{(\mu^\delta_-)} = -  {\bf e}_{\delta \gamma\epsilon}  \,  {\cal S}^\gamma\, A^\epsilon \,.
\ee
Likewise, the coupling  $S^{(J)}_{\rm bulk + matter}$ implies a  superpotential of the form
\be
 J_{(\mu^\delta_-)}    =   -  {\bf c}_{\delta \beta \epsilon  }    {\cal T}^\beta\, A^\epsilon \,,
\ee
in addition  to the pure bulk superpotential (\ref{BulkJ}). 


\subsection{Cubic Surface-Matter Interactions}

Apart from these cubic interactions with the bulk matter states, there are cubic interactions involving only the localised matter fields. As will be discussed more in section \ref{sec_holocouplings}, these interactions are localised at the intersection of matter surfaces over curves in the base, i.e. in codimension three in $B_4$.
We will summarize the resulting couplings here. The $E$-couplings have a contribution 
from three matter surfaces  intersecting over a curve, associated to three representations ${\bf R}_i$, as follows 
 \be \label{Etypematter1}
 E^{\left(\mu^{ {\bf R}_{a_1},\delta}_{-}\right)}=    - {\bf d}_{\delta \epsilon \gamma} ({ {\bf R}_{a_1} {\bf R}_{a_2} {\bf R}_{a_3}})    \left(  {\cal Z}_{a_2}^{ {\bf R}_{a_2},\epsilon} \, {\cal Z}_{a_3}^{ {\bf R}_{a_3},\gamma} \right) \,,
 \ee
 where ${\cal Z}$ can be a chiral superfield ${\cal S}$ or ${\cal T}$ such that the above coupling is gauge invariant.
Likewise, the bulk $J$-coupling induces a cubic matter surface superpotential
 \be\label{Jtypematter1}
 J_{\left(\mu^{ {\bf R}_{b_1},\delta}_{-}\right)} = -  {\bf h}_{\delta \epsilon \gamma}({ {\bf R}_{b_1} {\bf R}_{b_2} {\bf R}_{b_3}} )   \left(  {\cal Z}_{b_2}^{ {\bf R}_{b_2},\epsilon} \, {\cal Z}_{b_3}^{ {\bf R}_{b_3},\gamma} \right) \,.
 \ee
Both interactions  {are induced} by the bulk $E$- and $J$-type interactions (\ref{Sbulk1}) and (\ref{bulk-J-coupling}).
{Note that the supersymmetry requirement $\hbox{Tr} \, E\cdot J=0$ has to hold for the combination of all $E$- and $J$-couplings. }


\section{D3-brane Sector} 
\label{sec:D3Sector}

In addition to 7-branes on complex three-cycles, F-theory compactifications to 2d contain spacetime-filling D3-branes wrapping holomorphic curves in the base $B_4$ of the elliptic fibration.
In the dual M-theory compactification, such D3-branes correspond to spacetime-filling M2-branes.
In 2d compactifications these D3/M2-branes are of particular importance because of the appearance of chiral matter at the intersection with the 7-branes. This fundamentally distinguishes the 3-7 sector from its analogue in higher-dimensional theories.

The theory on a D3-brane is 4d $N=4$ SYM. To properly describe its coupling to the 7-brane sector derived in the previous sections, we must perform a compatible topological twist for this theory, similarly to the coupling of 6d $(1,0)$ theory at the intersection of two 7-branes. 
This analysis will be presented in \cite{Lawrie:2016axq}.
For the purpose of this article it suffices to get a handle on the matter in the 3-7 sector, and we here take the following simplified approach.
The DBI part of the 3-brane action
\be
S_{\rm D3} = \frac{2  \pi}{\ell_s^4} \int_{\rm D3} e^{-\phi} \sqrt{{\rm det} (g + \ell_s^2 F)}\,,
\ee
identifies the 2d gauge coupling for the effective gauge theory of  a D3-brane compactified on a curve $C^B_{\rm M2}$ as
\be \label{D3gym}
\frac{1}{g^2_{\rm D3}} = e^{-\phi} \,  {\rm Vol}(C^B_{\rm M2}) \, \ell_s^2 \,,
\ee
with the volume ${\rm Vol}(C^B_{\rm M2})$ measured in units of $\ell_s$.
The gauge theory on the 3-brane is therefore weakly coupled as long as the {\it product} of the string coupling $e^{-\phi} = {\rm Im}(\tau)$ times the volume of the wrapped curve is sufficiently large. 
Let us first assume that the 3-brane admits such a weakly-coupled description.
In perturbative string theory, a single spacetime-filling 3-brane contributes a $U(1)$ gauge group factor to the total gauge group in $2$ dimensions. Massless matter charged both under the 7-brane and the 3-brane gauge group arises from the spectrum of massless strings at the intersection of the two types of branes. 
Generically, the complex three-cycle wrapped by the 7-brane and the complex 3-brane curve $C^B_{\rm M2}$    intersect in an isolated number of points on $B_4$. In perturbative string theory, the open strings  in the 3-7 sector are subject to mixed Dirichlet-Neumann boundary conditions in all eight internal real dimensions. The vacuum energy for the Neveu-Schwarz ground state is thus $a_{NS} =  - \frac{1}{2} + \frac{8}{8} = \frac{1}{2}$. 

Consequently, the massless string spectrum contains only the fermionic excitations from the Ramond-Ramond sector with $a_R = 0$. In the 2d $(0,2)$ theory this gives rise to a negative-chirality spinor $\nu_-$ which forms the lowest-lying component of a Fermi multiplet. Apart from subtleties from $SL(2,\mathbb Z)$ monodromies to be discussed momentarily the number of such 3-7 Fermi multiplets is given by the number of intersection points
\be
\int_{B_4}  [M_G] \wedge [C^B_{\rm M2}] \,.
\ee
The Fermi multiplets transform in the fundamental representation  of the non-abelian gauge group $G$ realized on the 7-brane and carry charge $-1$ under the abelian gauge group on the 3-brane. We will denote this representation as ${\bf R}_{3-7}$.
The 3-7 brane matter can be summarized as follows:
\be \label{37multiplets}
\begin{array}{c|c}
\hbox{Cohomology}  & \hbox{Fermions } \oplus    \hbox{Anti-Fermions }\cr\hline\vspace{-4mm} \cr 
H_{\bar \partial}^{0}(M_G \cap C^B_{\rm M2}) \oplus H_{\bar \partial}^{0}(M_G \cap C^B_{\rm M2})^*  & \nu_-^{{\bf R}_{3-7}}  \oplus    \bar\nu_-^{{\bf \bar R}_{3-7}}
\end{array}
\ee
\smallskip

The assignment of representation ${\bf R}_{3-7}$ to the Fermi multiplet component, as opposed to its conjugate, is a matter of convention. 
As will be discussed at the end of section \ref{sec_2dAbelian}, the appearance of this matter induces a gauge anomaly for the D3-brane $U(1)$ factor, which is cancelled by a Green-Schwarz mechanism 
rendering the $U(1)$ massive.

In addition, there is matter from the bulk sector of the D3-branes in the adjoint representation of the D3-brane gauge group \cite{Lawrie:2016axq}. For single D3-branes with a (massive) $U(1)$ gauge symmetry, this matter is uncharged under the 2d gauge group and we therefore do not consider it further here.

Generically, the 3-7-matter sector cannot interact with the 7-7 matter via supersymmetric cubic non-derivative couplings of the form (\ref{LJint}) or (\ref{LFint}). Such interactions would require two chiral field insertions, which must come from the 7-7 sector as the 3-7 sector only contains Fermi multiplets. But generically, the D3-branes intersect the 7-brane cycle $M_G$ away from the codimension one matter surfaces so that the only gauge invariant interactions would be of the form ${\rm \bf Adj} \otimes {\bf R}_{3-7} \otimes  {\bar{\bf R}}_{3-7}$, in contradiction with the required structure of the couplings.
The same argument prevents such couplings between the modes from the 3-7 sector and from the D3-D3 sector.

 In general F-theory compactifications the axio-dilaton varies over the base $B_4$ and strong coupling effects become relevant, when ${\rm Im}(\tau) = {\cal O}(1)$ even though we stress again that it is the combination (\ref{D3gym}) rather than ${\rm Im}(\tau)$ itself which controls the gauge coupling on the 3-brane.
 In particular, the above perturbative derivation of the spectrum is expected to remain valid as long as the volume of $C^B_{\rm M2}$ is large enough and/or the 3-branes do not extend into regions of 
 small ${\rm Im}(\tau)$. A detailed analysis of D3-branes including non-perturbative regimes will appear in \cite{Lawrie:2016axq}.
However, even in such situations $SL(2,\mathbb Z)$ monodromies in $\tau$ do leave their imprint on the 3-7 sector: 
In particular,  the number of multiplets in the 3-7 sector is in general only a fraction $\frac{1}{{\rm ord}(g)}$ of the number of geometrical intersection points $\int_{B_4}  [M_G] \wedge [C^B_{\rm M2}]$ due to the appearance of monodromies of order ${\rm ord}(g)$ around the 7-brane locus. While the effect of these monodromies is automatically taken into account in the description of the 7-branes in the language of the elliptic fibration, it needs to be accounted for separately for the 3-7 sector, which after all is not geometrised in F/M-theory. We will encounter examples of this effect in sections \ref{sec:SOGlobal} and \ref{sec:E6Global}, where we consider the global consistency of an $SO(10)$ and an $E_6$ model, respectively, and test our description of the 3-7 sector by computing the contribution to the 7-brane gauge anomalies. 
We view this computation as a non-trivial check of our approach. This being said, when the D3-brane itself becomes strongly coupled an analysis in the spirit of \cite{Heckman:2010fh,Heckman:2010qv} is more appropriate and will be part of \cite{Lawrie:2016axq}.


\section{Elliptic Five-folds and 2d Gauge Theories} 
\label{sec:Singularfibers}

So far we have described 2d $(0,2)$ F-theory compactifications from the perspective of  the topologically twisted field theory realized on stacks of 7-branes and their intersections. This captures the local properties of the F-theory compactification, in the sense of decoupled gravity and without taking into account global consistency of the theory.
We now embed this construction into a globally consistent  compactification of F-theory to two dimensions.
The effective theory of such compactifications  is conveniently approached via duality with M-theory compactified on the very same elliptically fibered Calabi-Yau five-fold $Y_5$, {via M/F-duality (\ref{MFDuality}).}
In the sequel we summarize some of the salient features of such compatifications. 
Much of the material in sections \ref{sec:Dictionary} (apart from the specific relation to the dual 1d M-theory compactification) and \ref{sec:Codim12} {follows in close analogy} with F-theory compactifications to six and four dimensions, and we review this here for the reader's convenience.
In section \ref{sec_holocouplings} we develop the structure of surface localised couplings, which is very specific to compactifications to two dimensions, and in the remaining sections we put 
 special emphasis on some peculiarities on five-folds as compared to their lower-dimensional cousins. 


\subsection{Dictionary} 
\label{sec:Dictionary}

The setup we consider was already outlined in section \ref{sec:FCY5}. The 7-brane gauge theory  lives on a complex three-cycle (divisor) $M_G$ in the base $B_4$ of the elliptic Calabi-Yau five-fold $Y_5$, which is characterized in terms of the vanishing of the discriminant $\Delta$ of the elliptic fibration to  order $n>0$, i.e.
\be
\Delta = O(\zeta_0^n) \,,\qquad  \hbox{where} \quad \zeta_0=0:    \quad M_G\subset B_4\,.
\ee
The singularity type in the fiber above such codimension one loci, and thus the gauge algebra $\mathfrak{g}$ on the 7-brane world-volume, is characterized in terms of the Kodaira type of the fiber. 
 One way to determine this is to consider the $[p,q]$ 7-brane composition of such singularity and the resulting monodromy of the axio-dilaton. The $[p,q]$-strings give rise to precisely the adjoint of the gauge algebra $\mathfrak{g}$ \cite{Gaberdiel:1997ud, DeWolfe:1998zf,Grassi:2013kha,Grassi:2014sda}. Somewhat more directly, the gauge degrees of freedom can be understood from the dual M-theory picture in terms of the dimensional reduction of $C_3$  and wrapped M2-branes \cite{Witten:1996qb}. 
To characterize these degrees of freedom, it is useful to determine the fiber type by means of resolving the singularities. The resolved fibers are collections of rational curves, i.e. $\mathbb{P}^1$s, which intersect in (up to a few low rank oddities) affine Dynkin diagrams of an ADE Lie algebra $\mathfrak{g}$ and can be associated to its simple roots $\alpha_i$,
\be
\hbox{$\mathbb{P}^1$s above codim 1 loci $M_G$}  \qquad \leftrightarrow \qquad \hbox{Simple roots $\alpha_i$ of $\mathfrak{g}$}\,.
\ee
 This Kodaira fiber type in turn determines the gauge algebra of the 2d gauge theory. In M-theory the non-abelian gauge bosons arise from M2-branes wrapped on the $\mathbb{P}^1$s and the gauge bosons associated with the Cartan subalgebra of $\mathfrak{g}$  stem from reduction of $C_3$ along the $(1,1)$ forms $\omega^i$  related to these fibral curves
 \be\label{C3omega}
C_3 = A_i \wedge \omega^i + \ldots \,.
\ee
The $(1,1)$ forms are dual to the divisors which are obtained by fibering the rational curves $\mathbb P^1_i$ over the discriminant component and which  intersect with the fibral curves in the negative Cartan matrix of the gauge algebra $\mathfrak{g}$. 
In turn, each fibral curve is associated with a simple root of $\mathfrak{g}$. In the M-theory compactification to one dimension
the resolution of the singular fiber corresponds to moving onto the `Coulomb branch' along which the wrapped M2-brane modes become massive. The structure of this Coulomb branch will have a similarly elegant description as in 6d and 4d \cite{Hayashi:2013lra, Hayashi:2014kca,  Lawrie:2015hia, Braun:2014kla, Braun:2015hkv}.
 In the F-theory limit, which takes the volume of the fibral curves to zero, these wrapped M2-branes become massless gauge degrees of freedom. 

{Before discussing this point further, let us turn to the charged matter fields arising from  singularities above codimension two loci, i.e. complex surfaces $S_{\bf R}\subset M_G$ in the base $B_4$.}
These can be thought of as 7-brane intersections, or loci of enhancements of the singularities in the elliptic fibration. The geometric process that characterizes the matter fields is the splitting of rational curves in the fiber above codimension two loci in the base, along which the order of vanishing of the discriminant increases. 
Representation-theoretically, this means that some of the $\mathbb{P}^1$s  associated to simple roots become reducible above codimension two loci, and split into weights of representations ${\bf R}$ of the gauge algebra $\mathfrak{g}$. From the point of view of the 2d theory, the states originating in M-theory from wrapped M2-branes on these fibral curves correspond to matter fields in the associated representation
\be
\mathbb{P}^1\hbox{s above codim 2 loci $S_{\bf R}$} \quad \leftrightarrow \quad \hbox{Matter in Representation ${\bf R}$}\,.
\ee
The fibers above such codimension two matter surfaces $S_{\bf R}\subset M_G$ will be described in section \ref{sec:Codim12}, {where we characterize the fibral curves associated to matter in terms of their intersections as carrying charges associated to the weights of the representation ${\bf R}$. }

We now discuss in more detail the relation between the 2d field theory and the $(1+0)$-dimensional theory obtained from M-theory compactified on $Y_5$, {which} is a supersymmetric quantum mechanics (SQM) with two supercharges.
The SQM resulting from M-theory compactification on smooth, not necessarily elliptically fibered, Calabi-Yau five-folds has been studied in \cite{Haupt:2008nu}.
In our context, we need to implement the fibration structure of $Y_5$ and in addition uplift the $(1+0)$-dimensional theory to a 2d field theory by taking the F-theory limit. We reserve a detailed analysis to  \cite{Lawrie:2016rqe} and for the purpose of this paper it suffices to summarize simply the identification between these theories. 
The 1d SQM has two types of `bosonic' multiplets \cite{Haupt:2008nu}: the 2a multiplet, which has a real scalar, fermion and auxiliary field $f$, and the 2b multiplet, comprised of a complex scalar and fermion (in this case the auxiliary field is not an independent degree of freedom). In addition we will need a fermionic 2b multiplet with a fermion as its lowest component and otherwise only auxiliary fields \cite{Haupt:2008nu}.
This 1d super-mechanics is related to the 2d $(0,2)$ field theory obtained from F-theory by dimensional reduction of the latter on a circle $S^1$.
Upon circle reduction, a 2d $(0,2)$ Fermi descends to a fermionic 2b multiplet in the super-mechanics. 
A 2d chiral superfield can either descend to a 2b multiplet or to a 2a multiplet together with a 1-form potential in the super-mechanics theory.
A 2d vector multiplet either descends to a 2a multiplet plus 1-form, or to a 2b multiplet. All these possibilities are indeed at work.}

Consider first an off-shell vector multiplet in the 2d F-theory associated with one of the Cartan $U(1)$ gauge factors. The vector component along the compactification $S^1$ becomes a real scalar in a 2a multiplet, which is precisely 
the volume modulus of the associated resolution $\mathbb{P}^1$ in the elliptic fiber. 
Their number is given by $h^{1,1}(Y_5) - h^{1,1}(B_4)-1$, where we are subtracting the base K\"ahler moduli and the modulus associated with the generic fiber class.\footnote{In presence of extra rational sections, this quantity counts the number of Cartan and extra non-Cartan $U(1)$ gauge groups, as in higher-dimensional reductions \cite{Morrison:1996pp}.} 
Resolving the fiber gives vevs to these 2a scalar fields, which corresponds to moving onto a Coulomb branch of the 1d SQM.
On the other hand, we can reduce $C_3$ along $\omega^i$ as in (\ref{C3omega}), which in one dimension gives rise to a `1d vector' $A_i$. Despite being non-dynamical, this field will play an important role in our discussion of Chern-Simons couplings and global consistency conditions in section \ref{sec_tadpole}. 
Lifting this to F-theory, the $h^{1,1}(Y_5) - h^{1,1}(B_4)-1$ vectors $A_i$ yield the second off-shell vector degree of freedom of the 2d $(0,2)$ vector multiplets.
This is summarized in the first line of  table \ref{tab:MF}.

\begin{table}\begin{center}
\begin{tabular}{|c|c|c|}\hline
$\#$ & 1d SQM from M-theory & 2d (0,2) SYM from F-theory\cr\hline\hline
$h^{1,1}(Y_5) - h^{1,1}(B_4) -1$  & 2a multiplet + $A_i$ & gauge multiplet\cr \hline
$h^{1,1}(B_4) - 1$   &  2a multiplet + $A_a$ & chiral multiplet     \cr \hline
{$\frac{1}{2}(b^{3}(Y_5) - b^{3}(B_4))$ }  &  2b multiplet   & chiral multiplet     \cr \hline
$\frac{1}{2}b^{3}(B_4) $   &  2b multiplet   & Fermi multiplet    \cr \hline
$h^{4,1}(Y_5) $      &  2b multiplet   & chiral multiplet     \cr \hline
$h^{3,1}(Y_5) $      &  fermionic 2b multiplet   & Fermi multiplet     \cr \hline
\end{tabular}
\caption{{Partial} identification of multiplets in the 1d SQM obtained from M-theory on an elliptically fibered Calabi-Yau five-fold $Y_5$ with those in the 2d $(0,2)$ theory obtained form F-theory on $Y_5$. {The complete spectrum can be found in \cite{Lawrie:2016rqe}}.}\label{tab:MF}
\end{center}
\end{table}

For completeness let us also give the identification of the remaining moduli fields which are uncharged under $\mathfrak{g}$ and which are thus not part of the gauge theory considered so far: 
The  K\"ahler moduli of the F-theory compactification organize into 2d $(0,2)$ chiral multiplets with complex scalar fields {$\int_{D_a} (J \wedge J \wedge J + i J \wedge C_4)$}.
Here $D_a$ denote the independent divisor classes of $B_4$.
In the 1d super-mechanics obtained from M-theory these chiral multiplets dualize into 2a multiplets plus vectors $A_a$ from reduction of $C_3$ along the dual 2-forms  $\omega^a$. If the F-theory allows for a perturbative IIB limit defined on a Calabi-Yau four-fold, this number equals the number of orientifold even divisors $h^{1,1}(B_4) = h^{1,1}_+(X_4)$. In fact, the number of resulting 2d $(0,2)$ chiral multiplets is $h^{1,1}(B_4) - 1$ as the overall volume enters the gravity multiplet \cite{Lawrie:2016rqe}.
The modulus associated with the universal fiber volume and the associated 1-form $A_0$ uplifts to components of the gravity multiplet in 2d. 
Among the $\frac{1}{2} b^3(Y_5)$ 2b multiplets whose scalar components  combine the degrees of freedom from reduction of $C_3$ along the independent 3-forms \cite{Haupt:2008nu} in M-theory, 
{$\frac{1}{2} (b^3(Y_5) - b^3(B_4))$} 2b multiplets uplift to 2d $(0,2)$ chiral multiplets associated with reduction of the F-theory  $C_2$ and $B_2$-fields along $h^{1,1}_-(X_4)$.
. The complex structure moduli arise as $h^{4,1}(Y_5)$ 2b multiplets  \cite{Haupt:2008nu} in M-theory, which become chiral multiplets in F-theory. Finally, there are $h^{3,1}(Y_5)$ fermionic degrees of freedom sitting in a fermionic 2b multiplet in M-theory \cite{Haupt:2008nu}  and in a Fermi multiplet in F-theory. There are additional Fermi multiplets which will be discussed in \cite{Lawrie:2016rqe}.


\subsection{Geometry of Singular Fibers}
\label{sec:Codim12}

We will now give a more in-depth characterization of the singular fibers in elliptic five-folds. 
The fibers in elliptic Calabi-Yau $n$-folds in codimension one have a canonical description in terms of Kodaira fibers \cite{Kodaira, Neron}, which associate to the singular fibers a Lie algebra $\mathfrak{g}$. Likewise the situation in codimension two is by now very well understood -- see \cite{EY,MS,Krause:2011xj} for early discussions in explicit resolutions of Calabi-Yau four-folds and \cite{Morrison:2011mb} for an analysis of codimension two in Calabi-Yau three-folds. In fact, the general characterization of the fibers  is in terms of representation-theoretic data of $\mathfrak{g}$ \cite{Hayashi:2014kca}. What will be crucial in our analysis is the precise relation between curve classes above codimension one and two loci in the base. The notation in this section will be that of \cite{Hayashi:2014kca, Lawrie:2015hia}. 

Above codimension one, along a component $M_G$ of the discriminant $\Delta$ in the base $B_4$ of the elliptic fibration $Y_5$, the rational curves associated to the simple roots $\alpha_i$ of the non-abelian Lie algebra $\mathfrak{g}$ will be denoted by $F_i$, $i=1, \ldots, {\rm rk}({\mathfrak g})$. The so-called Cartan divisors, obtained by fibering these rational curves over $M_G$, will be denoted by $D_i$ with the following intersection property
\be
D_i \cdot_{Y_{5}} F_j = - C_{ij} \,,
\ee
in terms of the Cartan matrix  $C_{ij}$ of $\mathfrak{g}$. The curve $F_0$ associated to the extended node $\alpha_0$ will be intersected by the section of the elliptic fibration, and we define the singular limit $\pi: Y \rightarrow Y_{sing}$ as  the limit where all fiber components are shrunk to zero volume, except for $F_0$, which intersects the section.  
We furthermore define the relative Mori cone $NE(\pi)$ as the cone containing all curves that are contracted by the singular limit. 

This setup in codimension one gets modified along codimension two loci in the base, where the singularity of the elliptic fibration gets enhanced.  The main effect is that rational curves in the codimension one fiber can become reducible. These rational curves intersect with the Cartan divisors 
in terms of the weights of representations of $\mathfrak{g}$. The simplest instances is that of an $I_n$ (or $SU(n)$) fiber in codimension one, with fundamental matter ${\bf n}$, which corresponds to a splitting of the fibers along the matter locus as
\be\label{FCpCm}
F_i \quad \rightarrow \quad C^{+}_{i} +C^{-}_{i+1} \,,
\ee
for some $i$. Here $C_{i}^{\pm}$ are rational curves which correspond to fundamental weights $L_i$ and $-L_{i+1}$. 
What will be relevant in our context is that M2-branes wrapping $F_i$ along codimension one and M2-branes wrapping the curves $C_i^{\pm}$ are in fact not going to be independent states. The relevant notion is that of the generating set of the relative cone of curves. 

More generally (\ref{FCpCm}) is replaced by a splitting into curves $C_{\lambda^{\bf R}_a}^\pm$ for a representation ${\bf R}$ and associated weight $\lambda^{\bf R}_a$, $a=1,\ldots,{\rm dim}({\bf R})$. The effective curves are either associated to the simple roots $F_i$ or to weights with specific sign assignments 
\be\label{epsDef}
\ba
\varepsilon:\qquad {\bf R} \quad &\rightarrow \quad  \{\pm \} \cr 
\lambda^{\bf R} \quad & \rightarrow \quad \varepsilon(\lambda^{\bf R} ) \,,
\ea
\ee
and the associated curves are characterized by the weight as well as a sign
\be\label{Clambdas}
C_{\lambda^{\bf R}_a}^{\varepsilon(\lambda^{\bf R}_a )} \,,\qquad a= 1, \cdots, \dim {\bf R}\,.
\ee
For each of the $\dim {\bf R}$ states in representation ${\bf R}$ let  $\lambda_a^{\bf R}$ be the rk($\mathfrak{g}$)-component  weight vector in the Cartan Weyl basis.  
Intersecting these with the Cartan divisors $D_i$ results in 
\be\label{DCmat}
D_i \cdot_{Y_5} C_{\lambda^{\bf R}_a}^{\varepsilon(\lambda^{\bf R}_a)} =  \varepsilon(\lambda^{\bf R}_a) \, \lambda^{\bf R}_{ai} \,,\qquad i=1, \cdots, \hbox{rk}\, \mathfrak{g}\,.
\ee
Here $\lambda^{\bf R}_{ai}$ denotes the $i$th component of the weight vector of $\lambda^{\bf R}_a$. The consistent sign assignments (\ref{epsDef}) are encoded in the box graphs. 
The physical significance of these sign functions is that for $\varepsilon(\lambda^{\bf R}_a) = \pm 1$, the state   with weight $\lambda^{\bf R}_{a}$ arises from an M2-brane (anti-M2-brane) wrapping the effective curve $C_{\lambda^{\bf R}_a}^{\varepsilon(\lambda^{\bf R}_a)}$.

It was shown in \cite{Hayashi:2014kca} that the extremal generators of the relative cone of effective curves in codimension two are obtained in terms of data encoded in the so-called box graphs,and that this relative cone takes the form\footnote{For each codimension two locus there is a well defined cone of this kind. But if there are codimension three or four loci, there can be identifications and the set of extremal generators may be reduced \cite{Hayashi:2014kca}.}
\be\label{Nepi}
NE (\pi) = \bigoplus_{\ell_k \in \mathcal{K}_{fib}} \mathbb{Z}^+ \ell_k \,. 
\ee
The set of extremal generators $\ell_k\in \mathcal{K}_{fib}$ is given by those rational curves $F_i$ which remain irreducible above the codimension two loci,  as well as the curves $C_{\lambda^{\bf R}_a}^{\varepsilon(\lambda^{\bf R}_a )} $ which arise in the splitting along codimension two loci associated to representations ${\bf R}$ with weights $\lambda$. From this analysis, it follows that the fibers in codimension two can be either of standard Kodaira type or monodromy-reduced Kodaira fibers \cite{Hayashi:2014kca}.

So far we have only assumed the existence of a zero-section $\sigma_0$, but in general an elliptic fibration can have extra rational sections. These generate the Mordell-Weil group MW($Y_5$). Its rank $M$ counts the number of non-Cartan $U(1)$ gauge group factors on $Y_5$ \cite{Morrison:1996pp} as will be reviewed momentarily.
In the presence of extra rational sections $\sigma_m$ additional curves in the fiber arise over codimension two loci in $B_4$. At the level of homology, a basis of $H_2(Y_5)$ is therefore composed of a basis of $H_2(B_4)$ together with
the class of the generic fiber ${\mathfrak F}$,  a basis of the effective curves $\mathcal{K}_{fib}$ {of} the fiber, as well as a basis of effective curves $C^\sigma_m$, $m=1, \ldots M$, in the presence of $M$ additional independent sections $\sigma_m$. 

We also introduce a  dual basis of divisors
\be \label{divisor-list}
\ba
D_a^{(B)}, \qquad  &a= 1, \ldots, h^{1,1}(B_4) \cr
D_i,  \qquad &i =1, \ldots, {\rm rk}({\mathfrak g})  \cr
S_m, \qquad &0=1, \ldots, M = {\rm rk}({\rm MW}(Y_5)).
\ea
\ee
Here $D^{(B)}_a$ denotes the pullback of a basis of divisors from $B_4$, $D_i$ are the Cartan divisors associated with the non-abelian gauge algebra  $({\mathfrak g})$ and $S_0$ represents a divisor
whose only 
non-trivial intersection number with the above set of curves is 
\be \label{S0Fint}
S_0 \cdot_{Y_5} {\mathfrak F} = 1 \,.
\ee
If $Y_5$ is an elliptic fibration, $S_0$ is the class of the divisor defined by the zero-section $\sigma_0$, but a divisor $S_0$ can be defined also in absence of a zero-section.
To each additional section $\sigma_m$ the Shioda maps  associates a divisor 
\be \label{Shiodamap}
S_m=\sigma_m-\sigma_0 -   D_B -  \sum_i  n_i D_i \,,
\ee
where the coefficients  {$n_i$} are determined such that $S_m$ has trivial intersection with the $F_i$, and $D_B$ denotes a suitable base divisor. 
The significance of this divisor $S_m$ is that expansion of the M-theory 3-form $C_3$ in terms of its dual 2-form gives rise to the gauge potentials of extra, non-Cartan $U(1)_m$ gauge group factors, as studied for explicit fibrations recently e.g. in \cite{Grimm:2010ez,Braun:2011zm,Krause:2011xj,Morrison:2012ei,Cvetic:2012xn,Mayrhofer:2012zy,Braun:2013yti,Borchmann:2013jwa,Cvetic:2013nia,Braun:2013nqa,Cvetic:2013uta,Borchmann:2013hta,Cvetic:2013jta,Cvetic:2013qsa,Braun:2014nva,Kuntzler:2014ila,Klevers:2014bqa,Lawrie:2014uya,Lawrie:2015hia, Cvetic:2015ioa, Krippendorf:2015kta}.

Note that the Shioda-divisors $S_m$ have non-trivial intersections with the fibral curves $C_{\lambda^{\bf R}_a}^{\varepsilon(\lambda^{\bf R}_a )}$. These intersection numbers compute the $U(1)_m$ charges of the matter fields associated to $C_{\lambda^{\bf R}_a}^{\varepsilon(\lambda^{\bf R}_a )}$. 
The intersection possibilities for fibers in codimension two with the Shioda-divisors $S_m$, i.e. the $U(1)$ charges of matter fields, can be characterized comprehensively in terms of the splitting of the fibers in codimension two \cite{Lawrie:2015hia}. {The models with additional rational sections provide the framework for realizing GLSMs with abelian gauge groups, as will be discussed in section \ref{sec:NLSM}.}


\subsection{Cubic Matter Couplings} \label{sec_holocouplings}

Finally we are in a position to complete the general discussion of supersymmetric cubic couplings in the 2d theory.
In sections \ref{sec_BulkCouplings} and \ref{sec_BulkSurfaceCouplings} we had analyzed such interactions between the bulk matter states and, respectively, between bulk and surface matter.
In addition, the sector of holomorphic couplings includes triple interaction terms involving only matter fields localised on matter surfaces. Such interactions can arise when two or more matter surfaces $S_{{\bf R}_i}$ in $B_4$ intersect such that the internal wavefunctions describing the matter zero-modes can overlap and produce a gauge invariant coupling.  
{The triple coupling originates in the bulk Yukawa interactions (\ref{Sbulk1}) and (\ref{bulk-J-coupling}) by again treating the internal wavefunctions of the surface matter as localised bulk zero modes in the presence of a non-trivial Higgs bundle.}
This way we anticipate that the possible Yukawa couplings can only be of the form (\ref{Etypematter1}) or (\ref{Jtypematter1}). 

Generically, triple intersections of matter surfaces occur already in complex codimension three, i.e. over complex curves $\Sigma$. 
The significance of these codimension three loci is that here fibral curves $C_{\lambda_{{\bf R}_i}}$ associated to matter in representations with weights $\lambda_{{\bf R}_i}$ split, {i.e. become reducible}
\be
C_{\lambda_{{\bf R}_1}} \rightarrow C_{\lambda_{{\bf R}_2}} + C_{\lambda_{{\bf R}_3}} \,.
\ee
When this happens the singularity {of the fiber} enhances further. 
Such a splitting (or, viewed in the reverse, joining) is a necessary condition for a coupling between matter associated with M2-branes wrapped on $C_{\lambda_{{\bf R}_i}}$ to occur. 
Indeed, above a codimension three curve  $\Sigma_{{\bf R}_1 {\bf R}_2 {\bf R}_3} = \cap_i S_{{\bf R}_i}$ the fiber enhancement is compatible with a gauge invariant contraction
\be \label{gauge-inv-assump}
{{\bf R}_1 \oplus {\bf R}_2 \oplus {\bf R}_3\  \rightarrow  \ {\mathbb C} }\,.
\ee
To realize a coupling over a curve $\Sigma$ it must be possible to produce a $(1,1)$ form from the internal matter wave-functions, which can then be integrated over $\Sigma$. 
Let us again perform a decomposition of the form (\ref{zeromodedecomp}) for the zero-modes of the surface matter.
{As for a superpotential, a coupling of the form $\int_{\Sigma }\hat \mu^{\bf R_1} \hat T^{\bf R_2} \hat \tau^{\bf R_3}$, where for field $\hat\mu$ we dualized with (\ref{conjugate-dual}),  corresponds to a map }
\be \label{Jcouplingsequence}
H^1\left(\Sigma, L_{\bf R_1} \otimes   K^{1/2}_{S_{\bf R_1}} |_{\Sigma}\right) \oplus 
H^0\left(\Sigma,  L_{\bf R_2} \otimes  K^{1/2}_{S_{\bf R_2}} |_{\Sigma}\right) \oplus 
H^0\left(\Sigma, L_{\bf R_3} \otimes   K^{1/2}_{S_{\bf R_3}} |_{\Sigma}\right) \ \rightarrow \ \mathbb{C}\,,
\ee
which exists by assumption of gauge invariance (\ref{gauge-inv-assump}) for a suitable assignment of representations to the fields.
This leads to a superpotential coupling arising from the deformation of the bulk coupling (\ref{BulkJ}) of the form
 \be \label{Jcoupling-gen1}
 \int  d^2y \, {\bf h}_{\delta \epsilon \gamma} ({ {\bf R}_{1} {\bf R}_{2} {\bf R}_{3}}) \,  \mu^{ {\bf R}_{1},\delta}_{-} \left(  \tau^{ {\bf R}_{2},\epsilon}_{+} \, T^{ {\bf R}_{3},\gamma}_{}  +  \tau^{ {\bf R}_{3},\gamma}_{+} \, T^{ {\bf R}_{2},\epsilon}_{} \right)\,
\ee
with
\be \label{mattercubef1}
{\bf h}_{\delta \epsilon \gamma}({ {\bf R}_{1} {\bf R}_{2} {\bf R}_{3}} )    = 
\int_{\Sigma} \hat \mu^{\bf R_1} \wedge \hat T^{\bf R_2}  \wedge \hat \tau^{\bf R_3} \,.
\ee
The associated superpotential is topological and takes the form 
 \be
 J_{\left(\mu^{ {\bf R}_{1},\delta}_{-}\right)} = -  {\bf h}_{\delta \epsilon \gamma}({ {\bf R}_{1} {\bf R}_{2} {\bf R}_{3}} )   \left(  {\cal T}^{ {\bf R}_{2},\epsilon} \, {\cal T}^{ {\bf R}_{3},\gamma}_{} \right) \,.
 \ee
Note that as long as allowed by gauge invariance, similar couplings exist for, more generally, $\mathcal{Z}$ given by either ${\cal T}$ or ${\cal S}$.  For instance there can be any such couplings 
 \be
 J_{\left(\mu^{ {\bf R}_{1},\delta}_{-}\right)} = -  {\bf h}_{\delta \epsilon \gamma} ({ {\bf R}_{1} {\bf R}_{2} {\bf R}_{3}} )   \left(  {\cal Z}_2^{ {\bf R}_{2},\epsilon} \, {\cal Z}_3^{ {\bf R}_{3},\gamma} \right) \,.
 \ee
Furthermore, there is an $E$-type coupling induced by (\ref{Sbulk1}).  For a matter spectrum consistent with (\ref{Jcoupling-gen1}) the possible gauge invariant $E$-terms are 
\be
\int  d^2y \, {\bf d}_{\delta \epsilon \gamma}({ {\bf R}_{1} {\bf R}_{2} {\bf R}_{3}}) \,  \bar\mu^{ {\bf R}_{2},\delta}_{-} \left(  \sigma^{ {\bf R}_{1},\epsilon}_{+} \, T^{ {\bf R}_{3},\gamma}_{}  +  \tau^{ {\bf R}_{3},\gamma}_{+} \, S^{ {\bf R}_{1},\epsilon}_{} \right) 
\ee 
 with coupling constant 
 \be \label{mattercubed1}
 {\bf d}_{\delta \epsilon \gamma}( {\bf R}_{1} {\bf R}_{2} {\bf R}_{3})   =  \int_{\Sigma}   \hat{\overline{\mu}}^{ {\bf R}_{2}}_\delta \wedge   \hat\sigma^{ {\bf R}_{1}}_\epsilon \wedge  \hat T^{ {\bf R}_{3}}_\gamma  
    \ee
  plus an analogous term with  ${\bf R}_{2}$ and  ${\bf R}_{3}$ exchanged.
   To see these couplings we must use   (\ref{conjugate-dual}) for the mode associated with the superfields of type ${\cal S}$ and apply a similar logic as in (\ref{Jcouplingsequence}).
    These couplings imply an extra term in the $E$-auxiliary field
 \be
 E^{\left(\mu^{ {\bf \bar R}_{2},\delta}_{-}\right)}   \supset    - {\bf d}_{\delta \epsilon \gamma} ({ {\bf R}_{1} {\bf R}_{2} {\bf R}_{3}})    \left(  {\cal S}^{ {\bf R}_{1},\epsilon} \, {\cal T}^{ {\bf R}_{3},\gamma}_{} \right) 
 \ee
and similarly for $E^{\left(\mu^{ {\bf \bar R}_{3},\delta}_{-}\right)}$.
As before, more generally, there can be $E$-type interactions 
\be
E^{\left(\mu^{ {\bf \bar R}_{2},\delta}_{-}\right)}   =  - {\bf d}_{\delta \epsilon \gamma} ({ {\bf R}_{1} {\bf R}_{2} {\bf R}_{3}})    \left(  {\cal Z}_1^{ {\bf R}_{1},\epsilon} \, {\cal Z}^{ {\bf R}_{3},\gamma}_{3} \right) 
\ee
as long as gauge invariance allows it. 

In five-folds, additional fiber splittings occur in complex codimension four, when two or more of the codimension three curves $\Sigma$ intersect in a set of isolated points. 
Here the fibre splittings allow for a combination of two types of couplings into quartic couplings. While such couplings are always allowed field theoretically, here they localised at a point in the base. 
The precise structure of these couplings will be exemplified in section \ref{sec:SU5Ex}.

 Supersymmetry requires that the final structure of $J$- and $E$-type couplings must be such that the constraint 
 \be\label{EJConstraint}
{\hbox{Tr}} \left(\sum_a   E^a \,  J_a \right)= 0 
 \ee
 is satisfied with the index $a$ running over all massless Fermi multiplets.


\subsection{Monodromy and Non-minimality}

There are several effects which make the structure of higher-codimension fibers more intricate for five-folds. Particularly relevant for later considerations are the existence of additional monodromies in the fibers as well as non-minimality arising in codimension four. 

Monodromy is the effect that locally two curves may appear independent, but globally are identified. 
As already observed \cite{Hayashi:2014kca} in codimension two, monodromies can yield non-Kodaira fibers, which was shown to always occur whenever the local enhancement is to an algebra $\tilde{\mathfrak{g}}$ such that the commutant of the gauge algebra in $\tilde{\mathfrak{g}}$ is non-abelian, e.g. $\mathfrak{su}(2)$
 for $\mathfrak{su}(6)\subset \mathfrak{e}_6$, corresponding to $\Lambda^3{\bf 6}$ matter. 
 
 This continues to hold in higher codimension. In particular, for five-folds new monodromy reductions of the fibers can occur in codimension three. The effect can be easily explained by considering for instance an $SU(n)$ model with $I_n$ fiber in codimension one, which has $I_m^*$ fibers in codimension three. In \cite{Lawrie:2012gg} \footnote{See (3.27) of \cite{Lawrie:2012gg}, which is the equation for the fiber in codimension three for all $I_n$ to $I_m^*$ enhancements.} it was shown that these always have fiber components that are quadratic equations above codimensions three loci
\be\label{MonoReduxEq}
 b_2x^2 + b_4 \zeta_{2k-1} x + b_6\zeta_{2k-1}^2  =0 \,,
\ee
where $\zeta_{2k-1}$ is one of the resolution divisors and $b_i$ are certain sections on the base.
For a four-fold, this happens over a point, such that the quadratic can be factored into two irreducible components as first observed in \cite{EY}\footnote{This absence of monodromy-reduction was already observed in the four-fold case for $SU(5)$ in \cite{EY, MS}, for general $I_n^*$ in \cite{Lawrie:2012gg} and lower rank cases in \cite{Lin:2014qga}. The relevance of these non-monodromy-reduced fibers for the  generation of couplings was elucidated in  \cite{Martucci:2015dxa}.},  and the fiber is a Kodaira $I_m^*$. In higher-dimensional elliptic varieties the quadratic does not factor globally, and the two components are generically identified under monodromy of the quadratic equation. This results in non-Kodaira $I_n^*$ fibers in codimension three in five-folds. For a generic $SU(n)$ model in particular, the fiber in codimension three will be a monodromy reduced fiber with two of the multiplicity one curves getting identified. The condition for the monodromy to be absent is simply the vanishing of the discriminant $b_4^2 - 4 b_2 b_6 =0$. 
We will show this explicitly for $SU(5)$ in section \ref{sec:Examples}.

A second point to note is that in order for the singular five-fold to admit  a  flat Calabi-Yau resolution there must be no non-minimal singular loci. Non-minimal fiber enhancement occurs for vanishing orders of the Weierstrass model of the form $\hbox{ord}(f, g, \Delta) \geq (4,6,12)$.
We will oftentimes re-express the Weierstrass model in Tate-form (\ref{Tate}), with the role of $f$ and $g$ taken by the Tate coefficients $b_i$ transforming as certain sections on $B_4$.
In terms of these, the condition for non-minimality is that 
$\hbox{ord}(b_i; \Delta) \geq (1,2,3,4,6; 12)$.
Compared to Calabi-Yau three- and four-folds, for five-folds new constraints arise from requiring that no such non-minimal enhancement occurs in codimension four.
In numerous models this implies  that certain intersection loci in the base need to be trivial. 

As an  example consider a $IV^*$ fiber in codimension one, realizing an $E_6$ gauge theory with a Tate model with vanishing orders
$\hbox{ord}(b_i; \Delta)=(1,2,2,3,5;8)$. The fiber enhances in codimension two to $E_7$ realized by a type $III^*$ fiber, and in codimension three to $E_8$ with vanishing orders $(1,2,3,4,5;10)$. The only codimension four locus is $b_6=0$, which results in a non-minimal fiber.
In the following we will always remove such non-minimal loci by excluding such intersection points, in addition to the known non-minimal fibrations in codimension two  and three as listed in \cite{Lawrie:2012gg}. Removing such loci plays in particular a role, e.g in computing the anomalies and tadpole conditions.


\subsection{Singularity of Higher Codimension Loci}
\label{sec:chiSing}

Singularities of matter surfaces, i.e. codimension two loci in the base, can lead to corrections to the matter chiralities $\chi(S_{\bf R}, {\bf R}) $ in (\ref{chiS2}). The expression in (\ref{chiS2}) is applicable if there are no singularities on the matter surfaces. 
We will now explain where these effects come from, and provide examples for the corrected chirality formulas in the context of $SU(n)$ models. 
Without fluxes, i.e. for $L_{\bf R}= \mathcal{O}$, the expression (\ref{chiS2}) for the chirality can be rewritten via the adjunction formula  as 
\be\label{chiSmooth}
\chi(S_{\bf R}, {\bf R}) = {1\over 24} M_G \cdot [S_{\bf R}] \cdot \left(2 c_2 - c_1^2 + [S_{\bf R}]^2 + M_{G}^2\right) \,,
\ee
where the matter surface $S_{\bf R}$ on the 7-brane divisor  $M_G$ is written as the intersection of $M_G$  with a divisor whose class  we denote by slight abuse of notation also by $[S_{\bf R}]$ 
and $c_i = c_i(B_4)$. 
However, this expression does not account for contributions from singular matter surfaces. 
Let us define the elliptic fibration via a Tate form as in  (\ref{Tate}).
Generically, singular matter surfaces arise whenever the  divisor defining the codimension two locus on $M_G$ is the vanishing locus of a non-trivial polynomial in the coefficients $b_i$ in the Tate form as opposed to merely a monomial of $b_i$. The classic example is $SU(2k+1)$, where the fundamental matter arises along a  surface $S_{{\bf 2k+1}}$ given by the intersection of $M_G$ with the vanishing locus of
\be\label{Ppoly}
P=  b_1^2 b_6 - b_1 b_3 b_4 + b_2 b_3^2  \,.
\ee
There are now two effects: First, 
the codimension two locus $P=0$ is singular along $b_1 = b_3=0$. This implies  a curve of double points along 
\be
C_1 = M_G \cdot [b_1] \cdot [b_3] 
\ee
in the base $B_4$. Second, the discriminant of $P$, viewed as  a degree 2 polynomial in $b_1$ and $b_3$, is 
\be
\delta= b_4^2 - 4 b_2 b_6 \,.
\ee
Whenever $\delta=0$, the polynomial $P$ becomes a perfect square $P = \pi^2$, and there is a double curve 
\be
C_2 = M_G \cdot [P] \cdot [\delta] \,.
\ee
To account for the contributions of the singular curves we need to compute the Euler characteristic $\chi(C_1)$ and $\chi(C_2)$ and add these to the naive chirality formula (\ref{chiSmooth}). {Note that these may not be smooth and thus computing these contributions in general will require a more extensive treatment of such singularities.}

Furthermore, these two curves can intersect above points $M_G \cdot [b_1] \cdot [b_3]  \cdot [\delta]$.
A related effect was observed in the context of four-dimensional IIB orientifold models with pinch-point singularities along Whitney divisors wrapped by orientifold-invariant 7-branes in \cite{Collinucci:2008pf}. The correction to the Euler characteristic was determined by a local resolution, which essentially determines the contributions by counting the number of such pinch points. 
In our context the situation is somewhat more refined. 
In our case the correct expression for the chiral index of the fundamental representation of $SU(2k+1)$ is 
\be\label{ChiSUodd}
\chi (S_{\bf  2k+1}, {\bf 2k+1})^{total} =\chi(S_{\bf 2k+1}, {\bf 2k+1}) - {1\over 8} M_G \cdot [b_1] \cdot [b_3] \cdot ( [P] +  [\delta] ) \,.
\ee
Likewise for $SU(2k)$ we find the fundamental matter at $P=b_4^2 + b_1 b_3b_4 - b_6 b_1^2=0$, 
which is singular along $b_1=b_4=0$ and has  discriminant $\delta= b_3^2 + 4 b_6$. The corrected chirality count is then
\be\label{ChiSUeven}
\chi (S_{\bf  2k}, {\bf 2k})^{total} =\chi(S_{\bf 2k}, {\bf 2k}) - {1\over 8} M_G \cdot [b_1] \cdot [b_4] \cdot ([P] + [\delta]) \,.
\ee
Another example is one with gauge group $SO(10)$: Here the matter surfaces are all smooth, given by $b_3=0$ for the spin representation and $b_2=0$ for the fundamental, however there is a contribution from singular curves in codimension three, $b_3=\delta=0$ where now $\delta= b_4^2 - 4 b_2 b_6$. This gives a contribution 
\be\label{ChiSO}
\chi (S_{{\bf 16}}, {\bf 16})^{total} = \chi (S_{{\bf 16}}, {\bf 16}) + {1\over 4} M_G \cdot [b_3] \cdot [b_4] \cdot  [\delta] \,.
\ee
These correction terms have been somewhat empirically determined, by checking consistency of the anomaly and Chern-Simons constraints derived in section \ref{sec:ExamplesGlobal}. It would be very interesting to determine the found expressions from first principles by for instance resolving these singularities such as in \cite{Collinucci:2008pf}.


\section{{Example Fibrations}}
\label{sec:Examples}

Before we proceed with an in depth characterization of global consistency conditions of F-theory on Calabi-Yau five-folds, it will be useful to have a few examples in mind. These examples will be developed further in view of their global consistency in the second example section \ref{sec:ExamplesGlobal}. We will discuss here mainly the geometry of fibrations with non-abelian ADE type groups, with a focus on the odd $SU(2k+1)$ gauge groups, and discuss the corresponding geometry and 2d field theory associated to them. A second class of examples have, in addition to non-abelian gauge group factors, one additional rational section. Further theories with $SU(2k)$, $SO(2n)$ and exceptional groups are discussed in appendix \ref{app:Examples} and section \ref{sec:ExamplesGlobal}.


\subsection{$SU(2k+1)$ Theories}
\label{sec:SU5Ex}

{We begin our exploration of examples with $SU(2k+1)$ theories}. Geometrically, an $SU(2k+1)$ gauge group is realized by $I_{2k+1}$ fibers in codimension one. 
In general, we will not work with the Weierstrass form (\ref{Weier}), but with the so-called Tate form \cite{Bershadsky:1996nh, Katz:2011qp}
\be\label{Tate}
y^2 + b_1 xy + b_3 y = x^3 + b_2 x^2 + b_4 x + b_6 \,,
\ee
which is a simple coordinate change away from the Weierstrass form. It has the crucial advantage that the vanishing orders of the coefficients $b_i$ (which are sections of suitable line-bundles over the base) imply, without further tuning, the singularity type of the fibration. 
For instance the $SU(2k+1)$ theories have vanishing orders $\hbox{ord}_{\zeta_0}(b_i) = (0,1,k, k+1, 2k+1)$, whereas the Weierstrass form would require a suitable tuning to arrive at $\hbox{ord}_{\zeta_0}(f, g, \Delta) = (0,0,2k+1)$.
This is what allows resolving these models by toric resolutions \cite{Candelas:1996su,Candelas:1997eh} as exemplified for Calabi-Yau
four-folds in \cite{Blumenhagen:2009yv,Grimm:2009yu}.
The resolutions for these general fibers have been discussed in detail in \cite{Lawrie:2012gg}, up to codimension three. In the resolutions for five-folds, however, interesting new effects occur due to additional monodromies in the fibers. 

\begin{figure}
\centering
\includegraphics[width=15cm]{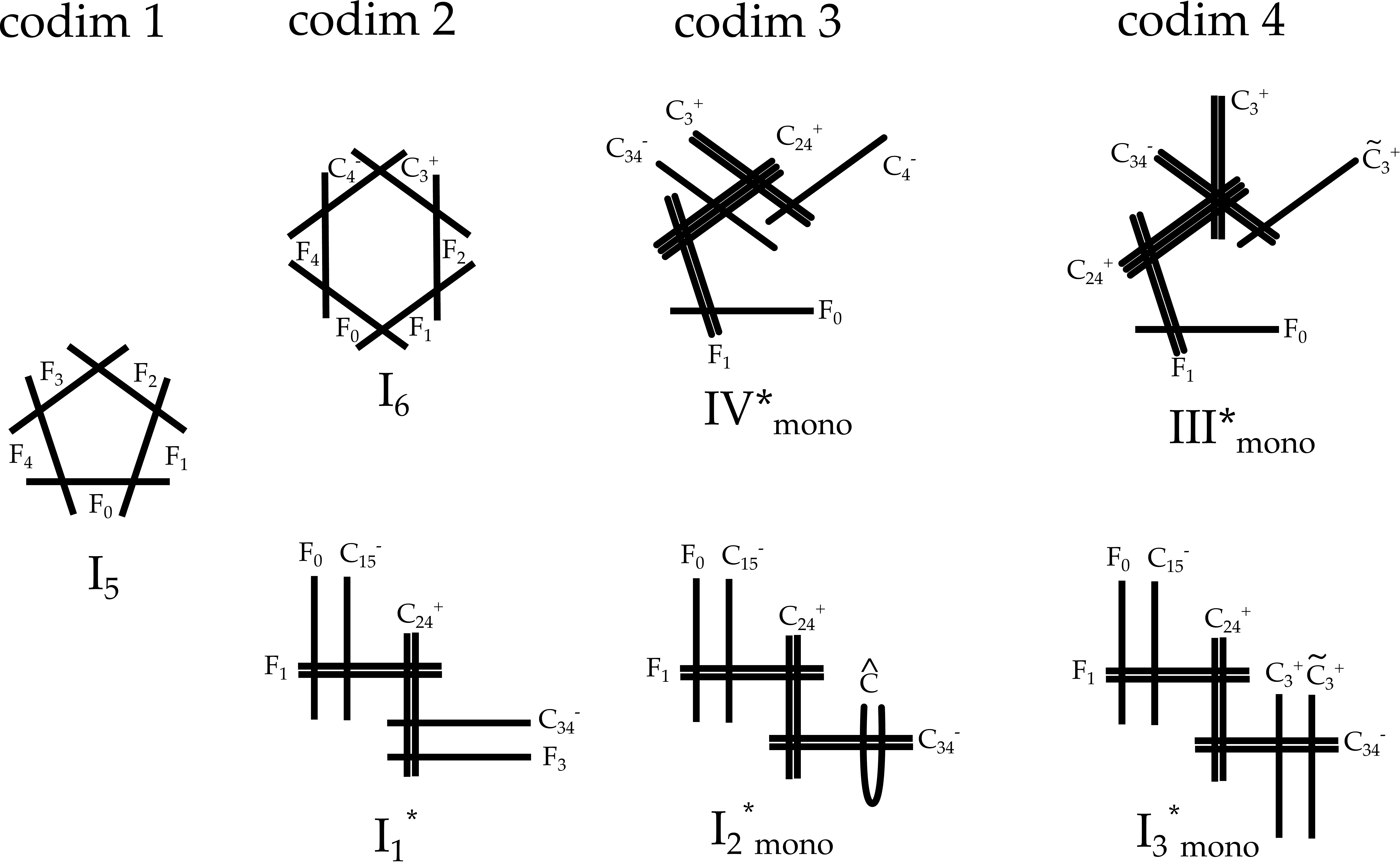}
\caption{Codimension one to  four fibers of an F-theory model with gauge group $SU(5)$
realized by  an $I_5$ fiber in codimension one. Lines correspond to rational curves, and multiple lines indicate the multiplicities of the fiber components. In codimension two, the fibers correspond to local enhancements to $SU(6)$ and $SO(10)$, respectively, and are given in terms of Kodaira fibers. All higher codimension fibers have monodromy reduction:  compared to the standard Kodaira fiber, components are absent due to monodromies. The resolution shown here is encoded in the box graph in figure \ref{fig:BoxGraphSU5}, and is realized in terms of the blowup sequence (\ref{ResSeq}). 
\label{fig:E6E7}}
\end{figure}

Here we start with the prototypical example of $SU(5)$ to illustrate the type of fibers that occur, {including the ones in} codimension four. {This example is particularly nice, as it has a rich class of higher codimension enhancements, including exceptional loci.}
The fibration is realized in terms of  a Tate model 
\be
y^2 + b_1 x y + b_3 \zeta_0^2 y = x^3 + b_2  \zeta_0 x^2 + b_4 \zeta_0^3x + \zeta_0^5 b_6\,,
\ee
with the singularity locus above $\zeta_0=0$. The classes of the coefficients $b_i$ are 
\be
[b_1]=c_1\,,\quad 
[b_2]= 2 c_1-M_G\,,\quad 
[b_3]=3 c_1-2 M_G\,,\quad 
[b_4]= 4 c_1-3 M_G\,,\quad 
[b_6]= 6 c_1-5 M_G \,.
\ee
The discriminant is
\be\ba
\Delta=& b_1^4 \left(b_2 b_3^2-b_1b_3 b_4+b_1^2 b_6\right) \zeta _0^5  + O(\zeta_0^6) \,.
\ea
\ee
Note that the discriminant locus $P\equiv b_2 b_3^2- b_1b_3 b_4+b^2_1 b_6=0$ is singular at $b_1 = b_3=0$, and there will be corrections to the chirality formulas as discussed in section \ref{sec:chiSing}.  
The complete enhancement patterns, including the putative unhiggsed gauge group in higher codimension, are summarized as follows:
\be\ba
&\ba
\hbox{Codim 2}: \quad \left\{
\ba
SO(10):&\quad b_1=0\cr 
SU(6): &\quad b_1^2 b_6 - b_1 b_3 b_4 + b_2 b_3^2=0
 \ea
 \right. 
\ea
\cr
&\ba
\hbox{Codim 3}: \quad \left\{
\ba
SO(12): &\quad b_1=b_3= 0 \cr  
E_6:&\quad b_1 =b_2=0 \ea
 \right. 
\ea
\cr 
&\ba
\hbox{Codim 4}:\quad  \left\{
\ba
SO(14): &\quad b_1=b_3=b_4^2-4 b_2 b_6= 0 \cr  
E_7:&\quad b_1 =b_2=b_3=0 
\ea
 \right. 
\ea\ea
\ee
To determine the actual fiber structure, as well as various topological quantities such as Chern classes, we resolve the model with the following resolution sequence\footnote{Using the notation in \cite{Lawrie:2012gg}, $(x, y, \zeta_1;\zeta_2)$ corresponds to the blowup of $x=y=\zeta_1=0$, where the exceptional section of the blowup is $\zeta_2$. Small resolutions are $(y, \zeta_1; \zeta_2)$ etc. }
\be\label{ResSeq}
\ba
(x,y,\zeta_0; \zeta_1) \,,\qquad (x,y,\zeta_1; \zeta_2) \,,\qquad (y,\zeta_1; \zeta_3) \,,\qquad (y, \zeta_2; \zeta_4) \,.
\ea
\ee
Applied to the standard Tate form the sections are associated to the simple roots via the correspondence $(\zeta_0, \zeta_1, \zeta_2, \zeta_3, \zeta_4) \leftrightarrow (\alpha_0, \alpha_1, \alpha_2, \alpha_4, \alpha_3)$. This corresponds to the toric triangulation introduced in \cite{Krause:2011xj,Krause:2012yh} as $T_{11}$ and in this algebraic form appeared in \cite{Lawrie:2012gg}. In the following, all resolutions and intersection computations are computed in {\tt Smooth} \cite{Smooth}. As introduced earlier, we denote the rational curves  in the fiber associated to the simple roots $\alpha_i$ by $F_i$.
The codimension two fibers correspond to the following splittings\footnote{To compare with the analysis of this splitting in the appendix of \cite{Krause:2011xj}, we give the map to the notation therein: $\zeta_0 \rightarrow w$, $b_1 \rightarrow a_1, b_2 \rightarrow a_{2,1}$, $b_3 \rightarrow a_{3,2}$, $b_4 \rightarrow a_{4,3}$, $b_6 \rightarrow a_{6,5}$, $F_i \rightarrow \mathbb P^1_i$, $C^+_{24} \rightarrow \mathbb P^1_{24}$, $C^-_{34} \rightarrow \mathbb P^1_{2B}$, $C_{15}^- \rightarrow \mathbb P^1_{4D}$, $C_3^+ + C_4^- \rightarrow \mathbb P^1_{3x} + \mathbb P^1_{3F}$. 
The toric resolution coordinates $e_i$ of \cite{Krause:2011xj} are associated with the Cartan divisors $E_i$, called $D_i$ in the present paper. Furthermore, to ease comparison with \cite{Braun:2014kla, Lawrie:2015hia}, this is the resolution/Coulomb phase 8 for the anti-symmetric representation and II for the fundamental. } 
\be
\begin{array}{c|c|c|l}
\hbox{Local Enhancement} & \hbox{Fiber Type} & \hbox{Codim 2 Locus} & \hbox{\quad Fiber Splitting}\cr \hline
SO(10) & I_1^* & b_1=0 & 
 \ba
F_2 \ &\rightarrow  \ C_{24}^+ + C_{34}^- \cr 
F_4 \ &\rightarrow  \ C_{24}^+ + F_1 +  C_{15}^- 
\ea
\cr\hline 
SU(6) & I_6 &  P=0&   F_3 \ \rightarrow \ C_3^+ + C_4^-
\end{array}
\ee
Here $C_i^\pm$ ($C_{ij}^\pm$) corresponds to the weight $\pm L_i$ ($\pm (L_i + L_j)$) of the fundamental (anti-symmetric) representation. In figure \ref{fig:E6E7}, the associated fibers are shown including multiplicities and intersections. The resolution corresponds to the Coulomb phases/box graphs in figure \ref{fig:BoxGraphSU5}.  

In codimension three these further split as follows (continuing the splitting from the codimension two locus $b_1=0$):
\be
\begin{array}{c|c|c|l}
\hbox{Local Enhancement} & \hbox{Fiber Type} & \hbox{Codim 3 Locus} & \hbox{\quad Fiber Splitting}\cr \hline
E_6 & {IV^*}_{mono} &  b_1=b_2=0  &
\ba
C_{15}^- \ &\rightarrow  \ C_{24}^+ + C_{3}^+ \cr 
F_3 \ &\rightarrow  \ C_{3}^+ + C_4^- 
\ea
\cr \hline
SO(12) & {I_2^*}_{mono} & b_1= b_3 = 0 & 
F_3 \ \rightarrow \  C_{34}^- + \hat{C}
\end{array}
\ee
Here $\hat{C}$ is defined by a quadratic equation, 
\be\label{QuadraticSO12}
\hat{C}:\qquad \zeta_3^2 b_6 + \zeta_3 x b_4+ x^2 b_2 =0\,,
\ee 
and corresponds to a curve intersecting $C_{34}^-$ twice.
Note that in four-folds, this would be a fiber above a codimension three locus, i.e. a point, where the quadratic factors into two lines \cite{EY}. 
These codimension three loci realize interactions of the type ${\bf 10} \times {\bf 10} \times {\bf 5}$ and $\overline{\bf 5} \times \overline{\bf 5} \times {\bf 10}$, respectively. We will determine the precise couplings in terms of $E$ and $J$ terms below. 

The codimension four fibers can be best understood by considering the enhancement from the $E_6$ locus with $IV^*_{mono}$ fibers. The extremal generators of the relative cone of effective curves there are
\be
\mathcal{K}_{IV^*_{mono}} = \{F_1, C_{24}^+, C_3^+, C_4^-, C_{34}^- \} \,,
\ee
which intersect in the monodromy reduced fiber shown in figure \ref{fig:E6E7}. 
Along the codimension four locus $b_1= b_2= b_3 =0$ the descriminant goes up to $\zeta_0^9$, and the $IV^*_{mono}$  fibers split as 
\be\label{IVIII}
III^*_{mono}: \qquad b_1=b_2=b_3=0 :\qquad 
C_{4}^- \ \rightarrow \ C_{34}^- + \tilde{C}_3^+ \,,
\ee
where the local enhancement is to $E_7$ and the  extremal generators of the relative cone of effective curves are
\be
\mathcal{K}_{III^{*}_{mono}} = \{F_1 , C_{24}^+, C_3^+,C_4^-, \tilde{C}_3^+ \} \,.
\ee
Equivalently, the splitting from ${I_2^*}_{mono}$ to $III^*_{mono}$ is 
\be
C_{15}^- \ \rightarrow \  C_{24}^+ + C_3^+ \,, \qquad \hat{C} \ \rightarrow \ C_3^+ + \tilde{C}_3^+ \,.
\ee
The fiber is shown in figure \ref{fig:E6E7}. It is a monodromy reduced $III^*$ fiber, which arises from the underlying Kodaira fiber by removing the component with multiplicity 4 and one of the multiplicity 3 ones. 
Note that $\tilde{C}$ has the same intersections as $C_3^+$, but these seem to be distinct curves.

These codimension four splittings of fibral curves have the following interpretation: starting with the codimension three coupling ${\bf 10} \times {\bf 10} \times {\bf 5}$ corresponding to the $E_6$ enhancement, we see that  the curve $C_{4}^-$ associated to the ${\bf 5}$ splits further  into the curves $C_{34}^-$ and $\tilde{C}_3^+$,  associated to states in the representations ${\bf 10}$ and ${\bf \bar{5}}$, respectively. This generates  a quartic coupling mediated by an M2-brane wrapping the interpolating 3-chain bounded by all these curves, much alike the codimension three case \cite{MS, Martucci:2015dxa}. Note that the 3-chain is localized above the codimension four point, and its volume vanishes in the F-theory limit.   
Similar reasoning applies to the splitting from $I_2^*$ to $III^*$. In summary, the two codimension four loci realize quartic interactions of the type 
\be
{\bf 10} \times{\bf 10} \times{\bf 10} \times {\bf \overline{5}} \qquad \hbox{and }\qquad {\bf \bar{5}} \times {\bf \bar{5}} \times {\bf \bar{5}} \times {\bf \overline{10}} 
\ee
localized over the point of $E_7$ enhancement. 

Finally, consider the splitting along $b_4^2-4 b_2 b_6= 0$ of the matter locus $b_1=b_3=0$. This is precisely the discriminant of (\ref{QuadraticSO12}), and thus all that happens at this locus is that the curve $\hat{C}$ in the enhancement from $I_{2\, mono}^{*}$ factors
\be
I_{3\, mono}^*:\qquad b_1=b_3=b_4^2-4 b_2 b_6= 0:\qquad \hat{C} \quad \rightarrow \quad  C_3^+ + \tilde{C}_3^+ \,.
\ee
Note that this is again a monodromy-reduced fiber, where one of the multiplicity two sets of curves is absent. All codimension three and four fibers follow the monodromy-reduction rules set out in \cite{Hayashi:2014kca, CLSSN} that they are given in terms of Kodaira fibers where nodes of the affine Dynkin diagram are deleted (irrespective of higher multiplicities). 
In the presence of singlets at this point, this would correspond to a coupling ${\bf 5} \,  \bar{\bf 5}\,  {\bf 1}$, but in absence of an extra $U(1)$ group no such singlet states are available. 

The chiralities for the two matter curves are 
\be\label{chisSU5}
\ba
\chi(b_1, {\bf 10}) &=\frac{1}{24} c_1 M_G \left(2 c_2+M_G^2\right) \cr 
\chi (P, {\bf 5}) & = \frac{1}{24} M_G \left(8 c_1-5 M_G\right) \left(-80 c_1 M_G+63 c_1^2+2 c_2+26 M_G^2\right) \cr 
&\quad + \frac{1}{8} c_1 M_G \left(16 c_1-11 M_G\right) \left(3 c_1-2 M_G\right) \cr 
&= \frac{1}{12} M_G \left(271 c_1 M_G^2-5 \left(76 c_1^2+c_2\right) M_G+180 c_1^3+8 c_2 c_1-65 M_G^3\right)\,,
\ea
\ee
where for the fundamental matter we included the correction due to the singular matter locus (\ref{ChiSUodd}). For later global considerations, note that the fourth Chern class in this resolution is
\be\label{c4SU5}
M_G \cdot_{Y_5}   c_4(Y_5) = M_G \cdot_{B_4} \left( 360 c_1^3-750 c_1^2 M_G+525 c_1 M_G^2+12 c_2 c_1-120 M_G^3\right) \,.
\ee
The even $SU(2k)$ theories proceed similarly, and we derive some of the details for the example $SU(6)$ in the appendix \ref{app:SU6}. 


\subsection{2d Gauge Theories} \label{sec_2dgaugetheory}

Let us exemplify the structure of the  2d $(0,2)$ theories {obtained} for the above $SU(5)$ model (without {any} gauge backgrounds turned on). 
We assume that there is only one D3-brane in the model wrapping a single curve in $B_4$. In this situation, the D3-sector contributes at best a massive $U(1)$ gauge multiplet to the 2d theory (see the discussion at the end of section \ref{sec_2dAbelian}). 
At the massless level, apart from the $SU(5)$ gauge multiplet, the theory contains charged massless matter fields as summarized in the following table:
\be \label{table-matterspecSU5} 
\begin{array}{c|c|c}
\hbox{{Matter Rep}}  & \hbox{Massless {Fields} } &  \hbox{Muliplicity}    \cr \hline \vspace{-4mm} \cr 
{\bf 24} &  \rho_-^{\bf 24} &  q_\alpha = 1, \ldots, h^2(M_{SU(5)}) \cr
{\bf 24} & A^{\bf 24} = (a^{\bf 24},\psi_{+}^{\bf 24})  & p_\beta = 1, \ldots, h^1(M_{SU(5)}) \cr
{\bf 24} &  \Phi^{\bf 24} = (\varphi^{\bf 24},\chi_+^{\bf 24}) &  r_\gamma = 1, \ldots, h^3(M_{SU(5)})   \cr \hline \vspace{-4mm} \cr 
 {\bf \overline{10}} & {  \mu_-^{\bf \overline{10}} }  & i_\alpha = 1, \ldots, h^1(S_{\bf 10}, \sqrt{K_{S_{\bf 10}}}) \cr
{\bf 10} & {\cal T}^{\bf 10} = (T^{\bf 10} , \tau^{\bf 10} _+)& j_\beta = 1, \ldots h^0(S_{\bf 10}, \sqrt{K_{S_{\bf 10}}}) \cr
{\bf \overline{10}} & {\cal S}^{\bf \overline{10}} = (S^{\bf \overline{10}} , \sigma^{\bf \overline{10}} _+)& k_\gamma = 1, \ldots, h^2(S_{\bf 10}, \sqrt{K_{S_{\bf 10}}}) \cr \hline \vspace{-4mm} \cr 
{\bf 5} & {  \mu_-^{\bf {5}}}  & l_\alpha = 1, \ldots, h^1(S_{\bf 5}, \sqrt{K_{S_{\bf 5}}}) \cr
{\overline{\bf 5}} & {{\cal T}^{\bf \bar{5}} = (T^{\bf \bar{5}}, \tau^{\bf \bar{5}}_+)} & m_\beta = 1, \ldots h^0(S_{\bf 5}, \sqrt{K_{S_{\bf 5}}}) \cr
{\bf{5}} & {{\cal S}^{\bf {5}} = (S^{\bf {5}}, \sigma^{\bf {5}}_+)} & n_\gamma = 1, \ldots, h^2(S_{\bf 5}, \sqrt{K_{S_{\bf 5}}}) \cr \hline \vspace{-4mm} \cr 
{\bf \bar{5}} & \nu_-^{{\bf \bar{5}}}  & s_\alpha = 1, \ldots, h_{\bar \partial}^{0}(M_{SU(5)} \cap C^B_{\rm M2})  \cr
\end{array}
\ee
The last line refers to the 3-7 matter discussed in section \ref{sec:D3Sector}.

At the level of cubic non-derivative couplings, the `bulk' matter in the ${\bf 24}$ interacts via pure bulk couplings of the from (\ref{Sbulk1})  and (\ref{bulk-J-coupling}) as well as via bulk-surface interactions (\ref{bulkmatterint}).
Let us choose the convention that the different types of matter fields localised along $S_{\bf 10}$ transform in representation ${\bf 10}$  versus ${\bf \overline{10}}$ as displayed in (\ref{table-matterspecSU5}), and similarly for ${\bf 5}$ versus ${\bf \bar 5}$.  With the assignment given above, the surface matter in the ${\bf 10}$ and ${\bf 5}$ representations couples in addition via $E$- {and $J$}-type interactions as discussed in  section  \ref{sec_holocouplings}.  At generic position of the D3-branes, no $E$- and $J$-type couplings are possible involving the Fermi multiplets   $ \nu_-^{\bf \bar 5} $ for the reasons given in section  \ref{sec:D3Sector}.    
The {allowed} couplings lead to the following {ans\"atze} for the auxiliary $E$-fields, 
\be
\ba
 - E^{( \rho_-^{{\bf 24},{q_\alpha}})} &= 
 {\bf f}_{q_\alpha p_\beta r_\gamma}  A^{{\bf 24},p_\beta} \Phi^{{\bf 24}, r_\gamma}  +  
{\bf b}^1_{q_\alpha j_\beta k_\gamma}  {\cal T}^{{\bf 10},j_\beta} {\cal S}^{{\bf \overline{10}}, k_\gamma}  + 
 {\bf b}^2_{q_\alpha m_\beta n_\gamma}  {\cal T}^{{\bf \bar{5}},m_\beta} {\cal S}^{{\bf 5}, n_\gamma}                            \cr
 - E^{( \mu_-^{{\bf \overline{10}},{i_\alpha}})} &= 
 {\bf d}^1_{i_\alpha j_\beta n_\gamma}  {\cal T}^{{\bf 10},{j_\beta}}   {\cal S}^{{\bf 5},{n_{\gamma} }}  +    {\bf d}^2_{i_\alpha m_\beta j_\beta}   {\cal T}^{{\bf {\bar{5}}},{m_\beta}}   {\cal T}^{{\bf {\bar{5}}},{j_\beta}}            +  \,    {\bf e}^1_{i_\alpha k_\gamma p_{\gamma}}   {\cal S}^{{\bf{\overline{10}}},{k_\gamma}}  {A}^{{\bf 24},{p_\gamma}}     \cr
 - E^{(\mu_-^{ {\bf 5},{l_\alpha}})} &= 
 {\bf d}^3_{l_\alpha m_\beta j_\beta}   {\cal T}^{{\bf {\bar{5}}},{m_\beta}}  {\cal T}^{{\bf  {10}},{j_\beta} }   +  {\bf d}^4_{l_\alpha m_\beta j_\beta}   {\cal S}^{{\bf {\overline{10}}},{m_\beta}}  {\cal S}^{{\bf  \overline{10}},{j_\beta} }    +  \,  {\bf e}^2_{l_\alpha m_\gamma p_{\gamma}}   {\cal S}^{{\bf{ 5}},{m_\gamma}}  {A}^{{\bf 24},{p_\gamma}}   \,,
\ea
\ee
and for the superpotential $J$
\be
\ba
 - J^{( \rho_-^{{\bf 24},{q_\alpha}})} & =  
 {\bf g}_{q_\alpha p_\beta p_\gamma}  A^{{\bf 24},p_\beta}  A^{{\bf 24},p_\gamma}      \cr
 - J_{\mu_-^{{\bf \overline{10}},{i_\alpha}}}  &=
  {\bf h}^1_{i_\alpha k_\gamma m_{\gamma}}    {\cal S}^{{\bf \overline{10}},{k_\gamma}}  {\cal T}^{{\bf \bar{5}},{m_\gamma}}  
{+ {\bf h}^2_{i_\alpha m_\beta m_{\gamma}}   {\cal S}^{{\bf {5}},{m_\beta}}  {\cal S}^{{\bf {5}},{m_\gamma}}  }
   + \,    {\bf c}^1_{i_\alpha j_\beta p_{\gamma}}   {\cal T}^{{\bf{10}},{j_\beta}}  {A}^{{\bf 24},{p_\gamma}}   
\cr
 - J_{\mu_-^{{\bf  5},{l_\alpha}}} &= 
{  {\bf h}^{3}_{l_\alpha j_\beta j_\gamma}  {\cal S}^{\overline{\bf 10},{j_\beta}   }{\cal S}^{{\bf 5},{j_{\gamma} }} }   +  {\bf h}^{{4}}_{l_\alpha j_\beta j_\gamma}  {\cal T}^{{\bf 10},{j_\beta}   }{\cal T}^{{\bf 10},{j_{\gamma} }} 
   + \,  {\bf c}^2_{l_\alpha m_\beta p_{\gamma}}   {\cal T}^{{\bf{\bar 5}},{m_\beta}}  {A}^{{\bf 24},{p_\gamma}} \,.
\ea
\ee
The bulk couplings ${\bf f}$ and ${\bf g}$ and the bulk-surface couplings  ${\bf b}^i$, ${\bf c}^i$ and ${\bf e}^i$ are computed as wavefunction  overlaps as in (\ref{Sbulk2}), (\ref{gcoupling}) and (\ref{bulkoverlaps}), while the surface matter couplings ${\bf d}^i$ {and ${\bf h}^i$} arise from the overlaps at codimension three curves, (\ref{mattercubed1}) and  (\ref{mattercubef1}). In addition we have seen quartic interactions from the codimension four points in the base of the fibrations.

Supersymmetry requires that the explicit form of the couplings as determined from the wavefunction overlaps must be compatible with the constraint
\be \label{JESU5}
\sum_{i_\alpha}  E^{( \mu_-^{{\bf \overline{10}},{i_\alpha}})}  J_{\mu_-^{{\bf \overline{10}},{i_\alpha}}}   + \sum_{l_\alpha}  E^{(\mu_-^{ {\bf 5},{l_\alpha}})} J_{\mu_-^{{\bf 5},{l_\alpha}}} = 0 \,.
\ee
Since the precise information about the couplings is encoded entirely in the geometry of the internal wavefunctions, consistency of the compactification will ensure that the constraint (\ref{JESU5}) is indeed satisfied.


\subsection{$SU(2k+1)\times U(1)$ Theories}

We have seen that the monodromy in (\ref{QuadraticSO12})  is due to the quadratic equation describing the fiber above the codimension three locus $b_1=b_3=0$. We can force the monodromy to be reduced by considering $b_6=0$ or $b_2=0$. The former is exactly the so-called $U(1)$-restricted Tate model of \cite{Grimm:2010ez}, which has gauge group $SU(5)\times U(1)$. In addition to the resolutions in (\ref{ResSeq}), we also blow up $(x, y; s)$, where $s=0$ corresponds to the additional rational section that gives rise to the abelian gauge factor. Resolutions of this $SU(5)\times U(1)$ model including the complete set of curve splittings in codimension two and three have been discussed torically in \cite{Krause:2011xj} and from algebraic resolutions in \cite{Kuntzler:2014ila}, to which we refer for more details. The complete set of fiber splittings, i.e. all resolutions, determined in terms of Coulomb phases for models with $U(1)$s can be found in \cite{Lawrie:2015hia}.

In fact we can state more generally that any model with $I_{2k+1}$ singularity in codimension one with a $U(1)$-restriction 
\be
b_6=0 
\ee
 in the Tate model guarantees that the monodromy in the $I_m^*$ codimension three fiber is absent, as can be readily seen from the factorization of the locus (\ref{MonoReduxEq}). We will show later on that this class of models is globally consistent and anomaly free. Furthermore, the matter loci are all smooth and there are no singular contributions to the chiral index of the surface matter.


\section{The Flux Sector} \label{sec_G4fluxes}

\subsection{Four-form Fluxes on Five-folds} \label{fluxes5folds}

An important ingredient in F/M-theory compactifications on Calabi-Yau five-folds is the flux background, which is the vacuum expectation value for the field strength $G_4 =d C_3$ of the M-theory 3-form potential.
Let us first briefly review the situation on a general Calabi-Yau five-fold as studied in \cite{Haupt:2008nu}.
The M-theory flux is described by an element $G_4 \in H^{4}(Y_5)$ subject to the Freed-Witten quantization condition \cite{Witten:1996md}
\be
G_4 + \frac{1}{2} c_2(Y_5) \in H^4(Y_5,\mathbb Z)\,.
\ee
Important aspects of this quantization condition have been discussed in detail for elliptically fibered four-folds in \cite{Collinucci:2010gz,Collinucci:2012as}.
On a Calabi-Yau five-fold  $H^{4}(Y_5)$ splits into $H^{3,1}(Y_5)$, $H^{2,2}(Y_5)$ and $H^{1,3}(Y_5)$.  As shown in \cite{Haupt:2008nu}, in order for the M-theory compactification on $Y_5$ to preserve 
 two supercharges, the $(3,1)$ and $(1,3)$ flux components must vanish and thus
\be
G_4 + \frac{1}{2} c_2(Y_5) \in H^4(Y_5,\mathbb Z) \cap H^{2,2}(Y_5)\,.
\ee
The remaining $(2,2)$ flux induces, in the effective ${N}=2$ super-mechanics, a scalar potential for the K\"ahler moduli of $Y_5$ that derives from the superpotential  \cite{Haupt:2008nu}
\be \label{M-theorysuperpot}
W_{\rm flux} = \int_{Y_5} G_4 \wedge J \wedge J \wedge J\,.
\ee

Let us now specialise to M-theory compactifications on elliptically fibered Calabi-Yau five-folds. By {M/F-theory duality (\ref{MFDuality})},  $G_4$ flux encodes both the analogue of the Type IIB/F-theory closed string Neveu-Schwarz and Ramond-Ramond fluxes and the gauge fluxes on the 7-branes. 
In order for $G_4$ to uplift to these types fluxes in the F-theory vacuum, it must satisfy the transversality constraints
\be \label{transverse1}
\int_{Y_5} G_4 \wedge S_0 \wedge \omega_4 = 0 \quad 
{\hbox{and}}
\quad  \int_{Y_5} G_4 \wedge \omega_6 = 0\,, \qquad \forall  \, \omega_4 \in H^4(B_4), \, \, \omega_6 \in H^6(B_4)\,.
\ee
These are the direct analogue of the familiar constraints first discussed in \cite{Dasgupta:1999ss} for $G_4$-fluxes in M/F-theory compactifications on Calabi-Yau four-folds to $3/4$ dimensions.
In the first condition, $S_0$ denotes the divisor defined around (\ref{S0Fint}) associated with the zero-section  of the elliptic fibration.\footnote{The part of $S_0$ in genus-one fibrations without a section is taken by a suitable modification of the divisor class describing the embedding of the base $B_4$ into $Y_5$ as analysed in  \cite{Lin:2015qsa}.} 
It rules out fluxes with all legs in the base as these would not survive the M/F-theory scaling limit. This is consistent with the absence of 4-form fluxes on the compactification space of F-theory/Type IIB vacua.
The second constraint ensures that the flux does not have two legs along the generic fiber as such flux would break Poincar\'e invariance in the dual F-theory.
If we insist that the flux do not break the non-abelian gauge symmetry on the 7-branes, we demand in addition that 
\be \label{transverse2}
\int_{Y_5} G_4 \wedge D_i \wedge \omega_4 =0 \qquad \quad \forall  \, \omega_4 \in H^4(B_4) \,.
\ee 
In this work we are primarily interested in $G_4$ fluxes which uplift to gauge flux along the 7-branes in F-theory.
The constraint that $G_4$ be of  $(2,2)$ type reproduces the BPS condition (\ref{20Fterm}) that the associated gauge flux be of (1,1) type. The supersymmetry condition induced by the superpotential (\ref{M-theorysuperpot}) in M-theory uplifts in F-theory to the requirement that 
\bea \label{fluxDterm1}
\int_{Y_5} G_4 \wedge S_m \wedge J_B \wedge J_B = 0
\eea
for all classes $S_m$ generating a $U(1)_m$ gauge symmetry via the Shioda map (\ref{Shiodamap}). $J_B$ is the K\"ahler potential on the base $B_4$. To see this expand the K\"ahler form of $Y_5$ as
\bea
J = t_0 S_0 + \sum_m t_m S_m + \sum t_i D_i + J_B
\eea
 and require that the derivative of  (\ref{M-theorysuperpot}) with respect to the K\"ahler moduli $t_0, t_m, t_i$ and the K\"ahler moduli on the base vanish. In the F-theory limit, where $t_0 \rightarrow 0$, $t_m \rightarrow0 $ and $t_i \rightarrow 0$, the only non-trivial constraint for fluxes satisfying (\ref{transverse1})  and (\ref{transverse2}) is (\ref{fluxDterm1}). 
 Note that (\ref{fluxDterm1}) corresponds to the BPS condition of vanishing D-term (\ref{DtermBPS}) (for trivial  charged matter field VEVs). From the perspective of the 2d $(0,2)$ theory obtained from F-theory this amounts to the vanishing of  the flux-induced field-dependent $U(1)_m$ Fayet-Iliopoulos term as will be discussed after (\ref{phikform}). 


\subsection{Extracting gauge bundles from $G_4$}\label{chiralitywithflux}

 The flux associated with a non-trivial gauge background has been described in sections \ref{sec:8dSYM} and \ref{sec:Matter} as the field strength of a line or in general vector bundle on the complex three-cycle $M_G$ wrapped by the 7-brane. Suitable powers of this bundle enter the cohomology groups (\ref{cohomologies-bulk}) and (\ref{table-matterspec}) counting, respectively, charged bulk matter along $M_G$ and charged matter at the intersection $S_{\bf R}$ of two 7-branes. The description of fluxes in terms of a gauge bundle sharply localised along individual 7-brane cycles is correct when the structure group of the associated bundle is contained in one of the non-abelian gauge groups of the model. 
Most gauge fluxes, however, are not of this form. This is because they are either associated  
to massless non-Cartan abelian gauge symmetries, or given in terms of even more general elements of $H^{2,2}(Y_5)$ with no connection to a massless gauge symmetry at all. As argued in \cite{Krause:2012yh}, the latter type of flux is to be interpreted as the F-theory analogue of gauge flux associated with geometrically massive $U(1)$ symmetries in the sense of \cite{Grimm:2011tb}.
While such fluxes are not localised in any way along the non-abelian 7-brane cycles $M_G$, the matter spectra (\ref{cohomologies-bulk}) and (\ref{table-matterspec}) only depend on the restriction of the gauge flux to the matter loci in question. It is therefore sufficient to extract these gauge data from a globally specified 4-form flux. 
Quite generally, since we are working on the Coulomb branch of the M-theory, i.e. on a resolved Calabi-Yau five-fold, we can only access the abelian gauge data. The bundles we can extract from $G_4$ are therefore necessarily line as opposed to higher rank vector bundles. Possible extensions to including non-abelian gauge data were obtained in \cite{Cecotti:2010bp,Donagi:2011jy,Marsano:2012bf,Anderson:2013rka, Collinucci:2014taa}.

Consider matter in representation ${\bf R}$ localised on a surface $S_{\bf R}$. The line bundle $L_{\bf R}$ whose cohomology groups count this matter as in (\ref{table-matterspec}) is related to the gauge flux $G_4$ as follows: Pick a fibral curve $C^{\varepsilon(\lambda^{\bf R}_a)}_{\lambda^{\bf R}_a} $ associated with one of the weights $\lambda^{\bf R}_a$ and assume for definiteness that $\varepsilon(\lambda^{\bf R}_a)=1$. Integration of $G_4$ over  this fibral curve gives rise to a 2-form on $S_{\bf R}$ which precisely describes the gauge flux to which the matter states in representation ${\bf R}$ couple.
This intuitive notion can be formalized as in \cite{Bies:2014sra} by describing the gauge data on $Y_5$ in terms of an element ${\mathcal G}$ of ${\rm CH}^2(Y_5)$, the rational equivalence class of complex codimension two cycles on $Y_5$. The cohomology class associated with ${\mathcal G}$ is precisely the gauge flux $G_4 \in H^{2,2}(Y_4)$, but viewed as an element of ${\rm CH}^2(Y_5)$  ${\mathcal G}$ contains considerably more information including that of the 'Wilson line' backgrounds of the 3-form $C_3$.
The fibration of $C^{\varepsilon(\lambda^{\bf R}_a)}_{\lambda^{\bf R}_a} $ over $S_{\bf R}$ describes by itself an element ${\mathcal C}_{\bf R}$   of ${\rm CH}^2(Y_5)$. 
At the level of intersection theory within the Chow ring, the notion of integrating $G_4$ over the fiber curve $C^{\varepsilon(\lambda^{\bf R}_a)}_{\lambda^{\bf R}_a} $ amounts to taking the pullback of ${\mathcal G}$ to ${\mathcal C}_{\bf R}$ and projecting onto $S_{\bf R}$. This gives rise to an element in ${\rm CH}^1(S_{\bf R})$, the group of line bundles on $S_{\bf R}$, which we identify with $L_{\bf R}$.\footnote{More precisely viewing ${\mathcal C}_{\bf R}$  and ${\cal G}$ as elements of ${\rm CH}_3(Y_5)$, this intersection-theoretic process defines an element of ${\rm CH}_{3+ 3 -5}(Y_5|_{S_{\bf R}})$, whose projection to $S_{\bf R}$  is an element of  ${\rm CH}_1(S_{\bf R}) \simeq {\rm CH}^1(S_{\bf R})  = {\rm Pic}(S_{\bf R})$. See sections 2.4 and 3.1 in   \cite{Bies:2014sra} for more details on the analogous construction on Calabi-Yau four-folds.} The result is independent of the choice of  $\lambda^{\bf R}_a$ inside the weight system of ${\bf R}$. The cohomology groups of $L_{\bf R}$ then count the massless matter according to (\ref{table-matterspec}) in presence of gauge data encoded in ${\cal G}$. This procedure will be exemplified in section \ref{sec:SU5U1consistency}. A similar construction extracts the line bundles relevant for the bulk sector in (\ref{cohomologies-bulk}). 

 The chiral index  (\ref{chiS2}) associated with these cohomology groups can be written as 
 \be
 \chi(S_{\bf R}) =  \left.\chi(S_{\bf R})\right|_{c_1(L_{\bf R}) = 0} +  \left.\chi(S_{\bf R})\right|_{\rm flux} \,.
 \ee
In absence of singularities of the type discussed in section \ref{sec:chiSing}, the flux-dependent part takes the form
 \be \label{chirality-fluxdep}
\left. \chi(S_{\bf R})\right|_{\rm flux} = \frac{1}{2} \int_{S_{\bf R}}c_1^2(L_{\bf R}) \,.
 \ee
 In section \ref{sec:CScouplings} we will see that 
 this piece is now directly related to the integrals $\frac{1}{2} (G_4 \wedge G_4) \cdot_{Y_5} D_i$ with $D_i$ the Cartan divisors given by fibering the {resolution $\mathbb P^1$s, $F_i$,} over the 7-branes or, in the presence of abelian gauge groups, to the integrals $\frac{1}{2} (G_4 \wedge G_4) \cdot_{Y_5} S_m$ with $S_m$ defined in (\ref{Shiodamap}).
 For instance suppose that a fibral $F_i$ associated with the simple root $\alpha_i$ splits into $C^+_i \cup C_{i+1}^-$ in the fiber over $S_{\bf R}$ such that the latter appear in the weight system of \emph{only} the representation ${\bf R}$ (and of no other representation). Then we will find that
\be
 -\frac{1}{2} G_4 \wedge G_4 \cdot_{Y_5} D_i = \frac{1}{2} \int_{S_{\bf R}} c_1(L_{\bf R})^2\,.
\ee
More generally, the methods developed in section \ref{sec:CScouplings} will allow us to systematically express the expression on the left as a linear combination of $\chi(S_{{\bf R}_i})|_{\rm flux}$ for several representations ${\bf R}_i$. We leave it as an interesting task for future work to 
derive these identities directly from the intersection theoretic relation between $L_{\bf R}$ and $G_4$ outlined above.


\section{Global Consistency Conditions and Anomalies} 
\label{sec:GlobalConsistency}

We are {now  in a} position to study the global consistency conditions for the construction of  2d F-theory vacua. 
The D3-bane tadpole, {which will be} analysed in section {\ref{sec_tadpole}} is crucial for cancellation of gauge anomalies in the 2d $(0,2)$ theory because of the chiral nature of matter from strings stretched between the D3{-} and 7-branes. The gauge anomalies will be discussed in detail in section \ref{sec_Anomalies}. 
In particular we will uncover a rich pattern of  
Green-Schwarz and St\"uckelberg type couplings, which are essential in the context of abelian gauge  anomalies.

\subsection{Tadpole Constraints}  \label{sec_tadpole}

The effective supergravity action for 11d M-theory on $\mathbb R \times Y_5$ contains two types of topological couplings of the 3-form potential  $C_3$ with field strength $G_4$,
\be
S_{\rm M} = {2 \pi} \left( \int_{\mathbb R \times Y_5} d^{11} x \sqrt{-g} R -  \frac{1}{2}\int_{\mathbb R \times Y_5} G_4 \wedge \ast G_4 \right)  + S_{\rm top} \,,
\ee
where $S_{\rm top}$ has the contributions 
\be\ba \label{C3couplings}
S_{\rm top} &= S_{\rm M2} + S_{\rm curv}  \cr
S_{\rm M2} &=  - {2 \pi} \int_{\mathbb R \times Y_5} C_3 \wedge \delta([C_{\rm M2}])  \cr 
S_{\rm curv} &=  {2 \pi} \int_{\mathbb R \times Y_5}  C_3 \wedge \left( \frac{1}{24} c_4(Y_5) - \frac{1}{6} G_4 \wedge G_4\right)\,.
\ea\ee
Here $[C_{\rm M2}]$ denotes the class of all curves on $Y_5$ wrapped by M2-branes. The non-compact part of the M2-brane worldvolume fills the time direction ${\mathbb{R}^{1,0}}$ of the effective super-mechanics theory. We are working in units in which the 11d Planck length $\ell_M = 1$. In general we allow for a non-trivial 4-form flux $G_4 \in H^{2,2}(Y_5)$ as introduced in section \ref{sec_G4fluxes}.

Given a basis $\{\omega_\alpha\}$, $\alpha = 1, \ldots, h^{(1,1)}(Y_5)$ of 2-forms on $Y_5$ we can expand $C_3$ as  
$C_3 = \sum_\alpha A_\alpha \wedge \omega_\alpha + \ldots$, where $A_\alpha$ denote 1-form potentials in $\mathbb R$.
Under this expansion the couplings (\ref{C3couplings})  induce the $(1+0)$-dimensional analogue of a Chern-Simons coupling \cite{Haupt:2008nu},
\bea
S_{\rm top} 		&=& 2\pi  \sum_\alpha   \int_{\mathbb R} A_\alpha \wedge (k_{\rm M2}^\alpha + k_{\rm curv}^\alpha) \label{CSgeneral} \\
k_{\rm M2}^\alpha 	& =& -  \int_{Y_5} \omega_\alpha \wedge   \delta([C_{\rm M2}])  \label{CSclassical} \\
k_{\rm curv}^\alpha  	&=&  \int_{Y_5} \omega_\alpha \wedge      \left(\frac{1}{24}   [c_4(Y_5)] - \frac{1}{2} G_4 \wedge G_4 \right) \label{CSloop}   \,.
\eea

In section \ref{AnomalyInflow} we will see that the Chern-Simons couplings for the 1-form fields $A_a$, $a= 1, \ldots, H^{1,1}(B_4)$, arise by dimensional reduction of a classical topological coupling in 2d F-theory upon circle reduction  to M-theory. By contrast, the Chern-Simons terms for the remaining 1-forms, studied in detail in section \ref{sec:CScouplings}, have no analogue in the 2d $(0,2)$ theory obtained from F-theory. They are induced at the quantum level in the process of this circle reduction.
Irrespective of their origin,  in (1+0) dimensions the Chern-Simons couplings constitute tadpoles for $A_\alpha$ and must therefore vanish.
This results in the (1+0)-dimensional analogue \cite{Haupt:2008nu} of the M2-brane tadpole cancellation condition familiar from higher-dimensional M-theory compactifications 
\bea \label{3-brane-tadpole1}
\delta([C_{\rm M2}]) = \frac{1}{24} c_4(Y_5) - \frac{1}{2} G_4 \wedge G_4 \,.
\eea
This tadpole condition can only be satisfied for $\delta([C_{\rm M2}]) \in H^8(Y_5, \mathbb Z)$. 
Otherwise, the compactification is inconsistent and must be discarded. 
In section \ref{sec_S2Matter} we had encountered another integrality condition for consistency of the spectrum:
The bundles $L_{\bf R} \otimes K_{S_{\bf R}}^{1/2}$ appearing in the cohomology groups in (\ref{table-matterspec}) counting massless matter states at the intersection of two 7-branes must also be integer quantized. We conjecture that this is guaranteed whenever $G_4 + \frac{1}{2} c_2(Y_5) \in H^4(Y_5,\mathbb Z)$ and
the right-hand side of (\ref{3-brane-tadpole1}) is integer-quantized. Indeed, $c_4(Y_5)$ is sensitive to the global details of the 7-brane configuration. 

We can now decompose the class $[C_{\rm M2}]$ appearing on the left-hand side of (\ref{3-brane-tadpole1})
into a base component $[C^B_{\rm M2}] \in H_2(B_4)$ and a remaining fibral part. M2-branes wrapping curves on the base $B_4$ dualize, upon M/F-theory duality, to D3-branes wrapping the same curve and filling the two spacetime dimensions of F-theory compactified on $Y_5$. 
 Supersymmetry requires that this base class be effective on $B_4$ 
\be \label{CM2B}
[C^B_{\rm M2}] = \frac{1}{24} [c_4(Y_5)]_{B} - \frac{1}{2} [G_4 \wedge G_4]_B \, \,  {\geq} \, \, 0 \,.
\ee
This ensures that the D3-brane tadpole can be canceled with D3-branes only, as opposed to anti-D3-branes.
As in higher-dimensional compactifications this implies a bound on the allowed {$G_4$-flux}.

By contrast, the components of $[C_{\rm M2}]$ along  the fiber dualize to matter particles in the 2d F-theory compactification. 
In fact, there exists an intriguing interpretation of the righthand side of the tadpole equation (\ref{3-brane-tadpole1}) regarding its uplift to F-theory, which is the subject of section \ref{sec:CScouplings}.


\subsection{Anomalies in 2d} \label{sec_Anomalies}

Two-dimensional gauge theories exhibit gauge and gravitational anomalies \cite{AlvarezGaume:1983ig}. These are generated by anomalous 1-loop diagrams with two exterior legs for the field strength $F$ or two exterior legs for the curvature tensor $R$.
In keeping with the general approach of this paper, we focus on the anomalies in the gauge sector, postponing a discussion of the more supergravity related questions concerning the gravitational anomaly to \cite{Lawrie:2016rqe}.

The gauge anomalies receive contributions from massless charged chiral fermions running in the loop and
from self-dual scalar fields which couple linearly to an abelian gauge potential. 
As lucidly reviewed e.g. in  \cite{Benini:2013cda} the contribution from a canonically normalised Weyl fermion in representation ${\bf R}$ with action
\bea
{\cal L}_{\rm Weyl} = \bar \psi (i \gamma^\mu \partial_\mu - i A_\mu T({\bf R})) \psi 
\eea
to the non-conservation of the gauge current takes the form
\bea
\partial_\mu J^{\mu a} = \frac{1}{8 \pi}  \, {\rm Tr} (\gamma^3 T^a _{\bf R} T^b _{\bf R}) \,   F^b_{\mu \nu} \epsilon^{\mu \nu} \,,
\eea
with $a, b$ Lie algebra indices and $\gamma^3$ the chirality matrix in 2 dimensions. Here we are working in the renormalisation scheme defined in appendix B of \cite{Benini:2013cda}.
Let us therefore define the anomaly coefficient of a single Weyl fermion of chirality $P= \pm 1$  in representation ${\bf R}$ to be
\bea \label{Acoefficient1}
{\cal A}({\bf R},P) = P \, C({\bf R}) \qquad {\rm with} \qquad \label{defCr}
{\rm tr} \, T_{\bf R}^a T_{\bf R}^b = C({\bf R}) \, \delta^{ab}.
\eea
It is worth noting that $C({\bf R}) = C({\bf \bar R})$ so that the anomaly contributions from chiral fermions in real representations do not automatically vanish, unlike in $4n$ dimensions.  
For example, for the group $G = SU(n)$ these anomaly coefficients take the form
\bea \label{SUnanomaly-factors}
G= SU(n): \qquad \quad C({\bf Adj}) = {n}, \qquad  C({\bf n}) = \frac{1}{2}, \qquad C({\bf \Lambda^2 n}) = \frac{n-2}{2}.
\eea

For completeness, let us add that the contribution from an  (anti-)self-dual scalar field of charge $q$ linearly coupled to a $U(1)$ gauge field via
\bea \label{Lscalar}
{\cal L}_{\rm scalar} = \frac{1}{2} \partial_\mu \phi \,  \partial^\mu \phi  + \frac{q}{\sqrt{\pi}} \partial^\mu \phi  A_\mu  \,,
\eea
to the gauge anomaly equals that of an (anti-)chiral Weyl fermion
\bea
\partial_\mu J^{\mu} = \frac{1}{8 \pi}  P\, q^2 \,   F_{\mu \nu} \epsilon^{\mu \nu} \,.
\eea
The anomaly is induced at tree-level by linear exchange of a scalar propagating between two gauge potential insertions.
Equivalently, after fermionisation of the current $\frac{1}{\sqrt{\pi}}~\partial_\mu\phi~\rightarrow~\bar \psi \gamma^\mu \psi$ the anomaly is induced  at the 1-loop level by the associated chiral fermionic degree of freedom. Since the scalar fields in the chiral multiplets of the $(0,2)$ theories under consideration comprise both a self-dual and an anti-self-dual contribution, the only contributions to the gauge anomalies  arise from the Weyl fermions.


\subsection{Non-abelian Gauge Anomalies from Charged Matter }

Consider first the gauge anomalies associated with the non-abelian gauge group $G$ realized on a 7-brane wrapping the divisor $M_G$ on the base of the Calabi-Yau {five-fold}.
The anomaly receives contributions from all charged chiral and anti-chiral fermions localised in the bulk of $M_G$, those at the intersection surfaces of $M_G$ with the other branes in the model and from the fermions at the intersection of $M_G$ with the 3-branes 
\be\label{AGen}
{\cal A}_{\rm total} = {\cal A}_{\rm bulk} + {\cal A}_{\rm surface} + {\cal A}_{3-7} \,.
\ee
Let us begin with the anomaly ${\cal A}_{\rm bulk}$ induced by the states in the bulk. The fermionic bulk matter content is given in table (\ref{cohomologies-bulk}).
To compute the contribution to the anomalies we take into account the Weyl fermions (as opposed to the anti-fermions) in every representation ${\bf R}$ appearing in the decomposition (\ref{adjointbreaking}). In absence of gauge flux, this is just the adjoint representation of $G$, but in general there will be contributions from all irreducible representations of the unbroken bulk gauge groups.
Taking into account the sign from the chirality of the matter states in (\ref{cohomologies-bulk}), the contribution from each representation ${\bf R}$ is 
\be \label{bulkRcontribution}
{\cal A}_{\rm bulk} ({\bf R}) = - C({\bf R}) \chi(M_G, L_{\bf R})\,,
\ee
with $\chi(M_G, L_{\bf R})$ given in (\ref{chibulk1}). 
For instance if the gauge flux on $M_G$ breaks $G \rightarrow H \times U(1)$ such that 
${{\rm \bf Adj}_G \rightarrow {\rm \bf Adj}_H \oplus \bigoplus ({\bf R} \oplus {\bf \bar R})}$, then the contribution to the anomalies of $H$ is

\bea
{\cal A}_{\rm bulk}  = {\cal A}_{\rm bulk}( {\rm \bf Adj}_H) + \sum_{\bf R} \left({\cal A}_{\rm bulk}( {\bf R})  +  {\cal A}_{\rm bulk}( {\bf \bar R}) \right)\,,
\eea
with 
\be \label{Anomadj}
\ba
{\cal A}_{\rm bulk} ({\rm \bf Adj}_H) &=  -  \frac{1}{24} \, C({\rm \bf Adj}_H) \,   \int_{M_G} c_1(M_G)\,  c_2(M_G) \\
\sum_{\bf R} \left({\cal A}_{\rm bulk}( {\bf R})  +  {\cal A}_{\rm bulk}( {\bf \bar R}) \right)&=  - \sum_{\bf R} C({\bf R}) 
\int_{M_G} c_1(M_G) \left(\frac{1}{12} \, {\rm rk}(L_{\bf R})\,c_2(M_G) + {\rm ch}_2(L_{\bf R})\right)
\ea
\ee

The anomaly contribution from the localised massless matter spectrum on a matter surface $S_{\bf R}$ is given by 
\bea
{\cal A}_{\rm surface}({\bf R}) = C({\bf R}) \chi(S_{\bf R},L_{\rm R})\,,
\eea
with $\chi(S_{\bf R},L_{\rm R})$ as in (\ref{chiS2}) (for smooth $S_{\bf R}$). Note the relative sign compared to (\ref{bulkRcontribution}).
This sums up to
\be \ba \label{Anomalysurface}
{\cal A}_{\rm surface} &= \sum_{\bf R} {\cal A}_{\rm surface}({\bf R})   \\
&= \sum_{\bf R} C({\bf R})  \int_{S_{\bf R}}  \left(c_1^2(S_{\bf R}) \left(\frac{1}{12} - \frac{1}{8} {\rm rk}(L_{\bf R}) \right) + \frac{1}{12} c_2(S_{\bf R}) + \left(\frac{1}{2} c_1^2(L_{\bf R}) - c_2(L_{\bf R})\right) \right)\,. 
\ea
\ee

Finally, the Fermi multiplets (\ref{37multiplets}) in the {3--7} sector in representation ${\bf R}$ yield a contribution to the gauge anomalies of the form
\be \label{Anomaly37}
\ba
{\cal A}_{3-7} &=  - \frac{1}{{\rm ord}(G)} \, C({\bf R})  \int_{B_4} [M_G] \wedge [C^B_{\rm M2}]     \cr
& =   - \frac{1}{{\rm ord}(G)} \, C({\bf R})\int_{B_4} [M_G] \wedge  \left(  \frac{1}{24} [c_4(Y_5)]_{B} - \frac{1}{2} [G_4 \wedge G_4]_B \right)\,.
\ea
\ee
The sign is a consequence of the negative chirality of the fermions. The integral counts the number of intersection points between the curve class $C^B_{\rm M2}$ on $B$ wrapped by the D3/M2-branes and the 7-brane cycle supporting the non-abelian gauge group in question, and the prefactor $\frac{1}{{\rm ord}(G)}$ accounting for $SL(2,\mathbb Z)$ monodromies was discussed at the end of section \ref{sec:D3Sector}. 
In the last equation we have implemented the result (\ref{CM2B}) for  $C^B_{\rm M2}$ assuming cancellation of the D3/M2-brane tadpole.
The contribution of the 3-7 string sector to the anomalies is a notable difference to F-theory compactifications to four dimensions, where the 3-7 spectrum is always non-chiral.


\subsection{Non-Abelian Anomaly Cancellation via Anomaly Inflow}
\label{AnomalyInflow}

In a consistent compactification all non-abelian gauge anomaly contributions must automatically cancel each other 
\be \label{AnomalyCancellation}
{\cal A}_{\rm bulk} + {\cal A}_{\rm surface} + {\cal A}_{3-7} = 0\,.
\ee
In fact, non-abelian anomaly cancellation is a direct consequence of the tadpole cancellation condition (\ref{3-brane-tadpole1}), thanks to the mechanism of anomaly inflow \cite{Green:1996dd,Cheung:1997az,Minasian:1997mm} applied to 7-branes in the F-theory/Type IIB setting:
Let us integrate both sides of (\ref{3-brane-tadpole1}) over the class $M_G$ of the 7-brane supporting the non-abelian gauge group $G$  and multiply by $- C({\bf R}_{3-7}) $ to find
\bea \label{3branetadpoleprojected}
 - C({\bf R}_{3-7}) \,  M_G \cdot_{Y_5} C^B_{\rm M2} =  - C({\bf R}_{3-7})   \,  M_G \cdot_{Y_5} \left(\frac{1}{24}   [c_4(Y_5)] - \frac{1}{2} G_4 \wedge G_4 \right).
\eea
Assuming an $SL(2,\mathbb Z)$ monodromy factor $\frac{1}{{\rm ord}(g)}=1$  {to begin with}, the lefthand side is the 3-7 anomaly contribution ${\cal A}_{3-7}$ determined in (\ref{Anomaly37}).
The righthand side uplifts, in F-theory, to the projection onto $M_G$ of the flux and curvature induced couplings of the IIB/F-theory  Ramond-Ramond  4-form $C_4$ in the presence of 7-branes.
This can be made precise if the F-theory vacuum admits a description in terms of a IIB orientifold on a Calabi-Yau four-fold $X_4$ (but is true more generally).
Such perturbative situations are discussed in detail in appendix \ref{app_IIB}.
The Chern-Simons couplings of a 7-brane and O7-plane on $X_4$ to $C_4$ are given by (\ref{CScouplingsFtheory}).
Upon dimensional reduction to 2d $C_4$ is expanded into a basis of orientifold even 2-forms of $X_4$ as $C_4 = c_2^a \, \omega_a$. 
Summing over all 7-branes in the vacuum results in a coupling of the top-forms $c_2^a$ in the 2d effective theory. Since the basis of orientifold even 2-forms uplifts to a basis of $H^{1,1}(B_4)$ in F-theory, we can directly identify these couplings with couplings in the 2d F-theory vacuum up to a factor of $\frac{1}{2}$ explained e.g. in \cite{Krause:2012yh}.\footnote{The simplifying assumption that $\frac{1}{{\rm ord}(g)}=1$ corresponds to a configuration where the D7-branes and image branes as well as the D3-branes and their images wrap cycles not invariant under the orientifold action so that no additional relative correction factor is necessary in comparing the D3 and and the D7-sector. In particular the non-abelian part of the 7-brane gauge group is $SU(n)$.}
Upon circle reduction to M-theory, the 2-forms $c_2^a$ with one leg along the compactifcation circle $S^1$ reproduce the 1-forms $A_a$, $a=1, \ldots, h^{1,1}(B_4)$ in M-theory obtained by reduction of the M-theory 3-form $C_3$ along a basis of $H^{1,1}(B_4)$. This identifies the 7-brane Chern-Simons couplings (\ref{CScouplingsFtheory}) as the origin of the curvature and flux-dependent part of the 1d Chern-Simons couplings (\ref{CSloop}) for this subclass $A_a$ of 1-forms. 
The particular choice $\omega_a = [M_G]$ singles out the projection of the K-theoretic Ramond-Ramond 4-form charges of the 7-branes and the O7-plane onto the 7-brane carrying non-abelian gauge group $G$. 
The role of these Ramond-Ramond charges in the worlvolume theory of the 7-branes  is to cancel the gauge (and gravitational) anomalies due to chiral fermions localised at the intersection of 7-branes in Type IIB/F-theory \cite{Green:1996dd,Cheung:1997az,Minasian:1997mm}.
This includes the bulk matter (\ref{cohomologies-bulk}) as a special case, viewed as matter at the intersection of the 7-brane with itself.  It follows that the righthand side of (\ref{3branetadpoleprojected}) equals minus the contribution of the full  7-7 sector to the non-abelian gauge anomalies, $-({\cal A}_{\rm bulk} + {\cal A}_{\rm surface})$, thereby establishing anomaly  cancellation.
The aforementioned factor of $\frac{1}{2}$ from the IIB/F-theory correspondence reproduces the factor $C({\bf R}_{3-7}) $ for 3-7 matter in the fundamental representation of $SU(n)$ as expected for F-theory models with a IIB limit. 

Despite our explicit reference to a weak coupling limit, we expect the correspondence between the correctly interpreted Ramond-Ramond charges and the 2d anomaly cancellation to hold more generally, where now also non-trivial $SL(2,\mathbb Z)$ monodromy factors  $\frac{1}{{\rm ord}(g)} \neq 1$ must be taken into account. Various examples including some with $\frac{1}{{\rm ord}(g)} \neq 1$ will be presented in section \ref{sec:ExamplesGlobal}.

\subsection{2d Abelian Anomalies and the {GSS} Mechanism} \label{sec_2dAbelian}

The structure of abelian gauge anomalies is considerably enriched by the possibility of a Green-Schwarz mechanism as described first in \cite{Mohri:1997ef,GarciaCompean:1998kh} for the $(0,2)$ worldvolume theory of D1-branes at singularities and studied in $(0,2)$ linear sigma models relevant for heterotic compactifications in \cite{Adams:2009av,Adams:2009tt,Adams:2009zg,Quigley:2011pv,Blaszczyk:2011ib}.
It is convenient to phrase the discussion in superspace: 
Under a $U(1)$ gauge transformation, the vector superfields $V$ and  $V_+$ defined in (\ref{defVfield}), (\ref{defVplusfield}) transform as
\be
\delta_{\Lambda} V_+ = \frac{1}{2i} (\Lambda - \Bar \Lambda), \qquad \delta_{\Lambda} V = - \frac{1}{2} \partial_- (\Lambda + \Bar \Lambda)\,,
\ee 
with $\Lambda$ a chiral superfield. A $U(1)$ gauge anomaly corresponds to a gauge variance of the quantum effective action $W$ of the form (see e.g. \cite{Quigley:2011pv} for a careful derivation)
\be
\delta_{\Lambda} W =   \frac{{\cal A}}{16 \pi} \int d^2 y \, d {\theta^+}  \Lambda \Upsilon  \, + \, {\rm c.c.}  \,.
\ee
The anomaly coefficient ${\cal A}$ is the specialization of (\ref{Acoefficient1}) to the case of a $U(1)$ gauge theory with charged Weyl fermions of chirality $P_i$ and charge $q_i$ {given by}
\be
{\cal A} = \sum_i  P_i q^2_i \,,
\ee
with obvious generalizations to mixed abelian anomalies.
In addition to this 1-loop induced quantum anomaly, the $U(1)$ anomaly can receive a contribution from the variation of a classically non-gauge invariant interaction term \cite{Adams:2009av,Adams:2009tt,Adams:2009zg,Quigley:2011pv,Blaszczyk:2011ib} of the form
\be \label{GSa}
S_{\rm GS} =  m \int d^2 y \, d \theta^+ \Phi  \Upsilon \, + \, {\rm c.c.}  \,. 
\ee
This is the generalization of the FI-term (\ref{FIterm-general}) with the the FI parameter promoted to a chiral $\Phi$ superfield. 
Such interaction induces an anomaly of the action provided $\Phi$ transforms under the $U(1)$ gauge symmetry as 
\be \label{gaugetraforPhi1}
\delta_{\Lambda} \Phi = \Phi +  q \Lambda  \,,
\ee
such that
\bea \label{SGS-superfield}
\delta_{\Lambda} S_{\rm GS} =   q\, m\,  \int d^2 y \, d {\theta^+} \Lambda \Upsilon \, + \, {\rm c.c.}  \,.
\eea
The scalar components of the Green-Schwarz interaction are given by
\bea \label{SGS_scalar}
S_{\rm GS} \supset 4 \, m \int d^2 y \, \Big( -{\mathfrak D}  \, {\rm Im} (\varphi) + F_{01} {\rm Re}(\varphi) \Big) \,,
\eea
with $\varphi$ the scalar component in $\Phi$. This identifies ${\rm Re}(\varphi)$ as an axionic scalar field whose linear coupling to the field strength $F$ is the hallmark of the Green-Schwarz mechanism.

For real $\Lambda$ the transformation (\ref{gaugetraforPhi1}) can be viewed as a gauging of the shift symmetry of the axion ${\rm Re}(\varphi)$.
In any event the gauging (\ref{gaugetraforPhi1}) requires a suitable modification of the kinetic term for $\Phi$ such as to keep the latter gauge invariant. In the present context we can take this kinetic term to be \cite{Blaszczyk:2011ib}
\bea \label{Stueckelberg1}
S_{\rm Stuckelberg} = \int d^2y \, d\theta^+ \left( \left(\frac{1}{2i} (\Phi - \bar\Phi) - q V_+\right)  \left( \partial_- \frac{1}{2} (\Phi + \bar \Phi) + q V  \right)     \right).
\eea
This gauge invariant coupling induces, amongst other things, a quadratic St\"uckelberg mass term for the gauge potential proportional to $q^2$. 
It also contributes interactions similar to (\ref{SGS_scalar}) of the form
\bea \label{Stueckelberg2}
S_{\rm Stuckelberg} \supset 2q \int d^2y \Big( {\mathfrak D}   \, {\rm Im} (\varphi) + F_{01} {\rm Re}(\varphi)  \Big).
\eea
Note the crucial relative sign difference between both terms in brackets compared to (\ref{SGS_scalar}). It is this sign which distinguishes the Green-Schwarz and the St\"uckelberg interactions. 
Furthermore, we would like to stress that the gauging (\ref{gaugetraforPhi1}) and the resulting gauge invariant modification of the kinetic term (\ref{Stueckelberg1}) do not require the existence of the anomalous Green-Schwarz coupling (\ref{GSa}). As in higher-dimensional theories, a St\"uckelberg massive $U(1)$ field need not be anomalous in the sense that the 1-loop fermionic gauge anomaly is cancelled by a tree-level gauge variance of Green-Schwarz type. By contrast, for the Green-Schwarz term to contribute to the anomaly, the gauging (\ref{gaugetraforPhi1})  and  (\ref{Stueckelberg1})   are of course required.

Combining (\ref{SGS_scalar}) and (\ref{Stueckelberg2}) we see that the sum of the Green-Schwarz and the St\"uckelberg couplings $4m + 2q$ can be determined from the coefficient of the axionic coupling 
${\rm Re} (\varphi) F_{01}$ in the effective action {-- which we will refer to as Green-Schwarz-St\"uckelberg (GSS) couplings}. To uniquely determine $m$ and $q$ individually further information is required, e.g. by inspecting also the D-term couplings of ${\rm Im} (\varphi)$.
A complete analysis of this type is beyond the scope of this paper and will appear in future work. {It suffices} here to outline the origin of the axionic couplings in the F-theory compactifications under consideration, to which we turn in the next section.

\subsection{Origin of the {GSS-couplings} in M/F-theory}

In 2d F-theory compactifications the axionic scalar fields $c^\rho = {\rm Re}(\varphi^\rho)$ participating in the Green-Schwarz mechanism arise by KK reduction of the F-theory/Type IIB 
Ramond-Ramond forms.
This can be made very precise in the special case of a Type IIB orientifold on a Calabi-Yau four-fold $X_4$ with stacks of  D7-branes along complex three-cycles $D_i$ with individual $U(1)$ gauge field strengths $F_i$.
As shown in appendix \ref{app_TadpolesGS}  there can in general be four different types of GSS-couplings in the 7-brane sector from reduction of $C_6$, $C_4$, $C_2$ and $C_0$. 
They take the form
\be \label{SGS2d}
S_{\rm GSS} \supset \sum_{\rho, i}  Q_{\rho i} \int_{\mathbb R^{1,1}} c^\rho \, F_i  =  - \sum_{\rho, i}  Q_{\rho i} \int_{\mathbb R^{1,1}}  d c^\rho \wedge A_i =  - \sum_{\rho, i}  Q_{\rho i} \int_{\mathbb R^{1,1}}  \ast d \tilde c^\rho \wedge A_i \,,
\ee
where we have introduced the dual axionic fields $\tilde c^\rho$.
The different types of couplings are listed in 
(\ref{S-Stgeom1}), (\ref{app_SGS2}),  (\ref{S2Stflgeom}) and  (\ref{app_SGS4}), {respectively}. 
Of these only  (\ref{app_SGS2}) possesses a straightforward derivation via M/F-theory duality.
This is the coupling to the 
 axions obtained by dimensional reduction of the IIB/F-theory self-dual 4-form $C_4$ along a basis of $H^{2,2}_+(X_4)$.
 The relevant $U(1)$ fields in this context are those linear combinations $U(1)_m$ of $U(1)_i$ gauge potentials which are massless in {the} absence of gauge flux. These geometrically massless $U(1)$ gauge fields   can be recovered in M-theory 
 by expansion of $C_3$ as 
 \bea  \label{c3m}
 C_3 = A_m \wedge S_m + \ldots
 \eea
 with $S_m$ the $U(1)_m$ generating divisor class (\ref{Shiodamap}). Under F/M-theory duality the GSS terms (\ref{SGS2d}) become couplings in the 1d super-mechanics obtained by dimensional reduction of M-theory of the form
\be \label{GS1d}
S^{\rm 1d}_{\rm GSS}  =    - Q_{\rho m}  \int_{\mathbb R}  \ast d \tilde c^\rho \,   A_m\,. 
\ee
One obvious source for such interactions is the $G_4$ dependent piece in the 11d Chern-Simons term (\ref{CSloop}),
\be \label{Sfluxa}
S_{\rm flux} = - \frac{2\pi}{6} \int C_3 \wedge G_4 \wedge G_4\,. 
\ee
The dual scalars $\tilde c^k$ are obtained by dimensional reduction of the M-theory 6-form $C_6$ magnetically dual to $C_3$ 
\be
C_6 =\sum_k  \tilde c^k \,   \tilde \omega_k, \qquad {\rm with} \, \,  \{{\tilde{\omega}_k} \} \,\,  {\rm a \, \, basis \,\,  of} \, \, H^{3,3}(Y_5) \,.
\ee
To make contact with (\ref{Sfluxa}) we express one copy of $G_4$ as  $G_4 = \ast_{\rm11d} G_7 = \ast_{\rm11d} d C_6$ and expand
\be
\ast_{\rm 11d} G_7 = \ast_{\rm 11d} d C_6 = \ast_{\rm 11d} d (\tilde c^k \wedge \tilde \omega_k) =  \ast_{\rm1d} d \tilde c^k \wedge \ast_{Y_5} \tilde \omega_k =  \ast_{\rm1d} d \tilde c^k \wedge \omega_k \,.
\ee
Here we have introduced the basis $\{ \omega_k \}$ of $H^{2,2}(Y_5)$ dual to $\{ \tilde \omega_k \}$. 
Reducing furthermore $C_3$ as in (\ref{c3m})  this results in a coupling
\be \label{GS2-Mtheory}
 - Q_{k m} \int_{\mathbb R}  \ast d c^k \,   A_m\, ,\quad \hbox{{where}}\quad  Q_{km} =    \, {2 \pi} \int_{Y_5} S_m \wedge G_4 \wedge \omega_k \,.
\ee
The couplings $Q_{km}$ in (\ref{GS2-Mtheory}) from M-theory are to be identified with the eponymous objects from the IIB/F-theory reduction obtained in (\ref{app_SGS2})  if we specify
\bea
\omega_k \in H^{2,2}(B_4) \,.
\eea
Indeed the basis of $H_+^{2,2}(X_4)$ involved in the reduction of $C_4$ in the Type IIB derivation uplifts to a basis of $H^{2,2}(B_4)$ with $B_4$ the base of the elliptic fibration $Y_5$. As in the Type IIB limit, there are no contributions from 4-forms in $H^{3,1}(B_4)$ because $G_4$ is of $(2,2)$-type.

In view of the general structure (\ref{SGS_scalar}) and (\ref{Stueckelberg2})    the axions $c^k$ must form the real part of a chiral multiplet in the 2d $(0,2)$ theory. 
The origin of $c^k$ as axionic modes of $C_4$ suggests that the imaginary part of the scalar component is related to the scalars $t^k$ obtained by reduction of $J \wedge J$ along the basis $\omega_k$.
Thus
\be \label{phikform}
\ba
{\rm Re}(\varphi)^k \sim  c^k  \  &\longleftrightarrow\  C_6 = \sum_k \tilde c^k \, \tilde \omega_k \cr 
{\rm Im}(\varphi)^k \sim  t^k \ &  \longleftrightarrow\  J \wedge J = \sum_k t^k  \omega_k  \,.
\ea\ee
The couplings of ${\rm Im \varphi}^k$ in (\ref{SGS_scalar}) and (\ref{Stueckelberg2}) yield a contribution to the scalar potential (\ref{scalar1}) which is minimized, for zero non-linearly charged matter fields, if the 
flux-induced D-term vanishes. 
The flux-induced D-term has already been derived from the supersymmetry variations as the first term in  (\ref{DtermBPS}). 
This expression translates into the object $\int_{Y_5} G_4 \wedge S_m \wedge J_B \wedge J_B$ from the M-theory perspective, and it is exactly this form which is in agreement with the proposal (\ref{phikform}) for the  ${\rm Im}( \varphi)^k$ moduli together with (\ref{GS2-Mtheory}). 
A full supergravity analysis of both the Green-Schwarz and the St\"uckelberg couplings and the relation to K\"ahler moduli will appear in \cite{TBA1}.

Let us briefly comment on the M-theory origin of the remaining Green-Schwarz couplings (\ref{S-Stgeom1}),  (\ref{S2Stflgeom}) and (\ref{app_SGS4}). 
First, Type IIB $U(1)$ symmetries which possess a geometric St\"uckelberg coupling (\ref{S-Stgeom1}) are massive already in absence of gauge flux. As in compactifications to 3/4 dimensions, their mass is at the KK scale and a description of their gauge potential requires the introduction of non-harmonic forms \cite{Grimm:2010ez,Grimm:2011tb}. In particular such $U(1)$ symmetries are not associated with extra rational sections on the elliptic fibration \cite{Braun:2014nva,Martucci:2015dxa}. What is new compared to the 3/4 dimensional situation is the appearance also of higher curvature geometric St\"uckelberg terms  (\ref{S2Stflgeom}). 
By contrast, the flux-dependent Green-Schwarz terms in (\ref{S2Stflgeom}) can be non-vanishing already in absence of a geometric  St\"uckelberg mass term. These interactions should therefore have a description in M-theory reduction with harmonic forms. In Type IIB, these terms must involve, for geometrically massless $U(1)$s, orientifold odd gauge fluxes, which are notoriously difficult to uplift to F-theory  \cite{Clingher:2012rg,Braun:2014nva,Martucci:2015dxa}. 
Finally the coupling (\ref{app_SGS4}) involves the axion $C_0$, which is geometrised in F-theory as the real part of the axio-dilaton $\tau = C_0 + \frac{i}{g_s}$. 
Couplings of this sort are particularly challenging to extract via M-theory (see e.g. \cite{Grimm:2012rg}), and we leave a derivation of all these Green-Schwarz couplings as an interesting challenge for future work. 

We conclude this section  by stressing that we have so far focused on the $U(1)$ gauge groups from the 7-brane sector. The D3-brane sector naively contributes a $U(1)$ gauge group from each single D3-brane wrapping a holomorphic curve $C_{\rm M2}^B$ as well. 
In Type IIB theory these $U(1)$s receive a Green-Schwarz-St\"uckelberg term (\ref{app_SGSD31}) from the coupling to $C_2$ provided the homology class of the wrapped curve in the IIB Calabi-Yau four-fold is not orientifold invariant. 
In fact, anomaly cancellation for the {D3-brane} $U(1)$ gauge group requires this {GSS} mechanism to be in work in order to cancel the anomalies from the 3-7 sector: The latter contains only charged Fermi multiplets (\ref{37multiplets}) at the intersection of the D3-branes with all 7-branes in the theory, and these contribute with the same sign to the anomaly. 
In absence of such homology-odd contributions to the curve class, the only other consistent option is that the $U(1)$ is projected out such that no $U(1)$ anomaly arises in the first place. This is the case if the D3-brane curve is invariant as a whole under the orientifold action. In both situations, the $U(1)$ is massive from the perspective of the low-energy effective action.
It will be interesting to study this more from the F-theory perspective \cite{Lawrie:2016axq}.


\section{Chern-Simons Couplings from M/F-duality} 
\label{sec:CScouplings}

We now come to an interpretation of the Chern-Simons couplings (\ref{CSgeneral}) in the light of F/M-theory duality {(\ref{MFDuality})}, which relates the supersymmetric quantum mechanics obtained by the reduction of M-theory 
on the resolved Calabi-Yau five-fold $Y_5$ to the 2d $(0,2)$ field theory obtained by F-theory compactification on the same space. 

Our first aim is to understand the Chern-Simons terms $ 2 \pi \int_{\mathbb R} A_{\alpha} \, k^\alpha_{\rm curv} $ involving the 1-forms $A_\alpha$ obtained by expanding $C_3 = A_\alpha \wedge  \omega^\alpha$, where $\omega^\alpha$ is a basis of 2-forms dual to the divisors listed in (\ref{divisor-list}). 
In section \ref{AnomalyInflow} we already identified the F-theory origin of the couplings involving the 1-forms $A_a$, $a= 1, \ldots, h^{1,1}(B_4)$,  in the M-theory effective super-mechanics with couplings of the Ramond-Ramond four-form in F-theory induced by the 7-branes of the system.
To understand the remaining Chern-Simons couplings, recall that 
at the level of effective field theories, the precise F/M-theory match is obtained by compactifying the 2d F-theory effective action on a circle $S^1$, similarly to the circle reduction relating F/M-theory in $d=6/5$ \cite{Intriligator:1997pq,Bonetti:2011mw,Bonetti:2013cza} and $d=4/3$ \cite{Aharony:1997bx,Grimm:2010ks,Grimm:2011fx,Cvetic:2012xn}      
 spacetime dimensions. As reviewed in section \ref{sec:Dictionary}, the fact that we are working {not on a} singular elliptic fibration, but on its resolution corresponds to the fact that the M-theory effective action is on its Coulomb branch, on which the non-abelian part  ${\mathfrak g}$ of the gauge algebra is broken to its Cartan subalgebra with massless gauge potentials $A_i$.

For notational simplicity let us first assume that this ${\mathfrak g}$ constitutes the full gauge algebra in F-theory, and analyze the couplings $ 2 \pi \int_{\mathbb R} A_{i} \, k^i_{\rm curv}$, $i=1, \ldots, {\rm rk}({\mathfrak g})$. The generalization including extra non-Cartan {$\mathfrak{u}(1)_m$} gauge potentials will be detailed momentarily.
Consider, as in the discussion around (\ref{DCmat}), a representation ${\bf R}$ of  ${\mathfrak g}$, described by the weight vector  $\lambda^{{\bf R}}_{a}$ for $a=1,\ldots, {\rm dim}({\bf R})$. The charge of the $a$-th state in this representation with respect to the Cartan ${\mathfrak{u}(1)_i}$ is given by 
\be \label{chargeqja}
q_{a i} = \lambda^{{\bf R}}_{a i} = \varepsilon(\lambda^{\bf R}_a)  \ D_i \cdot_{Y_5} C^{\varepsilon(\lambda^{\bf R}_a)}_{\lambda^{\bf R}_a} .
\ee
For $\varepsilon(\lambda^{\bf R}_a) =1 (-1)$ the fibral curve  $C^{\varepsilon(\lambda^{\bf R}_a)}_{\lambda^{\bf R}_a}$ is wrapped by the (anti-)M2-brane associated with the state under consideration, as explained in section \ref{sec:Codim12}.

Upon circle reduction, a charged particle in 2d  gives rise to a KK zero-mode in 1d plus a tower of KK-states. 
The mass of the fermionic KK zero-mode of the $a$-th state 
is given by
\be\label{m0qja}
\ba
m_0(\lambda^{\bf R}_a) 
&=  \sum_{j=1}^{\rm{rk}(\mathfrak{g})}  q_{a j }  \, \xi_{ j}  =   \sum_{j=1}^{\rm{rk}(\mathfrak{g})}  \lambda^{{\bf R} }_{a j} \, \xi_{ j}  \cr 
&=    \sum_{j=1}^{\rm{rk}(\mathfrak{g})}   \varepsilon(\lambda^{\bf R}_a)  (\xi_j D_j) \cdot_{Y_5} C^{\varepsilon(\lambda^{\bf R}_a)}_{\lambda^{\bf R}_a}     =  \varepsilon(\lambda^{\bf R}_a)  \int_{C^{\varepsilon(\lambda^{\bf R}_a)}_{\lambda^{\bf R}_a}}   J \,. \ea\ee
The $\xi_j$ denote the vacuum expectation values of scalar fields parametrizing the Coulomb branch of the supersymmetric quantum mechanics. From the discussion in section \ref{sec:Dictionary} we recall that these scalars are the volume moduli of the resolution $\mathbb P^1$s $F_i$. Correspondingly, the last equation relates this field theoretic expression to the volume of the fibral curve wrapped by the M2-brane associated {to} the state with charge $\lambda^{{\bf R}}_{a j}$. Note that this fermion mass can be positive or negative depending on the sign $\varepsilon(\lambda^{\bf R}_a)$. The mass at the $n$-th KK level is then given by 
\be
m_n(\lambda^{\bf R}_a)   = m_0(\lambda^{\bf R}_a) + n \int_{\mathfrak{F}} J \,.
\ee
Indeed, while the KK zero-modes originate from M2-branes wrapping fibral curves with vanishing intersection with the divisor $S_0$ characterized by $S_0 \cdot \mathfrak{F} = 1$ (see the discussion around (\ref{S0Fint})), the KK-tower arises by adding to this curve class $n$ powers of the class of the generic fiber $\mathfrak{F}$.

We are interested in the M-theory effective action at energies below the smallest mass of wrapped M2-brane states. At this energy, all massive M2-brane states have been integrated out and we are left with the massless fields only. The latter include the 1-form potentials $A_i$ in the Cartan subalgebra, which remain unbroken along the Coulomb branch. 
The effective action of the massless modes is to be compared with the circle reduction of the 2d F-theory effective action, {where} all massive modes {are} integrated out.
In this process the curvature and flux induced Chern-Simons terms $ 2 \pi \int_{\mathbb R} A_{i} \, k^i_{\rm curv}$ in (\ref{CSloop}) are reproduced by integrating out the massive fermionic modes charged under $A_i$ in the $S^1$-reduction of the F-theory effective action. The relevant diagrams arise  at 1-loop level only \cite{Witten:1996qb}.
This parallels the match of the Chern-Simons couplings in 5 \cite{Witten:1996qb,Intriligator:1997pq} and 3 \cite{Aharony:1997bx} dimensions obtained by M-theory compactifications on Calabi-Yau three-folds and four-folds, respectively, with the 1-loop terms obtained from F-theory in 6 \cite{Bonetti:2011mw,Bonetti:2013ela,Bonetti:2013cza,Grimm:2013oga} and 4 \cite{Grimm:2011fx,Cvetic:2012xn} dimensions reduced on an $S^1$.
In other words there exists a match
\be
\hbox{{\rm 1d {from}  M-theory}}:\  k_{\rm curv}^i \qquad \longleftrightarrow  \qquad  \hbox{{\rm 2d  {from} F-theory  on}}\   S^1:  \ k_{\rm 1-loop}^i \,,
\ee
where $k_{\rm 1-loop}^i$ denotes the 1-loop induced Chern-Simons term from integrating out massive states in F-theory reduced on an $S^1$.

Taking the chiral nature of the 2d theory into account, the result for the 1-loop amplitude we are encountering here is
\be\label{k1loop}
k_{\rm 1-loop}^i = - {1\over 2}
 \sum_{\bf R} \left( n^+_{\bf R} - n^-_{\bf R}\right)
 \sum_{a = 1}^{{\rm dim}({\bf R})}   q_{a i}  \,    {\rm sign}(m_0(\lambda^{\bf R}_a)) \,.
    \ee
{Indeed, a single massive fermion in representation ${\mathbf R}$ yields a correction 
\bea
\delta k_{\rm 1-loop}^i =  - \frac{1}{2} P \,  \sum_{a = 1}^{{\rm dim}({\bf R})}   q_{ai}  \,    {\rm sign}(m(\lambda^{\bf R}_a)) \,,
\eea
where $P = \pm 1$ denotes the 2d chirality of the fermion. 
This is the direct analogue of the higher-dimensional expressions determined in  \cite{Witten:1996qb,Intriligator:1997pq,Aharony:1997bx,Bonetti:2013ela}.
Assuming a mass hierarchy between the Coulomb-branch masses and {the masses of all KK-states}, $| m_0| < |m_n|$, each KK-state at level $n$ comes with an opposite sign compared to the KK-state at level $-n$ for $n \neq 0$. As stressed in \cite{Grimm:2013oga} this assumption corresponds to the zero-section of the fibration being holomorphic as opposed to rational. 
We henceforth assume that the fibration has a holomorphic zero-section, if possible at the expense of going to a birational model as demonstrated in \cite{Borchmann:2013hta}.\footnote{It would be interesting to determine under which conditions a {\it smooth} birational model with a holomorphic zero-section exists.}
Under this assumption all that remains is the contribution from the KK zero modes (\ref{k1loop}).
}

This field theoretic {relation} can be expressed in geometric terms with the help of (\ref{chargeqja}) and (\ref{m0qja}).
In particular from (\ref{m0qja}) we identify the sign contribution as\footnote{See \cite{Hayashi:2014kca} for the precise identification of the sign of the volume integral with the signs $\varepsilon$ in the box graphs.} 
\be \label{signexpl}
{\rm sign}\left(m_0(\lambda^{\bf R}_a) \right) = \varepsilon(\lambda^{\bf R}_a) \,.
\ee
Equating the resulting expression for $k_{\rm 1-loop}^i$  with the M-theoretic formula  (\ref{CSloop}) for $k_{\rm curv}^i$  we therefore conclude that 
\be
\ba
\label{chirality1}
D_i \cdot_{Y_5} \left(\frac{1}{24}   [c_4(Y_5)] - \frac{1}{2} G_4 \wedge G_4 \right)
&= -  {1\over 2}   \sum_{\bf R}  \left( n^+_{\bf R} - n^-_{\bf R}\right)  \left(
\sum_{a=1}^{{\rm dim}({\bf R})} \varepsilon(\lambda^{\bf R}_a)\,  \lambda_{a i}^{\bf R} \right) \cr
&=  - {1\over 2}   \sum_{\bf R}  \left( n^+_{\bf R} - n^-_{\bf R}\right)  \left(
\sum_{a=1}^{{\rm dim}({\bf R})}   D_i \cdot_{Y_5} C^{\varepsilon(\lambda^{\bf R}_a)}_{\lambda^{\bf R}_a} \right) \,,
\ea
\ee
{where the chiralities are related to the chiral indices $\chi$ as}
\be \label{signchibulk}
n^+_{\bf R} - n^-_{\bf R} = 
\left\{
\ba
 - \chi(M_G, {\bf R})       & \quad {\rm bulk \quad matter}     \cr 
 + \chi(S_{\bf R}, {\bf R})       & \quad {\rm {surface} \quad matter} \,.
\ea\right.
\ee
The chiral index $\chi(M_G, {\bf R})$  for a representation ${\bf R}$ in the bulk is given by (\ref{chibulk1}), while the chiral index $\chi(S_{\bf R}, {\bf R})$     for a representation localised on a smooth surface in codimension two is given by (\ref{chiS2}). 
Note that the sum runs over all particles which become massive along the Coulomb branch {and, for simple $\mathfrak{g}$, only receives contributions from those particles for which not all $\varepsilon(\lambda^{\bf R}_a) $ are of the same sign. This implies that the particles 
from the 3-7 strings do not contribute as will be discussed further in \cite{TBA1}.

This expression readily generalizes to situations with gauge algebra 
\be
{\mathfrak{g} \oplus \bigoplus_{m=1}^M \mathfrak{u}(1)_m} \,.
\ee
Every representation ${\bf R}$ carries in addition charge ${\bf Q} = (Q_1, \ldots, Q_m)$ under the non-Cartan $U(1)_m$. This includes singlets under ${\mathfrak g}$ with charges $Q_m \neq 0$. 
The charges can again be written as 
\be
{Q_m = \varepsilon(\lambda^{{\bf R}_{\bf Q}}_a) \ S_m \cdot_{Y_5} C^{\varepsilon\left(\lambda^{\bf R_Q}_a\right)}_{\lambda^{{\bf R_Q}}_a}  }\,,
\ee
 and are of course independent of the choice of $a$.
In particular, identifying the 1-loop CS coupling $k_{\rm 1-loop}^m$ with the M-theoretic expression for the couplings $k_{\rm curv}^m$ associated with the $U(1)_m$ gauge potentials $A_m$ results in 
\be \label{chirality2}
S_m \cdot_{Y_5} \left(\frac{1}{24}   [c_4(Y_5)] - \frac{1}{2} G_4 \wedge G_4 \right)
=  - {1\over 2}   \sum_{{\bf R}_{\bf Q}}     ( n^+_{\bf R_Q} - n^-_{\bf R_Q}  )    \left(
\sum_{a=1}^{{\rm dim}({\bf R})}  S_m  \cdot_{Y_5} C^{\varepsilon(\lambda^{\bf R_Q}_a)}_{\lambda^{{\bf R_Q}}_a}       \right) \,.
\ee
The terms in brackets are simply
\be
\sum_{a=1}^{{\rm dim}({\bf R})} \,  S_m  \cdot_{Y_5} C^{\varepsilon(\lambda^{\bf R_Q}_a)}_{\lambda^{{\bf R_Q}}_a}      = Q_m (2 N_+({\bf R}_{\bf Q}) - {\rm dim}({\bf R}_{\bf Q})) \,,
\ee
with $ N_+({\bf R}_{\bf Q})$ denoting the number of positive weights for representation ${\bf R}_{\bf Q}$.

In section \ref{sec:chiSing} we had seen that  the expression (\ref{chiS2}) for the chiral index on matter surfaces is a priori valid only if $S_{\bf R}$ is smooth and is in general modified in the presence of singularities. While in principle the correction terms can be derived on purely geometric grounds by passing to a suitable normalization of the singular surface, it is in fact simpler to 
 indirectly read off the chiralities by solving (\ref{chirality1}) for the individual $\chi(S_{\bf R})$. This is indeed what we have done to determine the correction factors presented in section \ref{sec:chiSing}.

What is left is a discussion of the CS-coupling involving the $U(1)$ potential $A_0$ associated with the expansion $C_3 = A_0 \wedge S_0$. As in higher-dimensional settings, this gauge potential $A_0$ corresponds to the KK $U(1)$ from the perspective of the circle reduction of F- to M-theory. This sector and its relation to the 2d gravitational anomalies will be discussed in \cite{Lawrie:2016rqe}.

So far we have only explained the origin of the couplings $k_{\rm curv}$ in (\ref{CSloop}) in the light of F/M-theory duality. By contrast, the couplings $k_{\rm M2}$ in  (\ref{CSclassical}) are induced, in M-theory, by the massive M2-branes wrapping fibral curves on $Y_5$. In the M-theory effective action at energies below the Coulomb branch scale, these massive states are not present any more, and consequently also their couplings $k_{\rm M2}$ to the massless gauge fields are to be discarded {below}  this energy scale. It is only at energies comparable to the Coulomb branch mass parameter that the massive M2-states become relevant and the couplings  $k_{\rm M2}$  complete the effective action.
At this mass scale the tadpole equation (\ref{3-brane-tadpole1}) follows from the effective action. 

Since it is the M-theory effective action at energies below the Coulomb branch mass parameter  which maps to F-theory on $S^1$, the couplings  $k_{\rm M2}$ cannot be reproduced from the F-theory circle reduction. Likewise, the tadpole constraint (\ref{3-brane-tadpole1}) evaluated along the fibral part of the homology of $Y_5$ has no analogue in F-theory: It is a consistency condition only of the M-theory compactification on $Y_5$.
An M-theory compactification violating this  does not give rise to a consistent vacuum to begin with and therefore has no F-theory dual. 
This is the F/M-theory analogue of the observation of \cite{Dasgupta:1996yh} (see also \cite{Gates:2000fj})  that Type IIA string theory compactified on a Calabi-Yau four-fold to 2 dimensions is subject to a tadpole constraint resulting from a term 
\be \label{TypeIIAcoupling}
S_{\rm IIA} \supset \int_{\mathbb R^{1,9}} B_2 \wedge X_8 \,.
\ee
This coupling  enforces the inclusion of a certain number of spacetime-filling fundamental strings in the 2d effective action to cancel the tadpole for $B_2$. By contrast, Type IIB string theory compactified on the same four-fold does not know of such a tadpole constraint because no corresponding coupling exists in the Type IIB effective action. 
Comparing the Type IIA and IIB vacua in 2 dimensions by T-duality maps the background strings required in {the} Type IIA theory to momentum modes of massless particles in Type IIB  \cite{Dasgupta:1996yh,Gates:2000fj}. 
While in this way the particle content of  {the} Type IIB theory automatically gives rise to the correct number of spacetime-filling Type IIA strings required to cancel the Type IIA tadpole, the couplings $(\ref{TypeIIAcoupling})$ are not reproduced in this 2d/2d T-duality map. Similarly the 2d/1d map between F/M-theory can reproduce only (\ref{CSloop}) but not (\ref{CSclassical}) for the reasons detailed above.

\section{Examples: Global Consistency}
\label{sec:ExamplesGlobal}

In this section we will exemplify the global consistency conditions derived in this paper for 2d F-theory compactifications.
 As a first step we will discuss the general computational method to determine the Chern-Simons terms in concrete models.
By explicitly matching both sides of our prediction (\ref{chirality1}) for these terms via F/M-theory duality we provide a strong general consistency check of the entire framework. In particular we will verify the corrections discussed in section \ref{sec:chiSing} for the chiral index of surface localised matter in presence of singularities. By analyzing the non-abelian gauge anomalies we will furthermore determine the subtle monodromy factors arising in coupling the D3-brane and the 7-brane sector. 

We begin with  $SU(2k+1)$ examples, with and without {an additional} $U(1)$, and in the appendix \ref{app:SU6}  show consistency of an $SU(6)$ class of models along the same lines. We then provide an example with $SO(10)$ and with $E_6$ gauge group.


\subsection{Intersections for $SU(2k+1)$ CS-terms}
\label{sec:AnoGenSU}

Before we can check the gauge anomaly and Chern-Simons terms (\ref{chirality1}), we determine some useful identities for the intersection ring in ${SU}(2k+1)$ fibrations with fundamental and anti-symmetric matter. We are interested in the expression in the last bracket of the righthand side of (\ref{chirality1}).

Let $L_i$, $i=1, \cdots, 2k+1$ be the weights of the fundamental representation, which satisfy $L_i \cdot L_j = - \delta_{ij}$
and the tracelessness condition $\sum L_i =0$. 
Then it is easy to verify that\footnote{We adopt here the convention, more appropriate for the geometric analysis, that the simple roots square to $-2$. This will directly give rise to the intersection ring, which is precisely $-1$ times the representation-theoretic convention. All group theory conventions are otherwise those in \cite{FultonHarris}.}
\be
\sum_{L_i}  C_{L_i}^{\varepsilon_i} \cdot_{Y_5} D_j 
=- \sum_{i=1}^{2k+1} \varepsilon_i (\delta_{i, j} -  \delta_{i, j+1}) =  \varepsilon_{j+1} - \varepsilon_{j}  \,.
\ee  
If $C_{L_j}$ and $C_{L_{j+1}}$ of this particular resolution have the same sign, then the result is 0, else, the result is $-2$. 
This has one non-trivial contribution coming from the simple root $\alpha_j$, and associated $D_j$, which splits in codimension two.  It is clear that the contributions arise precisely from the extremal generators of the cone of effective curves, i.e. elements of $\mathcal{K}_{fib}$. 
In summary for the fundamental we find 
\be
\sum_{L_i}   C_{L_i}^{\varepsilon_i} \cdot_{Y_5} D_j 
= \left\{ 
\ba
0 &\quad \hbox{if $F_j$ does not split} \cr 
-2 & \quad \hbox{if $F_j$ splits}.
\ea
\right.
\ee

Likewise for the anti-symmetric  representation 
\be\ba
\sum_{L_{i,j}}  C_{L_i + L_j}^{\varepsilon_{i,j}} \cdot_{Y_5} D_k 
&= -\sum_{i,j=1, i<j}^{2k+1} \varepsilon_{i,j} (\delta_{i, k} -  \delta_{i, k+1} + \delta_{j, k} - \delta_{j, k+1}) \cr 
&=  -\sum_{i<k} \varepsilon_{i,k} - \varepsilon_{i,k+1}  + \sum_{j>k} \varepsilon_{k,j} - \varepsilon_{k, j+1} \,.
\ea\ee
Again, the only non-zero contributions arise for those $k$ for which $F_k$ splits and the result is given by summing over the extremal generators.

Consider for example $SU(5)$. Then the intersections of the fundamental ${\bf 5}$ and anti-symmetric ${\bf 10}$ representations take the following form,
\be\label{GeneralSU5IntsCD}
\ba
\sum_{L_i} C_{L_i}^{\varepsilon_i} \cdot_{Y_5} D_k  &=
\{\varepsilon_2-\varepsilon_1,\varepsilon_3-\varepsilon_2,\varepsilon_4-\varepsilon_3,\varepsilon_5-\varepsilon_4\}_k\cr 
\sum_{L_{i,j}}   C_{L_i + L_j}^{\varepsilon_{i,j}} \cdot_{Y_5} D_k  &= 
\left\{
\ba
-\varepsilon _{1,3}-\varepsilon _{1,4}-\varepsilon _{1,5}+\varepsilon _{2,3}+\varepsilon_{2,4}+\varepsilon _{2,5} & \quad k=1 \cr 
-\varepsilon _{1,2}+\varepsilon _{1,3}-\varepsilon _{2,4}-\varepsilon_{2,5}+\varepsilon _{3,4}+\varepsilon _{3,5}& \quad k=2\cr 
-\varepsilon _{1,3}+\varepsilon _{1,4}-\varepsilon_{2,3}+\varepsilon _{2,4}-\varepsilon _{3,5}+\varepsilon _{4,5}& \quad k=3\cr 
-\varepsilon _{1,4}+\varepsilon_{1,5}-\varepsilon _{2,4}+\varepsilon _{2,5}-\varepsilon _{3,4}+\varepsilon _{3,5}& \quad k=4\,.
\ea\right.
\ea
\ee
For instance for $\varepsilon_{1, k} = 1$ and all others $-1$ this is $(-6, 0, 0, 0)$, and the contributions from $C^{+}_{1,k}$, $k=3, 4, 5$ equal those of $C^-_{2,k}$, consistent with the fact that only $F_1$ splits. For $\varepsilon_{1, k} = 1$ and $\varepsilon_{2,3}=1$ and all else $-1$, we obtain $(-4,0,-2,0)$ (or reversed order, depending on the assignment of roots to rational curves). In this case both $F_1$ and $F_3$ split.
We will determine similar relations for {the} $E_6$ and $SO(10)$ examples in the following.


\subsection{Global Consistency of $SU(5)\times U(1)$} 
\label{sec:SU5U1consistency}

After this {intersection-theoretic preparation} we can now explicitly address the global consistency of the  $U(1)$-restricted $SU(2k+1)$-models.
The reason why we begin with this class of fibrations is because here no subtleties due to singular matter surfaces or monodromy factors in the D3-brane sector arise, {nor are there any non-minimal loci in codimension three or four}.
For definiteness we specialise to $k=2$ in the geometric realization, including resolution, reviewed in section \ref{sec:Examples},\footnote{For all models we will only present one resolution and perform all computations therein. For $SU(5)\times U(1)$ we focus on the one introduced as $T_{11}$ in \cite{Krause:2011xj} and summarized in section \ref{sec:Examples}. Of course, there are generically several small resolutions. The complete network of resolutions for this model was determined in \cite{Hayashi:2013lra, Braun:2014kla}. Changing the resolution does not affect the singular F-theory limit, but it will change some of the details of our analysis such as $c_4(Y_5)$ and the expressions in (\ref{chirality1}). Of course the global consistency is independent of the resolution.} {however, the result holds more generally.}

As a warmup we check that the $SU(5)$ anomaly (\ref{AnomalyCancellation}) cancels. The first contribution is from the 3-7 strings. Their number is given by  $M_G \cdot_{Y_5} c_4(Y_5)$. This intersection can be computed in the specific resolution under consideration to be 
\be
M_G \cdot_{Y_5}   c_4(Y_4) = M_G \cdot_{B_4} \left(144 c_1^3-264 c_1^2 M_G+12 c_1 c_2+162 c_1 M_G^2-30 M_G^3 \right),
\ee
where all Chern classes without any specifications are taken for the base of the fibration, $c_i = c_i(B_4)$. Then the anomaly contribution from the 3-7 sector is
\be
\mathcal{A}_{3-7} = \frac{1}{2} \left(-{1\over 24} M_G \cdot_{Y_5}  c_4(Y_5)\right) \,,
\ee
where we have used that the 3-7 strings are in the fundamental representation of $SU(5)$ with anomaly coefficient $C({\bf 5}) = \frac{1}{2}$.

The matter surfaces $S_{\bf R}$ contribute as follows.
There are three $SU(5)$-charged matter loci corresponding to ${\bf 10}$ and two ${\bf 5}$ representations with classes 
\be
\ba
{\bf 10}_{-1}:\quad  &M_G \cdot [b_1]=  M_G \cdot c_1\cr 
{\bf 5}_{-3}:\quad& M_G \cdot [b_3]= M_G \cdot (3 c_1-2 M_G)\cr
{\bf 5}_{2}:\quad &  M_G \cdot[b_1 b_4 - b_2 b_3] = M_G \cdot (5 c_1-3 M_G) \,.
\ea
\ee
The subscripts denote the charges under the non-Cartan $U(1)_X$ associated with the divisor \cite{Krause:2011xj}
\be \label{SXDefinition}
S_X = 5 (\sigma_X - \sigma_0 -c_1) + 2 D_1 + 4 D_2 + 6 D_3 + 3 D_4 \,.
\ee
This is the image of the extra rational section $\sigma_X$ under the Shioda map (\ref{Shiodamap}).
In addition there exists a charged singlet localised at a matter surface away from the $SU(5)$ brane with class
\be
{\bf 1}_{-5}: \quad [b_3]\cdot[b_4] = (3 c_1 -2 M_G) \cdot (4 c_1 -3 M_G) \,. 
\ee
Let us first consider a configuration with vanishing gauge flux, $G_4 = 0$.
In this case, the chiral indices for the charged matter surfaces are
\be
\ba
\chi_{{\bf 10}_{-1}} &= \frac{1}{24} c_1 M_G \left(2 c_2+M_G^2\right)\cr 
\chi_{{\bf 5}_{-3}} &= \frac{1}{24} M_G \left(3 c_1-2 M_G\right) \left(-12 c_1 M_G+8 c_1^2+2 c_2+5
   M_G^2\right)\cr 
\chi_{{\bf 5}_{2}} &=   \frac{1}{12} M_G \left(5 c_1-3 M_G\right) \left(-15 c_1 M_G+12 c_1^2+c_2+5 M_G^2\right) \cr
\chi_{{\bf 1}_{-5}} & =    \frac{1}{24} \left(4 c_1 - 3 M_G\right) \left(3 c_1 - 2 M_G\right) \left(24 c_1^2 + 2 c_2 - 36 c_1 M_G + 13 M_G^2\right) .
\ea
\ee
They contribute
\be
{\cal A}_{\rm surface} = {3\over 2}  \chi_{{\bf 10}_{-1}} + {1\over 2}\chi_{{\bf 5}_{-3}} + {1\over 2}\chi_{{\bf 5}_{2}}
\ee
to the $SU(5)$ anomaly, with the numerical coefficients being the anomaly coefficients $C({\bf 10}) = \frac{5-2}{2}$, $C({\bf 5}) = \frac{1}{2}$.
Finally, the bulk contribution from the adjoint is, using $C({\bf 24}) =5, $\footnote{The factor of $-1$ is because  $\chi_{\rm bulk} \equiv \chi(M_G) = \sum_{i=0}^3 (-1)^i h^i(M_G)$ counts minus the number of chiral  plus the number of anti-chiral bulk fermions in the adjoint of $SU(5)$. }
\be
\mathcal{A}_{\rm bulk} =  - 5   \chi_{\rm bulk} = -\frac{5}{24} M_G \left(c_1-M_G\right) \left(M_G \left(M_G-c_1\right)+c_2\right) \,.
\ee
With the help of these expressions we verify the identity 
\be
\mathcal{A}_{3-7} +  \mathcal{A}_{\rm bulk} +   {\cal A}_{\rm surface} =0 \,,
\ee
which precisely reproduces the anomaly cancellation condition (\ref{AnomalyCancellation}). 

Likewise the relations (\ref{chirality1}) and (\ref{chirality2}) from the Chern-Simons analysis are automatically satisfied without any fluxes. For this note the following intersection relations in the resolution under consideration\footnote{Note that the specific expressions are resolution dependent, but the agreement with the F-theory predictions of course is not.} between $c_4(Y_5)$ and the Cartan divisors $D_i$ as well as the $U(1)_X$ divisor $S_X$,
 \be
\ba
c_4(Y_5) \cdot D_1 =& \, -4 c_1 M_G^3+2 c_1^2 M_G^2+2 c_2 M_G^2-2 c_1 c_2 M_G+2 M_G^4 \cr 
c_4(Y_5) \cdot D_2 =& \, -3 c_1 M_G^3+2 c_1^2 M_G^2+2 c_2 M_G^2+2 M_G^4\cr
c_4(Y_5) \cdot D_3 =& \, 175 c_1 M_G^3-272 c_1^2 M_G^2-8 c_2 M_G^2+144 c_1^3 M_G+14 c_1 c_2 M_G-38 M_G^4\cr
c_4(Y_5) \cdot D_4 =& \, -2 c_1 M_G^3+2 c_1^2 M_G^2+2 c_2 M_G^2+2 c_1 c_2 M_G+2 M_G^4 \cr
c_4(Y_5) \cdot S_X =&  \, 720 c_1^4 + 60 c_1^2 c_2 - 2016 c_1^3 M_G - 84 c_1 c_2 M_G + 2136 c_1^2 M_G^2 + 
 30 c_2 M_G^2 - \cr
 & \, -1011 c_1 M_G^3 + 180 M_G^4 \,.
\ea
\ee
These can be expressed in terms of the matter chiralities for $G_4=0$ as 
\be \label{Check-noflux}
\ba
{1\over 24} c_4(Y_5) \cdot D_1 & =  - 2   \chi_{\rm  bulk} \cr 
{1\over 24} c_4(Y_5) \cdot D_2  & = - 2  \chi_{\rm  bulk} + \chi_{{\bf 10}_{-1}} \cr 
 {1\over 24} c_4(Y_5) \cdot D_3  & = - 2 \chi_{\rm  bulk}  + \chi_{{\bf 5}_{-3}}+ \chi_{{\bf 5}_{2}} \cr
{1\over 24} c_4(Y_5) \cdot D_4  & = - 2 \chi_{\rm  bulk} + 2 \chi_{{\bf 10}_{-1}}  \cr
{1\over 24} c_4(Y_5) \cdot S_X  & =  - \frac{3}{2}  \chi_{{\bf 5}_{-3}}   +  \chi_{{\bf 5}_{2}}  + \frac{5}{2}   \chi_{{\bf 1}_{- 5}} \,.
\ea
\ee
The first four equations precisely reproduce the predicted relations (\ref{chirality1}),  given the splittings of the fibers in codimension two and the general relations (\ref{GeneralSU5IntsCD}). 
To see this for the bulk contribution we take into account that 
\be
\sum_{\alpha \in \Delta^+} D_{i} \cdot_{Y_5} F_{\alpha} =-2 \,,
\ee
where the sum is over all 10 generators associated to the positive roots $\alpha \in \Delta^+$ in the adjoint of $\mathfrak{su}(5)$.
This relation holds for all $D_i$, and a similar one exists for the negative roots $\alpha \in \Delta^-$. Together with the additional sign in  (\ref{signchibulk}) and the factor of $ - \frac{1}{2}$ in   (\ref{chirality1}) this reproduces (\ref{Check-noflux}). 
As for the matter contributions, 
$F_3$ splits along both ${\bf 5}$ matter loci, which thus contribute to the Chern-Simons term related to $D_3$, while $F_2$ and $F_4$ split for the ${\bf 10}$-representation. The associated box graphs are shown in figure \ref{fig:BoxGraphSU5}, from which one can read off the signs that enter into the general expressions for the Chern-Simons terms. More precisely, 
the geometric intersection numbers between the Cartan divisors and the curves are determined as explained in section \ref{sec:AnoGenSU}\footnote{This can also be seen directly from the explicit analysis in the appendix of \cite{Krause:2011xj}.}
\be
\ba
\sum_{a=1}^{10}   D_i \cdot_{Y_5} C^{\varepsilon\left(\lambda^{{\bf 10}_{-1}}_a\right)}_{\lambda^{{\bf 10}_{-1}}_a}  &= (0,-2,0,-4) \cr
\sum_{a=1}^{5}     D_i \cdot_{Y_5} C^{\varepsilon\left(\lambda^{{\bf 5}_{2}}_a\right)}_{\lambda^{{\bf 5}_{2}}_a}  &= (0,0,-2,0) = \sum_{a=1}^{5}     D_i \cdot_{Y_5} C^{\varepsilon\left(\lambda^{{\bf 5}_{-3}}_a\right)}_{\lambda^{{\bf 5}_{-3}}_a}  \,,
\ea
\ee
 in perfect agreement with (\ref{chirality1}) and (\ref{Check-noflux}).

\begin{figure}
\centering
\includegraphics[width=9cm]{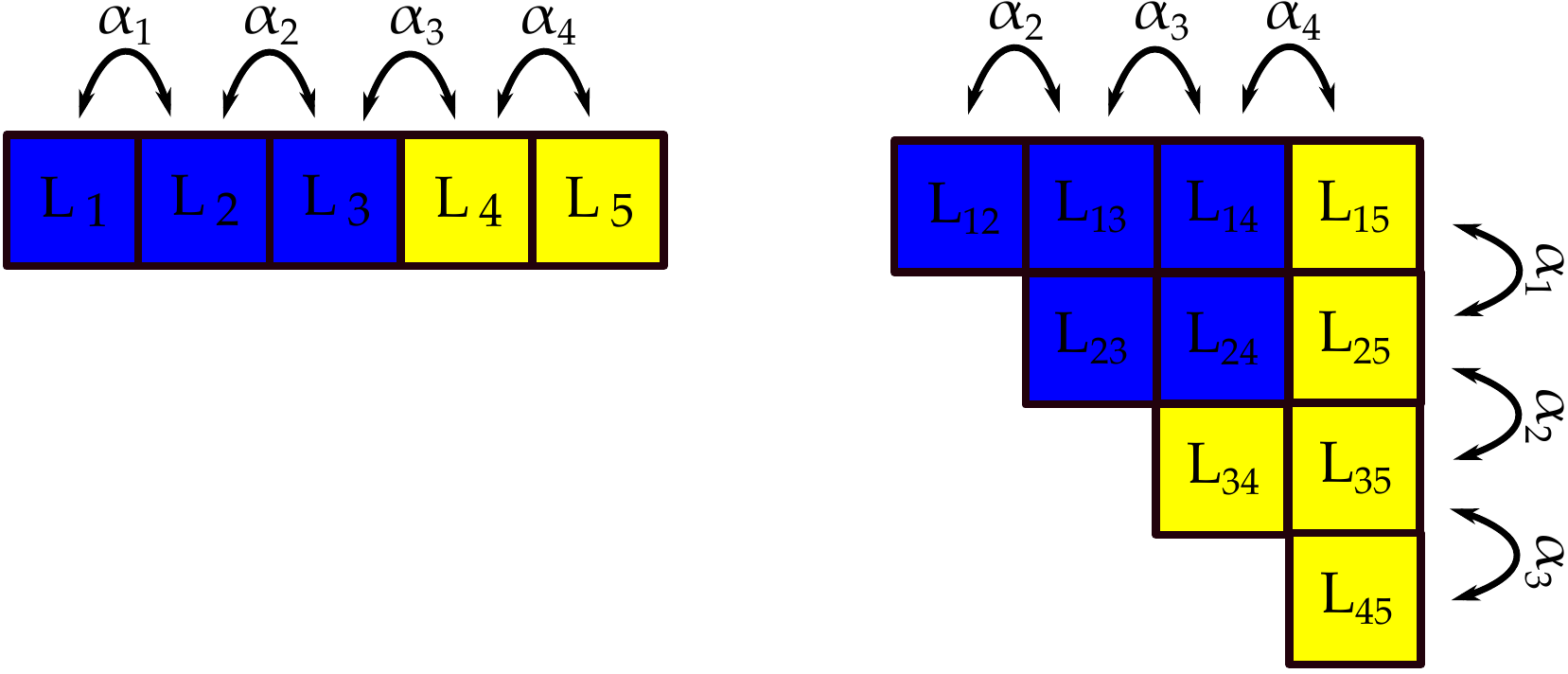}
\caption{Box Graph for the ${\bf 5}$ (on the left) and ${\bf 10}$ (right) representation of $SU(5)$ corresponding to the fiber in codimension two of the resolution discussed in section \ref{sec:SU5Ex}. Each box corresponds to a weight of the representations. The action of the roots $\alpha_i$ connects the weights into this representation graph.  The coloring corresponds to the signs blue $\varepsilon=+$ and yellow $\varepsilon=-$, indicating that $C^{\varepsilon(\lambda)}_{\lambda}$ is an effective curve. 
\label{fig:BoxGraphSU5}}
\end{figure}


Finally the last equation in (\ref{Check-noflux}) can be understood in terms of  (\ref{chirality2}).
For the ${\bf {10}}$-representation the number of positive and negative weights is equal in the resolution under consideration so that this state does not contribute, whereas for both ${\bf 5}$ representations
$N_{+}({\bf 5}) = 2$. As for the singlet, we have in fact $N_+(1_{\bf -5}) = 1$.\footnote{Indeed, the effective fibral curve wrapped by the rational section $\sigma_X$ over the singlet matter locus gives rise to a state of charge $-5$  \cite{Krause:2011xj}.}

Let us now exemplify the inclusion of $G_4$ flux by considering the gauge flux associated with the non-Cartan $U(1)$. It takes the form \cite{Krause:2011xj}
\be
G_4 = F \wedge S_X \,,
\ee
with $S_X$ defined in (\ref{SXDefinition}) and $F \in H^{1,1}(B_4)$ an arbitrary class parametrizing the flux. The procedure outlined in section \ref{chiralitywithflux} identifies the line bundles counting localised matter in representation ${\bf R}_{\bf Q}$ as \cite{Krause:2011xj}
\be
c_1(L_{\bf R}) = Q \, F|_{S_{\bf R_Q}}\,.
\ee 
Since the bulk matter is uncharged under the extra $U(1)$, the flux on the $SU(5)$ locus  $M_G$ induced by such $G_4$ is trivial. 
We can then check explicitly that the relations (\ref{chirality1}) continue to hold in the presence of $G_4$ by accounting for the flux-dependent correction in the chiralities of the matter states appearing on the lefthand side as in (\ref{chirality-fluxdep}). 
To be maximally explicit, we find
\be
\ba
- \frac{1}{2} (G_4 \wedge G_4) \cdot D_1 & = 0 \cr 
- \frac{1}{2} (G_4 \wedge G_4) \cdot D_2  & = \frac{1}{2} c_1  F^2 M_G    \cr 
- \frac{1}{2} (G_4 \wedge G_4) \cdot D_3 & =    \frac{47}{2} c_1 F^2 M_G - 15 F^2 M_G^2    \cr
- \frac{1}{2} (G_4 \wedge G_4) \cdot D_4 & =      c_1   F^2 M_G   \cr
- \frac{1}{2} (G_4 \wedge G_4) \cdot S_X & = 375 c_1^2 F^2 - \frac{1083}{2} c_1 F^2 M_G + 195 F^2 M_G^2 \,,
\ea
\ee
and therefore confirm that
\be
\ba
- \frac{1}{2} (G_4 \wedge G_4) \cdot D_1 & = 0 \cr 
- \frac{1}{2} (G_4 \wedge G_4) \cdot D_2  & =  \chi_{{\bf 10}_{-1}}|_{\rm flux}\cr 
- \frac{1}{2} (G_4 \wedge G_4) \cdot D_3 & =  \chi_{{\bf 5}_{-3}}|_{\rm flux}+ \chi_{{\bf 5}_2}|_{\rm flux} \cr
- \frac{1}{2} (G_4 \wedge G_4) \cdot D_4 & =  2 \,  \chi_{{\bf 10}_{-1}}|_{\rm flux}\cr
- \frac{1}{2} (G_4 \wedge G_4) \cdot S_X &=  - \frac{3}{2}  \chi_{{\bf 5}_{-3}}|_{\rm flux}   +  \chi_{{\bf 5}_{2}}|_{\rm flux}  + \frac{5}{2}   \chi_{{\bf 1}_{-5}}|_{\rm flux}\,,
\ea
\ee
with 
\be
 \chi_{\bf 10}|_{\rm flux} = \frac{1}{2} \int_{S_{{\bf 10}_{-1}}} F^2, \qquad \chi_{{\bf 5}_{2}}|_{\rm flux} = \frac{1}{2} \int_{S_{{\bf 5}_{2}}} 4 F^2\,, \qquad \chi_{{\bf 5}_{-3}}|_{\rm flux} = \frac{1}{2} \int_{S_{{\bf 5}_{-3}}} 9 F^2 \,.
\ee
Likewise, the flux-induced contributions to the $SU(5)$ gauge anomaly cancel automatically.

More generally, similar arguments hold for all  $SU(2k+1)\times U(1)$ gauge groups obtained from $U(1)$-restricted Tate forms, i.e. $b_6=0$. 


\subsection{Global Consistency of $SU(5)$ Models} 

Somewhat more subtle is the global consistency of the generic $SU(2k+1)$ model without $U(1)$-restriction. Let us illustrate this for $SU(5)$, assuming again that there are no gauge fluxes. 
Compared to the analysis for the model with $U(1)$-restriction in section \ref{sec:SU5U1consistency}, there are only two matter loci
$b_1=0$ and $P= b_1^2 b_6 - b_1 b_4 b_3 + b_2 b_3^2=0$. The chiral index of the fundamental matter acquires additional contributions, as discussed in section \ref{sec:chiSing}, and the expressions for $\chi (b_1, {\bf 10})$ and  $\chi(P, {\bf 5})$ are summarized in (\ref{chisSU5}). 
The contributions to the anomaly (\ref{AGen}) are as follows
\be\ba
\mathcal{A}_{3-7}  &=  - {1\over 48} M_G \cdot c_4(Y_5)\cr 
\mathcal{A}_{\rm bulk} &= - 5 \chi_{\rm bulk} = -{5\over 24}  M_G (c_1-M_G) (M_G (M_G-c_1)+c_2) \cr 
\mathcal{A}_{\rm surface} & = {3\over 2} \, \chi (b_1, {\bf 10}) + {1\over 2} \, \chi(P, {\bf 5}) \,.
\ea\ee
With the value for $c_4(Y_5)$ obtained in (\ref{c4SU5}), it is straightforward to check that these terms cancel to satisfy (\ref{AGen}). 
It is crucial for the cancellation that there are the additional contribution from the singularities of the matter surface $P$ that contribute to $\chi(P, {\bf 5})$. The additional term, which is given in (\ref{ChiSUodd}), is key to cancel both the anomaly as well for satisfying the relations arising from the Chern-Simons analysis (\ref{chirality1}) and (\ref{chirality2}). To check the latter, note that the intersection ring gives the following intersections between $c_4(Y_5)$ and the Cartan divisors $D_i$, again, in the resolution of section \ref{sec:Examples} 
\be
\ba
c_4(Y_5) \cdot D_1 &=  -4 c_1 M_G^3+2 c_1^2 M_G^2+2 c_2 M_G^2-2 c_1 c_2 M_G+2 M_G^4 \cr 
c_4(Y_5) \cdot D_2 &=  -3 c_1 M_G^3+2 c_1^2 M_G^2+2 c_2 M_G^2+2 M_G^4\cr 
c_4(Y_5) \cdot D_3 &=  -2 c_1 M_G^3+2 c_1^2 M_G^2+2 c_2 M_G^2+2 c_1 c_2 M_G+2 M_G^4  \cr
c_4(Y_5) \cdot D_4 &=538 c_1 M_G^3-758 c_1^2 M_G^2-8 c_2 M_G^2+360 c_1^3 M_G+14 c_1 c_2 M_G-128 M_G^4
  \,.
\ea
\ee
The Chern-Simons relations imply that these can be written in terms of linear combinations of the chiralities, with coefficients as dictated by the  general analysis in section \ref{sec:AnoGenSU}. Indeed, the following relations hold 
\be
\ba
{1\over 24} c_4(Y_5) \cdot D_1 &= - 2 \chi_{\rm bulk} \cr 
{1\over 24} c_4(Y_5) \cdot D_2 &= - 2 \chi_{\rm bulk} + \chi(b_1, {\bf 10} ) \cr 
{1\over 24} c_4(Y_5) \cdot D_3 &= - 2 \chi_{\rm bulk} + 2 \chi(b_1, {\bf 10} ) \cr 
{1\over 24} c_4(Y_5) \cdot D_4 &= - 2 \chi_{\rm bulk}+ \chi(P, {\bf 5} )  \,,
\ea
\ee
where we note again that the last equation crucially makes use of the additional contributions from the singularities in (\ref{ChiSUodd}). 
Similar relations hold for the remaining $SU(2k+1)$ models without $U(1)$ restriction. 


\subsection{Global Consistency of $SO(10)$ Models}
\label{sec:SOGlobal}

The models with $SO(10)$ or more general $SO(2n)$ gauge group can be studied along similar lines. The interest here is in 
exemplifying the subtleties in the D3-brane sector, as already discussed in section \ref{sec:D3Sector}. 
The Tate form for $SO(10)$ (or rather Spin$(10)$) has vanishing orders $(1, 1, 2, 3, 5)$, i.e. 
\be
y^2+x y b_1\zeta_0 +y b_3 \zeta _0^2= x^3 +x^2 b_2 \zeta _0+x b_4 \zeta _0^3+b_6 \zeta _0^5 \,,
\ee
whose discriminant is 
\be
\Delta= b_2^3 b_3^2 \zeta_0^3 + O(\zeta_0^4) \,.
\ee
We resolve the model as in appendix B.2 of \cite{Lawrie:2012gg}, and summarize here only the differences and additional information we need to study the consistency of the five-fold compactification. 
The enhancement patterns are as follows:
\be\ba
&\ba
\hbox{Codim 2}: \left\{
\ba
SO(12): &\quad b_3=0 \cr  
E_6: &\quad b_2=0 \ea
 \right. 
\ea
\cr 
&\ba
\hbox{Codim 3}: \left\{
\ba
SO(14): &\quad b_3=b_4^2-4 b_2 b_6= 0 \cr  
E_7: &\quad b_2=b_3=0 \ea
 \right. 
\ea
\cr 
&\ba
\hbox{Codim 4}: \left\{
\ba
SO(16): &\quad b_3=b_4^2-4 b_2 b_6= 2 b_4-b_1^2 b_2= 0 \cr  
E_8: &\quad b_2=b_3=b_4=0 \,. \ea
 \right. 
\ea
\ea
\ee
There are two matter loci given by intersection of $M_G$  with  $b_2=0$, which gives rise to the spin representation ${\bf 16}$, and with $b_3=0$, above which the fundamental matter ${\bf 10}$ is localized. 

To check the anomaly note first  
that the group theoretic anomaly contributions of the relevant representations of $SO(10)$  are
\be
C({\bf Adj}) = 4, \qquad C({\bf 10}) = \frac{1}{2}, \qquad C({\bf 16}) = 1 \,.
\ee
The key point, discussed in section \ref{sec:D3Sector}, is that the contribution from the D3-branes is only a fraction of what one would naively expect based on counting the number of geometrical intersection points with the 7-branes. The fraction is determined by ${1\over {\rm ord}(g)}$ where ${\rm ord}(g)$ is the order of the $SL(2,\mathbb Z)$ monodromy around the 7-branes. For $SO(2n)$ groups, this is $1/2$.\footnote{\label{footnote_Z2}This factor can be understood from the perspective of a Type IIB orientifold as follows: D7-branes producing a gauge group $SO(2n)$  necessarily lie on top of the O7-plane, while the D3-branes are generically not contained in the O7-plane. Matter between the D3-brane and the D7-brane stack is mapped to matter between the image D3-brane and the same D7-brane stack. This requires a factor of $\frac{1}{2}$ to avoid overcounting the D3-D7- matter compared to the matter in the D7-brane sector. Interestingly, this reasoning seems to remain valid even for $SO(2n)$ without a weakly coupled description such as the $SO(10)$ model with a spinor representation.}

Thus we have 
\be
 \mathcal{A}_{3-7} = {1\over 2 }  \, C({\bf 10}) \,  \left(-  \frac{1}{24} M_G \cdot c_4\right)  \,,
\ee
where
\be
M_G \cdot_{Y_5}   c_4(Y_5) = M_G \cdot_{B_3} (-756 c_1^2 M_G+528 c_1 M_G^2+360 c_1^3+12 c_2 c_1-120 M_G^3) \,.
\ee
The remaining contributions are 
\be\ba
\mathcal{A}_{\rm bulk} &= - C({\bf Adj})  \,  \chi_{\rm bulk} =  {-}{1\over 6} M_G \left(c_1-M_G\right) \left(M_G \left(M_G-c_1\right)+c_2\right) \cr 
\mathcal{A}_{\rm surface}& =  C({\bf 16})\, \chi(b_2, {\bf 16})  + C({\bf 10}) \, \chi(b_3, {\bf 10})  
\ea\ee
with matter chiralities 
\be\ba
\chi(b_2, {\bf 16}) &=\frac{1}{24} M_G \left(2 c_1-M_G\right) \left(-4 c_1 M_G+2 \left(c_2+M_G^2\right)+3 c_1^2\right) \cr 
& \qquad +  \frac{1}{4} M_G \left(4 c_1-3 M_G\right){}^2 \left(3 c_1-2 M_G\right)\cr 
\chi(b_3, {\bf 10}) &= \frac{1}{24} M_G \left(3 c_1-2 M_G\right) \left(-12 c_1 M_G+8 c_1^2+2 c_2+5 M_G^2\right)  \,.
\ea\ee
Note that the terms in the second line of $\chi(b_2, {\bf 16})$ are those arising from the singularities of the higher codimension loci (\ref{ChiSO}).
Putting these terms together, we see that the anomaly cancels 
\be
 \mathcal{A}_{3-7} + \mathcal{A}_{\rm bulk} + \mathcal{A}_{\rm surface} =0 \,.
\ee
Furthermore, we check the identities implied by the 1-loop Chern-Simons terms. Using the resolution, we find the intersections 
between the Cartan divisors and $c_4(Y_5)$
\be
\ba
 c_4(Y_5) \cdot D_1 &= -4 c_1 M_G^3+2 c_1^2 M_G^2+2 c_2 M_G^2-2 c_1 c_2 M_G+2 M_G^4   \cr 
 c_4(Y_5) \cdot D_2 &= -4 c_1 M_G^3+2 c_1^2 M_G^2+2 c_2 M_G^2-2 c_1 c_2 M_G+2 M_G^4   \cr 
 c_4(Y_5) \cdot D_3 &= 12 c_1 M_G^3-20 c_1^2 M_G^2-2 c_2 M_G^2+12 c_1^3 M_G+6 c_1 c_2 M_G-2 M_G^4  \cr 
 c_4(Y_5) \cdot D_4 &= -4 c_1 M_G^3+2 c_1^2 M_G^2+2 c_2 M_G^2-2 c_1 c_2 M_G+2 M_G^4  \cr 
 c_4(Y_5) \cdot D_5 &= 524 c_1 M_G^3-726 c_1^2 M_G^2-6 c_2 M_G^2+336 c_1^3 M_G+10 c_1 c_2 M_G-126 M_G^4  \,.
\ea
\ee
These formulae can be expressed in terms of chiralities of the matter surfaces as follows
\be\label{SO10CS}
\ba
{1\over 24}  c_4(Y_5) \cdot D_1 &= - 2 \chi_{\rm bulk}  \cr 
{1\over 24}  c_4(Y_5) \cdot D_2 &= - 2 \chi_{\rm bulk}  \cr 
{1\over 24}  c_4(Y_5) \cdot D_3 &=  - 2 \chi_{\rm bulk} +2  \chi(b_3, {\bf 16} )\cr 
{1\over 24}  c_4(Y_5) \cdot D_4 &=  - 2 \chi_{\rm bulk}\cr 
{1\over 24}  c_4(Y_5) \cdot D_5 &=  - 2 \chi_{\rm bulk} +2  \chi(b_2, {\bf 10} ) \,.
\ea
\ee


\begin{figure}
\centering
\includegraphics[width=8cm]{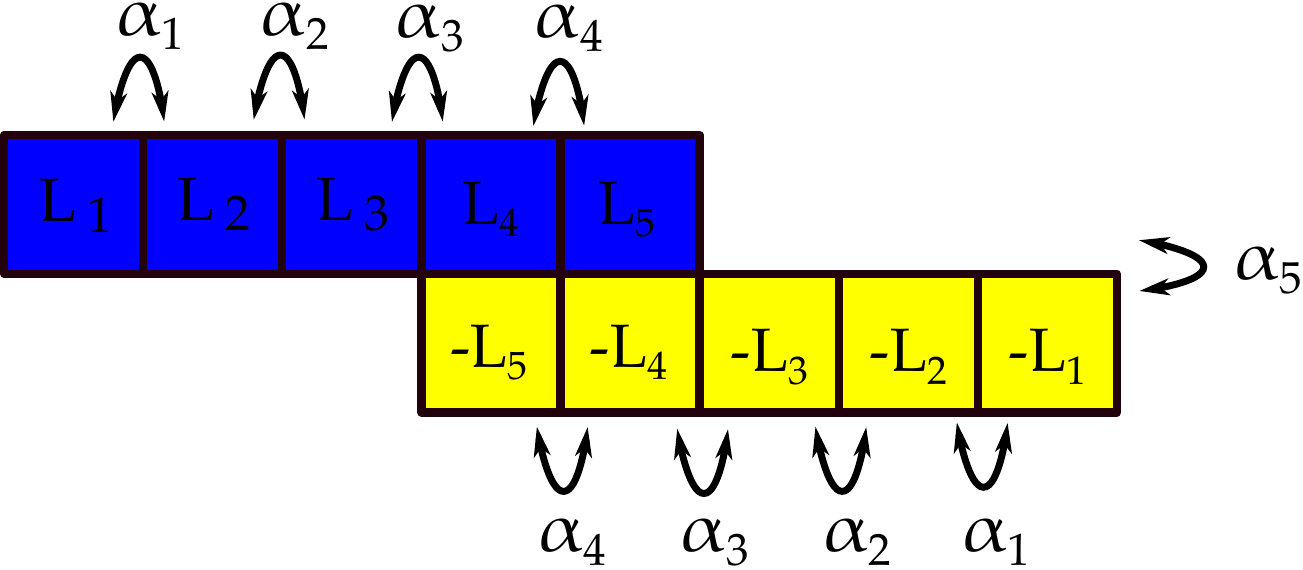}
\caption{Box Graph for the ${\bf 10}$ representation of $SO(10)$ corresponding to the fiber in codimension two of the $SO(10)$ model. Each box corresponds to a weight of the ${\bf 10}$. The action of the roots $\alpha_i$ connects the weights into this representation graph.  The coloring corresponds to the signs blue $\varepsilon=+$ and yellow $\varepsilon=-$, indicating that $C^{\varepsilon(\lambda)}_{\lambda}$ is an effective curve. \label{fig:BoxGraphSO10}}
\end{figure}

We now confirm these from the resolution of the fiber and the intersections of the effective curves associated to the matter represetations $C^\pm_\lambda$ with the Cartan divisors $D_k$. 
For the ${\bf 10}$ representation, it was shown in \cite{Hayashi:2014kca} that there are precisely two possible resolutions, for which either $\alpha_5$ or $\alpha_4$ split. In the resolution above, the former case is realized, and the associated box graph of the fiber in codimension two is shown in figure \ref{fig:BoxGraphSO10}. As for $SU(5)$, the $L_i$ are fundamental weights, and $\alpha_i=L_i-L_{i+1}$, for $i = 1,\cdots, 4$ and $\alpha_5= L_4 + L_5$. 
From this we compute the sum that enters the Chern-Simons couplings to be
\be\label{InterSO10}
\sum_{a=1}^{10}   D_k \cdot_{Y_5} C^{\varepsilon\left(\lambda^{{\bf 10}}_a\right)}_{\lambda^{{\bf 10}_a}  }=
 (0,0,0,0,-4)_k \,.
\ee
Finally, we need to check the intersections in the fiber realizing the ${\bf 16}$ of $SO(10)$. The box graphs for this case are determined in \cite{CLSSN}. 
The resolution is such that $F_3$ splits corresponding to the box graph 
and the only contribution arises from intersections with this, giving rise to
\be\label{InterSO16}
\sum_{a=1}^{16}   D_k \cdot_{Y_5} C^{\varepsilon\left(\lambda^{{\bf 16}}_a\right)}_{\lambda^{{\bf 16}_a}  }=
 (0,0,-4,0,0)_k \,.
\ee 
Combining these expressions, (\ref{InterSO10}) and (\ref{InterSO16}), we obtain precisely the desired result (\ref{SO10CS}), confirming our general analysis.


\subsection{Models with Exceptional Gauge Group}
\label{sec:E6Global}

For elliptic five-folds, the exceptional theories generically lead to non-minimal enhancement loci in codimension four (and already in codimension two and three for $E_7$ and $E_8$, respectively). 
We briefly discuss the salient properties of these models. The resolutions are summarized in appendix \ref{app:ExamplesE}.
For $E_6$ the codimension four locus $b_3=b_4=b_6=0$ is non-minimal and thus we impose that these intersection points are absent in the base four-fold: 
\be\label{NoNonMinE6}
M_G \cdot[b_4]\cdot[b_3]\cdot [b_6] =0 \,.
\ee
The anomaly and Chern-Simons relations can be checked and shown to be satisfied:
The group theoretic anomaly factors for $E_6$ are
\bea
C({\bf Adj}) = 2, \qquad C({\bf 27}) = \frac{1}{2}.
\eea
The anomaly from the 3-7 sector is then 
\be
\ba
\mathcal{A}_{3-7}&=  -  \frac{1}{{\rm ord}(g)}  \,  C({\bf 27})   \,   {1\over 24} \left( c_4 \cdot_{Y_5} M_G\right) \\
&= - \frac{1}{{\rm ord}(g)}    {1\over 48}
{M_G \cdot } \left(-774 c_1^2 M_G+549 c_1 M_G^2+360 c_1^3+12 c_2 c_1-126 M_G^3\right) \,,
\ea
\ee
and the single matter locus $b_3=0$ that gives rise to the ${\bf 27}$ representation has 
\be
\chi (b_3, {\bf 27}) =   \frac{1}{24} \, M_G \left(3 c_1-2 M_G\right) \left(-12 c_1 M_G+8 c_1^2+2 c_2+5 M_G^2\right) \,.
\ee
Then the anomaly indeed cancels  
\be
\mathcal{A}_{3-7} + \frac{1}{2} \chi (b_3, {\bf 27}) + 2 \chi_{\rm bulk} =0 \,
\ee
with $ \frac{1}{{\rm ord}(g)} = \frac{1}{6}$.
The naive expected value based on the $\mathbb Z_3$ monodromy around an $E_6$ locus would be $1/3$, but the monodromy is in general also sensitive to higher-codimension singular fibers.\footnote{The analogue of the perturbative orientifold reasoning sketched in footnote \ref{footnote_Z2} would be in terms of the non-perturbative $\mathbb Z_3$ `orientifold' of \cite{Dasgupta:1996ij}.}
It would be interesting to precisely understand the origin of this monodromy reduction in more detail. 

Finally, we can check also the relations from the Chern-Simons couplings. The intersections of $c_4(Y_5)$ with the Cartan divisors $D_i$ are summarized in (\ref{c4DiE6}). These satisfy the relations  
\be\label{E6CSCheck}
\ba
 {1\over 24} c_4(Y_5) \cdot D_1 &=  - 2 \chi_{\rm bulk} +2 \chi (b_3, {\bf 27}) \cr 
 {1\over 24} c_4(Y_5) \cdot D_2 &=  - 2 \chi_{\rm bulk} +\chi (b_3, {\bf 27})\cr 
 {1\over 24} c_4(Y_5) \cdot D_3 &=  - 2 \chi_{\rm bulk} \cr 
 {1\over 24} c_4(Y_5) \cdot D_4 &=  - 2 \chi_{\rm bulk}+\chi (b_3, {\bf 27}) \cr 
 {1\over 24} c_4(Y_5) \cdot D_5 &=  - 2 \chi_{\rm bulk} \cr 
 {1\over 24} c_4(Y_5) \cdot D_6 &=  - 2 \chi_{\rm bulk} 
\,. 
\ea
\ee


\begin{figure}
\centering
\includegraphics[width=8cm]{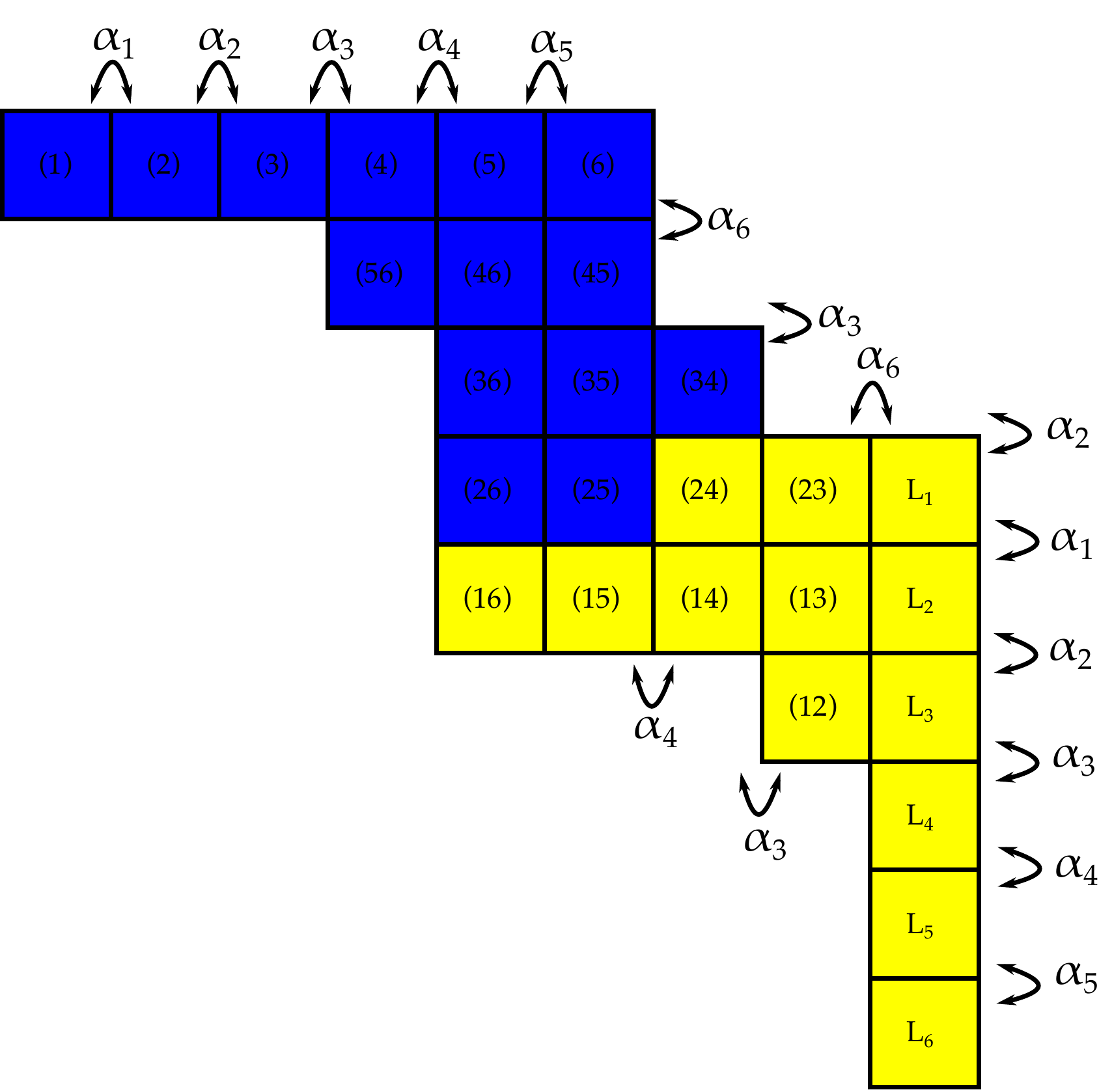}
\caption{Box Graph for the ${\bf 27}$ representation of $E_6$ corresponding to the fiber in codimension two of the $E_6$ model. Each box corresponds to a weight of the ${\bf 27}$, as listed in (\ref{E627Weights}). The action of the roots $\alpha_i$ connects the weights into this representation graph. 
The coloring corresponds to the signs blue $\varepsilon=+$ and yellow $\varepsilon=-$, indicating that $C^{\varepsilon(\lambda)}_{\lambda}$ is an effective curve.  
\label{fig:BoxGraphE6}}
\end{figure}


In the expression for $D_5$ we made use of the absence of the non-minimal loci (\ref{NoNonMinE6}). We now show that this is in agreement with the general analysis from the fiber under consideration. To determine the sign assignments of the weights of the effective curves $C^{\varepsilon\left(\lambda^{{\bf 27}}_a\right)}_{\lambda^{{\bf 27}_a}}$ of the ${\bf 27}$ representation, recall that these arrange in a representation graph as shown in figure \ref{fig:BoxGraphE6}. The resolution that we consider is given in terms of the sign assignments shown in the figure, where blue/yellow corresponds to $\varepsilon = \pm$.  
The notation here is as follows (for a more detailed exposition of these matters related to box graphs and fibers we refer the reader to \cite{Hayashi:2014kca}): $L_i$, $i=0, \cdots, 6$ as in $SU(n)$, and furthermore $3L_0 - (L_1 +L_2 + L_3 + L_4 + L_5 + L_6 )=0$, for these to represent the roots and weights of $E_6$. 
The simple roots are then $\alpha_i = L_i - L_{i+1}$ for $i=1, \cdots, 5$ and $\alpha_6= L_0 - L_1 -L_2 -L_3$. The weights of the ${\bf 27}$ can be written in terms of 
\be\label{E627Weights}
\lambda^{\bf 27}: \qquad \left\{ 
\ba
L_i &\qquad i= 1, \cdots, 6 \cr 
(i) &\qquad  2 L_{0} - \sum_{j\not= i} L_j \cr
(ij) & \qquad L_0 - L_i - L_j  
\ea\,,
\right.
\ee 
 and they are connected as in figure \ref{fig:BoxGraphE6} by the action of the simple roots. The coloring/sign-assignment of the graph dictates which curves are effective. Using this data, we can then compute the intersections relevant for the Chern-Simons couplings to be 
\be
\sum_{a=1}^{27}   D_k \cdot_{Y_5} C^{\varepsilon\left(\lambda^{{\bf 27}}_a\right)}_{\lambda^{{\bf 27}_a}  }=
{(-4,-2,0,-2,0,0)_k }\,.
\ee
These are precisely the values that enter into the linear combination in (\ref{E6CSCheck}), thus confirming our general expressions (\ref{chirality1}) and (\ref{Check-noflux}).

{Regarding models with gauge group $E_7$ and $E_8$, it must be ensured that all non-minimal loci in codimension two and three are absent. This requires that the  corresponding intersections of the discriminant components vanish. In these instances, we nevertheless obtain non-trivial gauge theories through bulk matter and its couplings. The resolutions and codimension two fiber properties for the $E_7$ model with matter in the ${\bf 56}$ representation as well as the $E_8$ model can be found in \cite{Lawrie:2012gg}, with the box graphs characterizing the fibers above the codimension two loci in the $E_7$ model classified in \cite{Hayashi:2014kca}. We expect there to be bulk-matter-surface interactions, and global consistency is ensured by restricting to models where potential non-minimal loci are absent. }


\section{Superconformal Theories and GLSM} 
\label{sec:SCFTGLSM}

There are many applications of the constructions obtained in this paper from the perspective of 2d field theory, of which we outline two in this section:
First we briefly comment on the relation between our models and $(0,2)$ superconformal field theories, including an outlook on the possible geometric realization of strongly coupled 2d theories from F-theory.
Second, we interpret the 2d $(0,2)$ models obtained from F-theory as heterotic worldsheet theories, relating in particular the celebrated Calabi-Yau - Landau-Ginzburg correspondence via 2d gauge theories of \cite{Witten:1993yc} to different Higgs bundle configurations in F-theory.


\subsection{{$(0,2)$ SCFTs}}

The 2d theories studied in this paper are $(0,2)$ supersymmetric, but in general not superconformal. 
In particular the gauge theory is super-renormalisable with a coupling $g_{YM}$ of mass dimension one. On general grounds, such theories become weakly coupled in the UV, where they flow to a trivial fixed-point, and strongly coupled in the IR. 
This raises the question of the existence of a strongly coupled superconformal fixed point in the IR. 
For $(2,2)$ gauged linear sigma models (GLSM), such a superconformal fixed-point is believed to exist in the IR and to describe the non-linear sigma model underlying Type II compactifications on a Calabi-Yau space \cite{Witten:1993yc,Witten:1993jg}. 
For $(0,2)$ GLSMs, superconformal invariance might be broken, corresponding to the appearance of a destabilizing superpotential in the  $N=1$ effective action describing the heterotic sigma model \cite{Silverstein:1995re}.

A {perturbative} criterion for existence of a superconformal fixed point in  $(0,2)$ GLSMs has been given in \cite{Silverstein:1994ih} (see also \cite{Distler:1993mk}): It involves the existence of a very specific non-anomalous $U(1)_R$ symmetry. This $U(1)_R$ symmetry can be constructed as the linear combination of the naive $U(1)_R$ symmetry associated with the $(0,2)$ supersymmetry algebra and any further global $U(1)$ symmetry present in the model. Let us parametrise the charges under this $U(1)_R$ symmetry of the various chiral multiplets $\Phi_i = (\phi_i, \psi_{+,i})$, the Fermi multiplets with fermions $\lambda_{-a}$ and the gauge multiplet with gaugino $\eta_-$  as
\be
Q_R(\varphi_i) = - \alpha_i, \qquad Q_R(\psi_{+i}) = 1 - \alpha_i, \qquad  Q_R(\lambda_{-a}) = -\alpha_a, \qquad Q_R(\eta_{-}) = -1 \,.
\ee
Then the criterion for existence of a superconformal IR fixed-point is that this $U(1)_R$ symmetry is free of mixed anomalies and that the charges must be related to the degrees of homogeneity of the superpotential $J_a$ and auxiliary $E^a$-fields as \cite{Silverstein:1994ih}
\be
\ba
\alpha_a J^a + \sum_i \Phi_i \frac{\partial J^a}{\partial \Phi_i} = J^a, \qquad   - \alpha_a J_a + \sum_i \Phi_i \frac{\partial E_a}{\partial \Phi_i} = E_a \,.
\ea
\ee 
An obvious first step in analyzing the possible superconformal IR fixed points in our context is therefore the study the existence of such a non-anomalous $U(1)_R$ symmetry \cite{TBA1}. This in particular exemplifies the importance of a complete and quantitative understanding of the Green-Schwarz mechanism for abelian symmetries. An alternative approach to determining the R-symmetry of the SCFT fixed-point is via $c$-maximisation as explored in \cite{Benini:2012cz,Benini:2013cda}.

In the 2d $(0,2)$ theories constructed in this {paper}, the gauge coupling is directly related to the volume of the complex three-cycle $M_G$ wrapped by the 7-branes, measured in string units,
$\frac{1}{g_{YM}^2} \simeq \ell_s^2 \,  {\rm Vol}(M_G)$. A similar relation holds for the Yang-Mills coupling for the gauge group factors associated with the D3-branes in the model, as summarized in (\ref{D3gym}). The flow to the strong coupling regime  $g_{YM} \rightarrow \infty$ can thus be engineered by taking the limit of shrinking complex three- and one-cycle volumes. 
The shrinking of a complex three-cycle $M_G$ to zero volume is compatible  with $M_G$ shrinking to a complex two-cycle, a one-cycle or even collapsing to a point on $B_4$.
 As this happens, M5-brane instantons wrapped along $M_G$ will become light and are expected to correct the dynamics of the $(0,2)$ theory. The engineering of the strong coupling regime for 6d $(0,1)$ theories by collapsing curves wrapped by 7-branes in F-theory has recently sparked a lot of interest \cite{Heckman:2013pva,Heckman:2015bfa}. The light modes associated with M5-instantons encountered in the 2d context are the analogue of the mysterious tensionless strings from wrapped M5-branes along the collapsing curves in 6d.  It will be interesting to study these effects, the relation to the existence of a strongly coupled superconformal sector and the possible classification of collapsing divisors on the base $B_4$ in {\cite{TBA1}}.


\subsection{GLSM Phases as T-branes/Gluing Data} 
\label{sec:NLSM}

Part of the fascination of 2d $(0,2)$  gauge theories {realized in terms of} GLSMs
is due to their role in interpolating \cite{Witten:1993yc} between non-linear sigma models (NLSMs) describing, for instance, the propagation of the heterotic string on a target space $X_{\rm het}$ with non-trivial gauge bundle $V_{\rm het}$ and a Landau-Ginzburg (LG) model, which can oftentimes be solved exactly. 
The GLSMs appearing in this context in principle  fall within the class of 2d $(0,2)$ F-theory models considered in this paper. 
By interpreting the heterotic worldsheet GLSM as an F-theory compactification on an elliptic fibration $Y_5$, we find a correspondence
\bea \label{correspondence1}
(X_{\rm het}, V_{\rm het})  \quad  
{\xleftrightarrow[{}]{\quad \text{F-theory}\quad } } 
 \quad  (Y_5, G_4)
\eea
between the heterotic target space {$X_{\rm het}$} and gauge bundle {$V_{\rm het}$} on the one hand,  and the F-theory five-fold $Y_5$ and extra gauge data $G_4$ on the other hand.
As one example of this correspondence, 
we will now relate the NLSM-LG-duality of  \cite{Witten:1993yc}  for the heterotic string to a change in the underlying F-theory Higgs bundle data.

The 2d F-theory models allow for two different regimes:
The first regime corresponds to a strict field theory limit with the 2d $(0,2)$ supergravity decoupled.
 The other is where we do not decouple supergravity, in which case we will have to integrate over all field configurations.
We now discuss both possibilities.
The decoupling limit is achieved by taking the base $B_4$ of the elliptic five-fold $Y_5$ to be non-compact while keeping the 7-brane volumes finite such that the 2d Newton constant goes to zero.\footnote{Note that Newton's constant in 2d is dimensionless. It is proportional to $1/{\rm vol}(B_4)$ with ${\rm vol}(B_4)$ measured in string units.} In this limit the geometric K\"ahler and complex structure moduli become non-dynamical fields and decouple from the gauge sector on the 7-branes. 
The resulting 2d gauge theory can then in principle be interpreted as a heterotic GLSM, in a fashion which we will discuss in more detail below.
The second regime corresponds to a finite base volume such that the 2d (0,2) supergravity sector remains dynamical. 
As it stands this sector differs from conventional heterotic worldsheet theories prior to gauge fixing because in the latter only an $N=(0,1)$ supersymmetry sector is local. It will therefore be interesting to study the implications of local $N=(0,2)$ supersymmetry, possibly in terms of a super-critical string theory as suggested in \cite{Apruzzi:2016iac}. Furthermore in this sector also the geometric moduli are fully dynamical. The relevance of this is\footnote{We thank Cumrun Vafa for discussions on this point. See also \cite{Apruzzi:2016iac}.} that
dynamical {\it massless} scalar fields in 2d quantum field theory cannot take a definite field value due to the well-known logarithmic infra-red divergence in their correlators. Rather, all field configurations must be integrated over in a quantum description of such theories. As far as the geometric moduli are concerned, this suggests that all regions of moduli space must be taken into account unless a dynamical stabilization mechanism gives rise to a mass term of the modulus in question.

After these general remarks let} us first briefly review the perhaps simplest example of a $(0,2)$ GLSM \cite{Witten:1993yc}. Its associated NLSM describes the heterotic string propagating on the quintic Calabi-Yau three-fold $X_{\rm het} = \mathbb P^4[5]$ coupled to a rank three vector bundle. The gauge group is just one $U(1)$, with fields charged as follows:
\be\label{QuinticEx}
\begin{array}{c|c|c}
\hbox{Field} & \hbox{Type} & U(1) \hbox{ Charge} \cr \hline
\Phi_i, \, i=1, \cdots, 5 & \hbox{Chiral} & +1 \cr 
P_i,  \, i=1, \cdots, 5 & \hbox{Fermi} & +1 \cr 
\Phi_0  & \hbox{Chiral} & -5 \cr 
P_0 & \hbox{Fermi} & -5 \cr  
\Sigma & \hbox{Chiral} & 0 
\end{array}
\ee
The Yukawa couplings in this model are determined by the auxiliary fields $E_i \equiv E^{(\rho_{- i})}$ and superpotentials $J^i \equiv J_{\rho_{- i}}$. These are taken to be the lowest order, but at least quadratic polynomials which are allowed by the gauge charges, subject to the constraint $E_i J^i= 0$ and otherwise generic.\footnote{Genericity implies transversality of the polynomials as detailed in \cite{Witten:1993yc}.}. Higher order terms compatible with the gauge charges are considered irrelevant in the RG sense \cite{Witten:1993yc} and are therefore discarded. This fixes
\be \label{EJQuintic}
\ba
E_i &= \Phi_i \, \Sigma \cr 
J^i &= \Phi_0  \, {\cal J}^i(\Phi_j) 
\ea
\qquad 
\ba
E_0 &= \Phi_0 \, \Sigma \cr
J^0 &= {\cal P}(\Phi_j) \,,
\ea
\ee
with ${\cal P}(\Phi_j)$ and   ${\cal J}^i(\Phi_j)$ homogeneous polynomials in $\Phi_i$ of degrees $5$ and $4$, respectively. These must obey the above constraint $E_i J^i= 0$.
The induced scalar potential takes the form
\be
\ba
V &= V_F + V_D \cr
V_F &= |{\cal P}|^2  + \sum_i  |\varphi_0|^2  | {\cal J}^i |^2    +  \sum_i |E_i|^2 + |E_0|^2 \cr
V_D &=  \frac{e^2}{2} \left( \sum_i |\varphi_i|^2 - 5 |\varphi_0|^2  -  r  \right)^2  \,,
\ea
\ee
with $r$ the Fayet-Iliopoulos (FI) parameter of the $U(1)$ gauge group with gauge coupling $e$.

The NLSM phase corresponds to the limit where $r \gg 0$: The D- and F-term constraints enforce  $\varphi_0=0$, but $\sum_i |\varphi_i|^2 =  r$ and ${\cal P}=0$. This suggests interpreting the charged scalar fields $\varphi_i$ as homogeneous coordinates of the space $\mathbb P^4 = ({\mathbb C}^5)^*/U(1)$. The NLSM target space is the hypersurface $X_{\rm het}: {\cal P}=0 \subset \mathbb P^4$. 
The gauge bundle is determined via the remaining $E-$ and $J$-fields (see e.g. \cite{McOrist:2010ae} for a review). Note that in the NLSM phase, the GLSM gauge group is completely broken by the VEV of  $\varphi_i$.
On the other hand, for $r \ll 0$ the GLSM flows to a Landau-Ginzburg orbifold model with $|\varphi_0|^2  =  - \frac{r}{5}$ and $\varphi_i=0$. Here the gauge group $U(1)$ is broken to the discrete remnant $\mathbb Z_5$ because of the charge $Q_0 = -5$ of $\varphi_0$.

To realize such GLSMs from F-theory, with only an abelian gauge group,  
our starting point is an elliptic Calabi-Yau five-fold $Y_5$ with Mordell-Weil group of rank one, realizing the $U(1)$ gauge group. 
This ensures the existence of one independent rational section $\sigma_1$
in addition to the zero-section $\sigma_0$. The $U(1)$ gauge group of the GLSM is then obtained by expanding the M-theory 3-form along the harmonic 2-form dual to the class $S_1$ obtained from $\sigma_1$ via the Shioda map, as reviewed around  (\ref{Shiodamap}).
This  makes direct contact with the recent advances  \cite{Grimm:2010ez,Braun:2011zm,Krause:2011xj,Morrison:2012ei,Cvetic:2012xn,Mayrhofer:2012zy,Braun:2013yti,Borchmann:2013jwa,Cvetic:2013nia,Braun:2013nqa,Cvetic:2013uta,Borchmann:2013hta,Cvetic:2013jta,Cvetic:2013qsa,Braun:2014nva,Kuntzler:2014ila,Klevers:2014bqa,Lawrie:2014uya,Lawrie:2015hia, Cvetic:2015ioa, Krippendorf:2015kta}  in the construction of elliptic fibrations with extra abelian gauge group factors for F-theory.  We will outline the fiber structure of $Y_5$ at the end of this section.

Of central importance in the NLSM-LG correspondence is the FI parameter $r$. 
As described in section \ref{fluxes5folds}, a {\it{field-dependent}} FI term arises in 2d F-theory models with $G_4$ flux from the  gauging of the axionic shift symmetry of axionic fields on $Y_5$. This term is simply the contribution of the charged K\"ahler moduli to the D-term and given by 
\be \label{rdefinition}
r \simeq \int_{Y_5} G_4 \wedge S_1 \wedge J_B \wedge J_B  =  \int_{B_4} F \wedge D_B \wedge J_B \wedge J_B \,,
\ee
with $J_B$ the K\"ahler form on the base $B_4$ of $Y_5$. The divisor class $D_B$ on $B_4$ describes the 7-brane effectively associated with the $U(1)$ gauge group and the class $F \in H^{1,1}(B_4)$ parametrises the flux on $D_B$. Both classes are determined by evaluating the expression after the first equality in the above equation \cite{Krause:2011xj}.

Let us first consider the decoupling limit, in which
\be \label{scaling-requirements}
{\rm vol}(B_4)= \int_{B_4} J_B^4 \rightarrow \infty\,, \qquad  \, \frac{1}{2 \kappa_2^2} \rightarrow \infty\,, \qquad  {\rm vol}(D_B) =  \int_{D_B} J_B^3 \,\, \, {\rm finite}\,.
\ee
The first two limits ensure that 2d gravity decouples and {some of the} K\"ahler moduli become non-dynamical, while by the last condition the $U(1)$ gauge coupling stays finite. The K\"ahler moduli dependent D-term remains finite in this limit.

As noted in \cite{Komargodski:2010rb } in the context for 4d $N=1$ theories, the existence of an FI term can be compatible with a decoupling limit, as long as one gauges the $\mathcal{S}$-multiplet introduced therein. We expect this result to hold similarly in 2d $(0,2)$ theories. In addition to the requirement of keeping a constant FI-parameter, and sending $\frac{1}{2 \kappa_2^2}$ and the volume of $B_4$ to infinity, we also have to require that the K\"ahler modulus coupling in the D-term $r$ becomes non-dynamical,  i.e. that its kinetic term diverges in this limit.\footnote{We thank Shamit Kachru for useful discussions on this point.} It would be particularly interesting to construct an example of a geometry which allows for such a decoupling, and a constant FI-parameter in the field theory.\footnote{A 4d example would for instance be realized in the context of the $\mathbb P_{11136}[12]$ geometry studied in \cite{Denef:2004dm}, where one can show that the D-term receives contributions from two moduli, one of which becomes non-dynamical in the limit of one of the cycles going to infinity whilst the other stays finite.}

By contrast, no such challenge arises for the interesting, though much less studied, class of GLSMs with simple gauge groups, which naturally arise from F-theory. Since the FI parameter in the GLSM is associated with the K\"ahler moduli of the target space, such models should correspond rather to non-geometric heterotic theories, and it would be very interesting to investigate these in the future.

Consider next the complementary regime, with 2d (0,2) supergravity not decoupled. As remarked above, a full description requires now integrating over all possible values of the moduli space (unless a dynamical stabilization mechanism is at work which induces a physical mass for the 2d scalars) \cite{Apruzzi:2016iac}. Despite this caveat, we see that for suitable fluxes\footnote{For $F = F_1 - F_2$ with $F_1$ and $F_2$ both effective the sign on $r$ depends on the K\"ahler moduli.} the K\"ahler moduli dependent D-term (\ref{rdefinition}) evaluates, in different regimes of the K\"ahler moduli space, to different values of $r$, interpolating between and including the  regions $r \gg 0$ and $r \ll 0$.     

These regions correspond to the NLSM and the LG phase 
and realize the two possible ways of Higgsing the $U(1)$ gauge group with the matter content (\ref{QuinticEx}). As the spectrum is chiral due to the fluxes, the vevs correspond to so-called Higgs bundle configurations, which are non-diagonalizable  and were termed gluing data or T-branes \cite{Cecotti:2010bp, Donagi:2011jy, Donagi:2011dv, Marsano:2012bf, Anderson:2013rka, Collinucci:2014taa}. Indeed, as pointed out in \cite{Marsano:2012bf}, globally, the different gluing configurations correspond to different types of fluxes. 
The NLSM-LG correspondence has then an interpretation in terms of different gauge theory backgrounds, given in terms of gluing data (or, equivalently, T-brane) configurations of the 7-brane theories in the five-fold:

\be
\begin{array}{ccccc}
{\rm NLSM-phase}             &  & {\rm GLSM}                                       &  &  {\rm LG-phase}   \vspace{2mm} \cr 
G = \emptyset                    &  \xleftarrow[{\ \rm data\ } ]{\ \rm gluing\ }  & G= U(1)                                            &  \xrightarrow[{\ \rm data\ }]{\ \rm gluing\ }  & G = \mathbb Z_5 \vspace{2mm}  \cr
(\tilde{A}, \tilde{\Phi})                           &   &  (A, \Phi)               &   &   (\hat{A}, \hat{\Phi})\vspace{2mm}  \cr
\end{array}
\ee

\begin{figure}
\centering
\includegraphics[width=12cm]{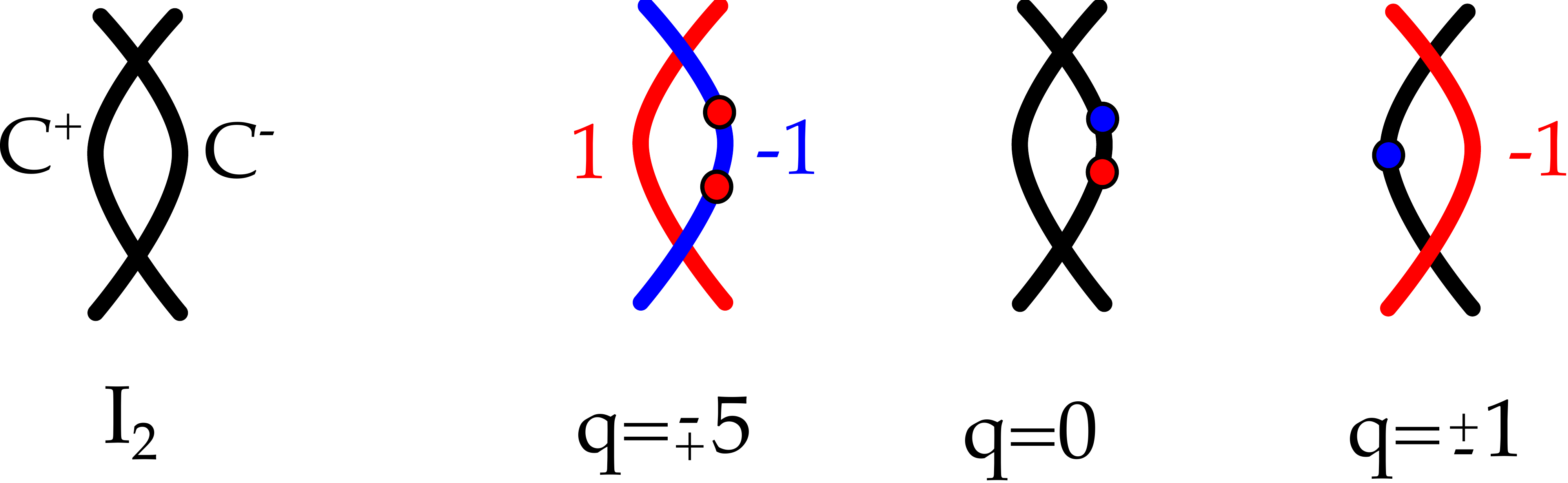}
\caption{The codimension two $I_2$ fibers, realizing  matter with charges $q= \mp 5, 0, \pm 1$. The left-most picture shows the 
$I_2$ fiber, with the two rational curves $C^\pm$ intersecting in two points. The remaining fiber diagrams show how the charges are realized in terms of sections intersecting or containing the curves $C^\pm$. Blue/red corresponds to the sections $\sigma_0$ and $\sigma_1$, respectively. The numbers next to fiber components contained (colored) in sections are the degrees of the normal bundle of the curve in the section.  
\label{fig:QuinticFib}}
\end{figure}

We now exemplify the fiber structure of the F-theory elliptic fibration $Y_5$ leading to the GLSM with matter content (\ref{QuinticEx}).
As noted already, to engineer such a model we require an elliptic fibration with one additional rational section, without any non-abelian enhancements in codimension one. The singular fiber in codimension one is {therefore a Kodaira $I_1$ fiber}. Along codimension two the singularity enhances to $I_2$ and generates the suitable charged matter. Let us denote the two fiber components in codimension two by $C^\pm$, which have the property that $C^+ \cdot C^- =2$. 
The enhacement from codimension one to two is in terms of the splitting 
\be
F_0 \quad \rightarrow \quad C^+ + C^- \,,
\ee
where $F_0$ is the single nodal fiber component of the $I_1$ fiber. 
The possible $U(1)$ charges for such models have been classified in \cite{Lawrie:2015hia}. For Calabi-Yau three-folds the constraints on the normal bundle degrees of contractible rational curves imply a finite range of matter charges\footnote{The normal bundle in the Calabi-Yau three-fold of a contractible rational curve can only be $N_{C/Y_3} = \mathcal{O}(p) \oplus \mathcal{O}(-2-p)$ for $p=-1,0,1$ \cite{Reid, Laufer}.}. For four- and five-folds, no analogous restriction on the normal bundle is known for contractible curves. Nevertheless, the charges can be determined as a function of the normal bundle degree \cite{Lawrie:2015hia}. Let us briefly summarize how the results therein would realize the spectrum in (\ref{QuinticEx}). Let $\sigma_0$ and $\sigma_1$ be the two sections of the model, with $\sigma_i \cdot F_0=1$. The $U(1)$ generator is given by $S_1= \sigma_1-\sigma_0 + D_{ B}$  for a suitable base divisor $D_B$. The charges of the singlet fields ${\bf 1}_{\pm q}$ are then obtained by 
\be
(\sigma_1 - \sigma_0) \cdot C^\pm = \pm q \,.
\ee
In codimension two, the section can either transversally intersect the fiber components $C^\pm$, or contain them $C^\pm \subset \sigma$. In such cases, the intersection number depends on the normal bundle of the curve $C^\pm$ in the divisor $\sigma$. The summary of this analysis is given in figure 17 in \cite{Lawrie:2015hia}, for 
\be
\ba
N_{C^+/Y_5} &= \mathcal{O} \oplus \mathcal{O} \oplus \mathcal{O}(1) \oplus \mathcal{O}(-3) \cr 
N_{C^- /Y_5} &= \mathcal{O} \oplus \mathcal{O} \oplus \mathcal{O}(-1) \oplus \mathcal{O}(-1) \,.
\ea
\ee
The charges $q=-5, 0, 1$ required for the GLSM in (\ref{QuinticEx}) can be obtained from the fiber configurations shown in figure \ref{fig:QuinticFib}.

To {obtain the exact spectrum including multiplicities in (\ref{QuinticEx}), we first need to find 
a realization of this model in terms of an explicit fibration giving rise to the codimension two fibers in figure \ref{fig:QuinticFib}. 
The above fibers are not unique in realizing these charges, and the complete set can be obtained from \cite{Lawrie:2015hia}. 
In addition multiplicities will be generated from fluxes.} 
{More precisely, we need to determine a gauge field background such that the cohomology groups counting matter with all charges other than the ones in (\ref{QuinticEx}) is trivial.}
{The construction of such an elliptic fibration realizing these fiber types will be an interesting challenge in the future. }

In a model with the exact matter content  (\ref{QuinticEx}), the $E$- and $J$-fields follow from the structure of matter interactions described in section \ref{sec_holocouplings}.
In that section, we focused on cubic $E$- and $J$-type interactions, assuming that suitable massless matter exists to form cubic gauge invariant interactions.
For the spectrum  (\ref{QuinticEx}), however, no such cubic $J$-type interactions are possible. Cubic interactions of the fields $\Phi_i$ with charge $1$ must necessarily involve fields of charge smaller than $5$, whose mass sits at the KK scale for suitable gauge flux, {as we assume here}. 
Integrating out these massive states will lead to higher-order effective couplings of the form (\ref{EJQuintic}) as these are the leading order gauge invariant couplings involving the massless spectrum  (\ref{QuinticEx}). Furthermore, if the original cubic couplings satisfy the constraint $E_i J^i =0$, this condition cannot be violated by integrating out massive states in a supersymmetric manner.

While we have focused on the simplest example of a GLSM, there are many generalizations to be explored.
For instance, GLSMs describing the heterotic string on hypersurfaces or complete intersections within toric spaces correspond, via the map (\ref{correspondence1}), to F-theory compactifications with a richer variety of charged matter and higher Mordell-Weil group rank.
Consider for example the GLSM whose associated NLSM has as its target space the CICY 
\be
X_{\rm het} = 
\Bigg[\begin{array}{c||ccc}
\mathbb P^2  & 1 & 1 & 1 \cr
\mathbb P^4  & 2 & 2 & 1
\end{array}\Bigg] \,.
\ee
This is a complete intersection of three hypersurfaces of degrees $(1,2)$,  $(1,2)$ and $(1,1)$ inside $\mathbb P^2 \times \mathbb P^4$.
The GLSM is a $U(1) \times U(1)$ gauge theory with the following fields:

\be\label{CICY}
\begin{array}{c|c|c}
\hbox{Field} & \hbox{Type} & U(1) \times U(1) \hbox{ Charge} \cr \hline
\Phi_i, \, i=1, \cdots, 3 & \hbox{Chiral} & (1,0) \cr 
P_i,  \, i=1, \cdots, 3 & \hbox{Fermi} & (1,0) \cr 
\tilde\Phi_m, \, m=1, \cdots, 5 & \hbox{Chiral} & (0,1) \cr 
\tilde P_m,  \, m=1, \cdots, 5 & \hbox{Fermi} & (0,1) \cr 
\Phi^{(A)}_0, A=1,2  & \hbox{Chiral} & (-1,-2) \cr 
P^{(A)}_0, A=1,2 & \hbox{Fermi} & (-1,-2) \cr  
\Phi^{(3)}_0 & \hbox{Chiral} & (-1,-1) \cr 
P^{(3)}_0 & \hbox{Fermi} & (-1,-1) \cr  
\Sigma & \hbox{Chiral} & (0,0)
\end{array}
\ee
In particular, the homogeneous coordinates of the  ambient space factors $\mathbb P^2$ and $\mathbb P^4$ are identified with the GLSM  fields $\Phi_i$, $i=1,\ldots,3$  and $\tilde \Phi_m$, $m=1,\ldots,5$, respectively. 
This GLSM can be obtained from F-theory compactified on a five-fold $Y_5$ with Mordell-Weil group of rank two.
In fact, the required type of fibration fits into the class \cite{Borchmann:2013jwa,Cvetic:2013nia,Cvetic:2013uta,Borchmann:2013hta}  constructed as an explicit hypersurface in a ${\rm Bl}_2 \mathbb P^2$-fibration, but now with a base four-fold $B_4$.
In this model there are six types of localised charged matter representations with charges $\pm(1,0)$, $\pm (0,1)$, $\pm(1,2)$, $\pm(1,1)$, $\pm(0,2)$, $\pm(1,-1)$ from fiber enhancements to $I_2$ along surfaces. The structure of the associated fibers is depicted e.g. in figure  2  of \cite{Borchmann:2013hta}. 
To interpret the 2d $(0,2)$ theory obtained from F-theory on this class of fibrations as the above GLSM, we must invoke suitable flux ensuring the precise spectrum  (\ref{CICY}) of massless fields, while all matter with charges  $(0,2)$, $(1,-1)$ must become massive.
The neutral field $\Sigma$ can be identified with a suitable supergravity mode.
The required $J$-couplings follow from the cubic gauge invariant couplings allowed by the fiber structure \cite{Borchmann:2013jwa,Cvetic:2013nia,Cvetic:2013uta,Borchmann:2013hta} upon integrating out the massive states of charge $(0,2)$ and $(1,-1)$.

More generally, for complete intersections in toric varieties, the number of scalings is reflected in the number of $U(1)$s of the GLSM and thus the rank of the Mordell-Weil group of the Calabi-Yau five-fold. The degree of each of the defining equations of the CICY gives a constraint on the $U(1)$ charges of the theory. The interplay between the charges both in concrete models such as e.g. \cite{Borchmann:2013jwa,Cvetic:2013nia,Cvetic:2013uta,Borchmann:2013hta,Cvetic:2013qsa,Klevers:2014bqa,Braun:2014qka,Cvetic:2015ioa}, as well as using the abstract classification of $U(1)^n$ charges in \cite{Lawrie:2015hia}, will be developed and explored in \cite{TBA1}.

{Complementary to this,} F-theory models with gauge group $U(n) \simeq SU(n) \times U(1)/\mathbb Z_n$ should describe the GLSMs underlying heterotic string propagation on Grassmannians \cite{Witten:1993yc}. Many more interesting possibilities which can be engineered from F-theory are described e.g. in  \cite{Donagi:2007hi,Jockers:2012zr,Halverson:2013eua} and references therein.  
It will be interesting to study the large class of GLSM gauge groups arising in F-theory from the perspective of the dual heterotic NLSM in the future.



\section{Conclusions and Future Directions} \label{sec_Concl}

F-theory on Calabi-Yau five-folds provides a rich class of $(0,2)$ supersymmetric string vacua in two dimensions. 
In this paper we have initiated the exploration of such 2d $(0,2)$ F-theory vacua by laying out
the 
correspondence between the field theoretic data of the 2d gauge theories and the geometry of the underlying elliptically fibered Calabi-Yau five-fold $Y_5$. 
We have applied two central tools in arriving at this dictionary: The first is the analysis of the 8d SYM theory on a stack of 7-branes, dimensionally reduced and topologically twisted along an internal K\"ahler three-cycle. This way we have determined the spectrum of charged massless $(0,2)$ multiplets and their non-derivative interactions in agreement with the structure of $(0,2)$ supersymmetry.
{Secondly},  in a global compactification this gauge and charged matter sector is encoded in the geometry of the elliptic fibers of $Y_5$ and their singularities. Non-trivial gauge backgrounds translate into M-theory four-form fluxes. Utilitizing M/F-theory duality, we {were} able to derive a rich set of global consistency conditions, and checked the validity of our approach in terms of models with ADE-type gauge groups, including also {additional} abelian gauge {group} factors. 
From these results, many exciting avenues for exploring this new class of 2d $(0,2)$ theories open up. 
\begin{enumerate}
\item Derivation of {the} Supergravity Spectrum: 
The main focus of this first analysis has been on the gauge theoretic data of the 2d $(0,2)$ theories and therefore on the charged sector. 
However, as demonstrated for instance by the intricate structure of Green-Schwarz terms in the presence of $U(1)$ gauge symmetry, the gauge sector cannot always be analyzed in complete isolation from the supergravity modes arising from the Calabi-Yau five-fold.
An identification of the spectra in M-theory and F-theory can be found in section \ref{sec:Dictionary}, but it wold be 
particularly interesting to derive the structure of superfields in the 2d $(0,2)$ supergravity in full detail and match these with the dual $N=2$ super-mechanics \cite{Haupt:2008nu} obtained from M-theory. 
\item Geometry of higher-dimensional elliptic Calabi-Yau varieties:   Our understanding of five-folds in this paper builds upon
the recent progress in describing the geometry of elliptically fibered three- and four-folds. Nevertheless, as we have seen, the higher-codimension fibers offer several {new effects}, and an in-depth analysis of these is mandatory in order to fully understand the gauge-geometry dictionary. 
Our analysis here has focused on the fiber structure without any reference to the specifics of the base $B_4$ of the elliptic five-fold.
However, fundamental questions such as the criteria for non-Higgsability of singularities, as analysed for three- and four-folds in \cite{Morrison:2012np, Morrison:2012js, Grassi:2014zxa, Morrison:2014lca, Halverson:2015jua, Taylor:2015isa}, depend on the specifics of the base. Understanding which four-folds $B_4$ can serve as base spaces for consistent elliptically fibered Calai-Yau five-folds is thus an important step towards classifying the resulting 2d $(0,2)$ theories.
\item  D3-brane sector:
Apart from the 7-brane sector, gauge and matter degrees of freedom arise from D3-branes wrapping holomorphic curves on the base $B_4$. These are particularly important because the matter in the 3-7 sector contributes to the gauge anomalies  of the chiral 2d $(0,2)$ theory.
Unlike the 7-branes, the D3-brane sector is not automatically encoded in the geometry of the elliptic fibration. It is considerably harder to approach via duality with M-theory, where the D3-branes dualize to M2-branes.
In this paper we have treated the D3-brane sector purely perturbatively. A priori this is only an accurate description for certain types of singularities. Interestingly, this approach nonetheless {gives} a consistent spectrum of 3-7 strings even for F-theory models without an orientifold limit {upon inclusion of appropriate $SL(2,\mathbb Z)$ monodromy factors}. 
Generalizing our treatment of the D3-brane sector to arbitrary monodromies of the axio-dilaton $\tau$ will be an important 
step towards {understanding} this sector completely and will be addressed in \cite{Lawrie:2016axq} on the basis of a topological twist similar to the analysis of the 7-brane sector.\footnote{Studies of related D3-brane setups with varying coupling can be found in \cite{Martucci:2014ema, Assel:2016wcr}.}
\item Relation to 2d SCFTs and strong coupling limit: As briefly recalled in section \ref{sec:SCFTGLSM}, the existence of a superconformal IR fixed point is far from trivial for 2d $(0,2)$ theories. It will be {interesting} to apply the techniques of \cite{Silverstein:1994ih} or \cite{Benini:2012cz,Benini:2013cda} in order to address this question for the 2d $(0,2)$ models obtained from F-theory \cite{TBA1}.
The results of these papers suggest that only a subclass of the 2d $(0,2)$ theories obtainable from F-theory may flow to a strongly coupled IR SCFT, and it would be exciting to develop methods for their classification.
Another, possibly related direction is to study the strong coupling regime of the 2d $(0,2)$ models in the limit of vanishing volume of the base three-cycles wrapped by the 7-branes  and likewise of  the holomorphic curves wrapped by the D3-branes. In this context the aforementioned study of the base properties will play a crucial role in pursuing the ambitious long-term goal of obtaining a classification of the 2d $(0,2)$ SCFTs obtainable via F-theory. 

\item
{Heterotic/F-theory duality}: While we have started exploring 2d F-theory vacua from the perspective of duality with M-theory as well as in their Type IIB description, another angle is via duality to the heterotic string. This requires the base $B_4$ to be $\mathbb P^1$-fibered over a three-fold $B_3$. The dual heterotic theory is defined by compactification on a Calabi-Yau  four-fold $Z_4$ which is elliptically fibered over $B_3$. It will be interesting to extend the construction of heterotic  gauge bundles via spectral covers known for Calabi-Yau three-folds \cite{FMW} to Calabi-Yau four-folds. 
More generally, one should systematically explore the construction of 2d $(0,2)$ gauge theories obtained via heterotic compactification on possibly not elliptically fibered Calabi-Yau four-folds.

\item Relation to $(0,2)$ worldsheet theories:
As the study of the 2d $(0,2)$ theories obtained from F-theory progresses, it will be crucial to determine the relation between this class of models and the $(0,2)$ theories considered in the literature as heterotic worldsheet theories. 
As {discussed} in section \ref{sec:NLSM}, 
engineering a  GLSM \cite{Witten:1993yc}  with only abelian gauge multiplets from F-theory  requires a fibration with a non-trivial Mordell-Weil group of rational sections as these are responsible for abelian gauge symmetries in the effective theory, but without 
{additional} non-abelian singularities. 
Since the heterotic target space geometry and gauge bundle are determined by the $J$ and $E$-type interactions of the GLSM
it will be important to understand the structure of couplings in more detail, including also non-perturbative corrections. 
The ease with which non-abelian gauge groups appear in F-theory suggests studying also the associated heterotic worldsheet interpretation of the associated GLSMs.

\end{enumerate}

The synthesis of the last three directions {laid out above may well} establish 2d $(0,2)$ theories as a link in a new duality between Calabi-Yau spaces of different dimensions:
  As we have seen, the geometry (plus extra M-theory data such as fluxes) of an elliptic Calabi-Yau five-fold defines a 2d $(0,2)$ gauge theory. If this theory has an IR SCFT fixed-point, it should admit an interpretation as the worldsheet theory of the heterotic string describing compactification on another Calabi-Yau space, together with a gauge bundle modulo the caveats we have described. The information of this effective heterotic compactification geometry must therefore be related to the geometry of the elliptic five-fold in a non-trivial manner. It will be exciting to explore this new connection in the future.

\subsection*{Acknowledgements}

We thank Benjamin Assel,  Andreas Braun, Stefano Cremonesi, Arthur Hebecker, Craig~Lawrie, Ling Lin, Christoph Mayrhofer, Eran Palti, Christian Reichelt, Eric Sharpe, Oskar Till and Jenny Wong for discussions. We thank the Aspen Center for Physics, where this project was started, and the Galileo Galilei Institute, Florence, for hospitality during the course of this work. The work of SSN is supported in part by STFC grant ST/J002798/1 and the work of TW in part by the DFG Transregio 'The Dark Universe'. The Aspen Center for Physics is supported by National Science Foundation grant PHY-1066293.

\appendix

\section{Conventions and Supersymmetry Variations}
\label{app:Gamm}

In this appendix we will summarize our conventions in the main text regarding the 8d SYM theory and its dimensional reduction and topological twist. 

\subsection{Conventions}

We construct the 8d SYM theory by dimensionally reducing 10d SYM. The twisted reduction of the 8d theory is then performed by further reducing on a (Euclidean signature) 6-cycle.  
It is therefore useful to build the 10D Gamma matrices $\Gamma^M$ for $SO(1,9)$ starting with the $SO(6)$ gamma matrices $\gamma^m$ as follows
\be\ba
\Gamma^0 &= \sigma \otimes (i \sigma^1) \otimes {\bf 1}_8 
\cr 
\Gamma^1 &= \sigma \otimes \sigma^0 \otimes {\bf 1}_8 \cr 
\Gamma^m &= \sigma \otimes \sigma \otimes \gamma^m \,,\qquad m=2, \cdots, 7\cr
\Gamma^8 &= \sigma^0 \otimes {\bf 1}_{16} 
\cr 
\Gamma^9 &= \sigma^1 \otimes {\bf 1}_{16} 
\,,
\ea\ee
where the abbreviation was used
\be
 \sigma = \left( \begin{array}{cc}
-1& 0\cr 0& 1
\end{array}\right) \,,\qquad 
 \sigma^0 = \left( \begin{array}{cc}
0& 1\cr 1& 0
\end{array}\right)\,,\qquad 
 \sigma^1 = \left( \begin{array}{cc}
0& -i\cr i& 0
\end{array}\right) \,.
\ee
These satisfy the standard 10d Clifford algebra 
\be
\{\Gamma^M, \Gamma^N\} = 2 \eta^{MN}\,.
\ee
The dimensional reduction from 10d to 8d is along $x^8$ and $x^9$, and the transverse directions after the reduction along $M_G$ are $x^{0}$ and $x^1$. The chirality operators in each of the relevant dimensions will be useful in the following and are 
\be\ba
\Gamma_{2d} &= \Gamma^0 \Gamma^1 \,,\quad  
\Gamma_{6d} = i \, \Gamma^2 \Gamma^3 \ldots \Gamma^7 \,, \quad   
\Gamma_{8d} = \Gamma_{2d} \,  \Gamma_{6d} = i \, \Gamma^0 \ldots \Gamma^7,  \\
\Gamma_{10d} &= \Gamma_0 \Gamma_1 \ldots \Gamma_9 \,. 
\ea\ee
The conventions for the Lorentzian chirality matrices are $\Gamma_d = i^{-k} \prod \Gamma_i$   with $d = 2k +2$. In the Euclidean chirality matrix  $\Gamma_{6d}$ we have chosen the prefactor $i$ in order to ensure that $\Gamma_{8d} = \Gamma_{2d} \,  \Gamma_{6d} $. Furthermore define the R-symmetry generator as 
\be
\Gamma_R =  - i \, \Gamma_8 \Gamma_9 \,,
\ee
which is the chirality matrix in the Euclidean $8-9$ plane, and  $\Gamma_{10d} = \Gamma_{8d} \,  \Gamma_R$.  
Reality conditions on spinors are imposed with 
\be
B =\Gamma^3 \Gamma^5 \Gamma^7 \Gamma^9 
\ee
with the properties 
\be
B^* B =  {\bf1}, \qquad B = B^T \,,
\ee
and the charge conjugation matrix in 10d is 
\be \label{10dConjugation}
C= B\,  \Gamma^0 .
\ee
The 10d 32-component spinor can be written as
\bea
\Psi_{10d} = (\psi^{++}, \bar\psi^{++},\psi^{-+},\bar\psi^{-+}, \psi^{+-},\bar\psi^{+-},\psi^{--},\bar\psi^{--})^T,
\eea
where the first superscript denotes the 2d chirality, i.e. the eigenvalue with respect to $\Gamma_{2d}$, and the second superscript denotes the R-charge.

The 10d positive and negative chirality spinors, defined with respect to $\Gamma_{10d}$, decompose into 8d spinors with R-charges $\pm 1$ according to
\be
\ba
{\bf 16} &= {\bf 8}^{c,+_R} + {\bf 8}^{s,-_R}  =  (\psi^{++}, 0,0,\bar\psi^{-+}, 0,\bar\psi^{+-},\psi^{--},0) \cr 
\overline{\bf 16} &= {\bf 8}^{c,-_R} + {\bf 8}^{s,+_R}  =  (0, \bar\psi^{++},\psi^{-+} ,0, \psi^{+-},0,0,\bar\psi^{--}) \,,
\ea
\ee
where
\be
\ba
{\bf 8}^{c,+_R} &= \Psi^{++} + \overline\Psi^{-+}= (\psi^{++}, 0,0 ,\bar\psi^{-+}, 0,0,0,0) \cr 
{\bf 8}^{c,-_R}  &=  \Psi^{+-} + \overline\Psi^{--}= (0, 0,0 ,0, \psi^{+-},0,0,\bar\psi^{--}) \cr 
{\bf 8}^{s,+_R} &=\Psi^{-+} + \overline\Psi^{++}= (0, \bar\psi^{++},\psi^{-+},0, 0,0,0,0) \cr 
{\bf 8}^{s,-_R} &=   \Psi^{--} + \overline\Psi^{+-}= (0, 0,0,0, 0,\bar\psi^{+-},\psi^{--},0) \,.
\ea
\ee
Each of the $\Psi$ ($\overline{\Psi}$) transform as ${\bf 4}$ ($\bar{\bf 4}$) under $SO(6)$. 
Let $\alpha =1, \cdots, 4$ and $\dot\alpha= \dot{1}, \cdots, \dot{4}$  be indices labeling the four components of ${\bf 4}$ and $\bar{\bf 4}$, respectively. 
Then for instance
\be
\Psi^{++} = (\psi^{++}_\alpha, 0,0 ,0, 0,0,0,0) \,, \qquad 
\overline\Psi^{-+}= (0, 0,0 ,\bar\psi^{-+ \dot\alpha}, 0,0,0,0)  \,.
\ee
The Majorana condition on the ten-dimensional spinors is 
\be
\Psi^* = B \Psi \,,
\ee
in particular
\be
\ba
B \Psi^{++} &= {\overline{\Psi}^{+-}}^* \cr 
B \Psi^{--} &= {\overline{\Psi}^{-+}}^* \,.
\ea
\ee
 The conjugate spinor is then defined to be 
\bea
\bar\Psi = \Psi^T B   \Gamma^0  = \Psi^T C\,.
\eea

Furthermore, acting with charge conjuation $C= B \Gamma^0$ yields 
\be \label{BarRels}
\overline{\Psi^{++}}  =\overline{\Psi}^{--} \,,\quad  
\overline{\overline{\Psi}^{--}} = \Psi^{++}\,,\quad 
\overline{\Psi^{-+}}  =\overline{\Psi}^{+-} \,,\quad 
\overline{\overline{\Psi}^{+-}} = \Psi^{-+}\,,
\ee
i.e. the conjugate spinor to $\Psi^{++}$ transforms in $\bar{\bf 4}$, and has $SO(1,1)$ and $U(1)_R$ charges $-1$. 
Using the block form of the charge conjugation matrix, the conjugate of a 32-component positive chirality Majorana-Weyl spinor is found to be given by
\be\ba
\bar \Psi &= \left( (0,\overline{\psi^{--}}), (\overline{\bar\psi^{+-}},0), ( - \overline{\bar \psi^{-+}},0), (0, - \overline{\psi^{++}}) \right) \cr 
&=:  \left( (0,{\bar\psi^{++}}), ({\psi^{-+}},0), ( - {\psi^{+-}},0), (0, - {\bar\psi^{--}}) \right) 
\ea
\ee
in terms of its constituent $SO(6)$ Weyl spinors. The latter are given for instance by
\bea
\psi^{--}_\alpha &=& (\psi_1, \psi_2,\psi_3,\psi_4) \rightarrow (\bar\psi^{++})_{\dot\alpha}  = (-\psi_4, \psi_3,-\psi_2,\psi_1) \label{conjugate1}, \\
{(\bar\psi^{+-})}^{\dot\alpha} &=& (\bar\psi^{\dot1}, \bar\psi^{\dot2},\bar\psi^{\dot3},\bar\psi^{\dot4})  \rightarrow  (\psi^{-+})^{\alpha}   =  (\bar\psi^{\dot4 }, - \bar\psi^{\dot 3},\bar\psi^{\dot 2}, -\bar\psi^{\dot 1}). 
\eea

We will make frequent use of the decomposition of the vector and spinor representations under $SU(4) \rightarrow SU(3) \times U(1)$, under which
\be
{ \bf 4} \rightarrow  {\bf 3}_{-1} + {\bf 1}_{3}\,, \qquad  
{ \bf \bar 4} \rightarrow  {\bf \bar 3}_{1} + {\bf 1}_{-3}\,, \quad 
{\bf 6} \rightarrow  {\bf 3}_{2} + {\bf \bar 3}_{-2} \,.
\ee
Let us fix the embedding of $SU(3)$ into $SU(4)$ by the convention that for the ${\bf \bar 4} $ representation, i.e. the anti-chiral spinor $\bar \psi^{\dot\alpha}$, we identify
\bea \label{4into31}
(\bar \psi^{\dot 1}, \bar \psi^{\dot 2}, \bar \psi^{\dot 3})    \longleftrightarrow {\bf \bar 3}_{1}, \qquad \quad \bar \psi^{\dot 4} \longleftrightarrow 1_{-3} .
\eea
Since the product of an anti-chiral spinor $\bar\psi^{\dot \alpha}$ with the conjugate of a chiral spinor, $\bar\psi_{\dot \alpha}  \bar\psi^{\dot\alpha}$, forms a singlet, this implies that for the conjugate spinor the components $\bar\psi_{\dot 4}$ and  $(\bar\psi_{\dot1}, \bar\psi_{\dot2}, \bar\psi_{\dot3})$ correspond to the singlet and the triplet, respectively. Remembering the relation (\ref{conjugate1})  defining the conjugate spinor 
we conclude that
\bea  \label{4into32}
(-\psi_4, \psi_3, -\psi_2)  \longleftrightarrow {\bf 3}_{-1}, \qquad  \psi_{1} \longleftrightarrow 1_{3}.
\eea

With conventions fixed like this, the decomposition of the vector representation of $SO(6)$  is determined uniquely. Given a vector $A_m$, $m=2,\dots, 7$, we interpret its components in terms of the antisymmetric ${\bf 6}$ of $SU(4)$ and its conjugate with the help of the conjugate $SO(6)$ gamma matrices
\be
\sum_m A_m  (\gamma_m)_{\alpha}^{\, \, \, \beta} =
\left( \begin{array}{cccc}
0 & A_6 + i A_7   & A_4 + i A_5  & A_2 + i A_3    \cr
 - A_6 - i A_7  & 0 & A_2 - i A_3 & - A_4 + i A_5  \cr
  - A_4 - i A_5 &    - A_2 + i A_3   & 0 & A_6 - i A_7 \cr
   - A_2 - i A_3   & A_4 - i A_5  & - A_6 + i A_7 & 0 
 \end{array} \right)
 \ee
 \be
 \sum_m A_m (\gamma_m^\dagger)^{\dot \alpha}_{\, \, \dot \beta}  = 
 \left( \begin{array}{cccc}
0 &  -A_6 - i A_7   &  - A_4 - i A_5  & -  A_2 + i A_3    \cr
  A_6 + i A_7  & 0 & - A_2 - i A_3 &  A_4 - i A_5  \cr
   A_4 + i A_5 &     A_2 + i A_3   & 0 &  - A_6 + i A_7 \cr
    A_2 - i A_3   &  - A_4 + i A_5  &  A_6 - i A_7 & 0 
 \end{array} \right) .  \label{6ofSU4dot}
\ee
In the decomposition ${\bf 6} \rightarrow  {\bf 3}_{2} + {\bf \bar 3}_{-2}$, the ${\bf \bar 3}_{-2}$ corresponds to the two-index anti-symmetric representation of $SU(3)$. With the anti-fundamental representation ${\bf \bar 3}_{1}$ fixed to be associated with spinors indices ${\dot 1, \dot 2, \dot 3}$ by (\ref{4into31}), there exists a singlet in the product
\bea
({\bf \bar 3}_2)^{\dot {\underline \beta}} \, \,  ({\bf \bar 3}_{-1})^{\dot {\underline \alpha}} \,  \, ({\bf \bar 3}_{-1})^{\dot {\underline \gamma}} \, \varepsilon^{{\dot {\underline \beta}} {\dot {\underline \alpha}} {\dot {\underline \gamma}}}, \quad \quad    {\dot{\underline \alpha}} , \dot{{\underline \beta}} , \dot{{\underline \gamma}}  \in {\dot1, \dot2, \dot3}.
\eea
For example, the component $({\bf \bar 3}_2)^{\dot { 3}} $ must correspond to the entry in the representation (\ref{6ofSU4dot}) of the antisymmetric ${\bf 6}$ of $SU(4)$ which contracts with the component ${\bf \bar 3}_{-1}^{\dot 1}$ and ${\bf \bar 3}_{-1}^{\dot 2}$, i.e. the entry associated with $ (\gamma_m^\dagger)^{\dot 2}_{\, \, \dot 1})$.
In all this yields  the identification 
\be
A^{\underline{\dot\beta}} 
= \left( \begin{array}{c}
 \sum_m A_m (\gamma_m^\dagger)^{\dot 3}_{\, \, \dot 2} \cr
  \sum_m A_m (\gamma_m^\dagger)^{\dot 1}_{\, \, \dot 3} \cr
   \sum_m A_m (\gamma_m^\dagger)^{\dot 2}_{\, \, \dot 1})\cr
    \end{array} \right)  
=      \left( \begin{array}{c}
A_2 + i A_3\cr
 - A_4 - i A_5\cr
   A_6 + i A_7\cr
    \end{array} \right)  
    \quad      \longleftrightarrow\quad  {\bf \bar 3}_{-2}\,.
\ee

Applying analogous reasoning to the representation ${\bf 3}_{2}$, or simply using that its components are the complex conjugate of those of ${\bf \bar 3}_{-2}$, we furthermore identify
\be
A_{\underline\beta} =
\left( \begin{array}{c} 
\sum_m A_m (\gamma_m)_2^{\, \, 3} \cr
 \sum_m A_m (\gamma_m)_2^{\, \, 4} \cr
  \sum_m A_m (\gamma_m)_3^{\, \, 4} \cr
  \end{array} \right)  
  =      \left( \begin{array}{c}
A_2 - i A_3\cr
 - A_4 + i A_5\cr
   A_6 - i A_7\cr
    \end{array} \right)  
\quad 
   \longleftrightarrow 
   \quad 
   {\bf 3}_{2} \,.
\ee


\subsection{Variations}

The variation of $\Phi_{8}$ and $\Phi_{9}$ is given by
\bea
i \delta \Phi_8 &=& \bar\varepsilon \,  \Gamma^8 \,  \Psi = \overline{\varepsilon^{--}}_{\dot\alpha} (\bar\psi^{+-})^{\dot\alpha} +   (\overline{\bar\varepsilon^{+-}})^{\alpha} \psi_\alpha^{--} -  (\overline{\bar \varepsilon^{-+}})^\alpha \psi^{++}_\alpha 
- (\overline{\varepsilon^{++}})_{\dot\alpha} {(\bar\psi^{-+})}^{\dot\alpha} \\
&=& (\bar\varepsilon^{++})_{\dot\alpha} (\bar\psi^{+-})^{\dot\alpha} +   ({\varepsilon^{-+}})^{\alpha} \psi_\alpha^{--} -  (\varepsilon^{+-})^\alpha \psi^{++}_\alpha - 
 (\bar\varepsilon^{--})_{\dot\alpha} (\bar\psi^{-+})^{\dot\alpha} 
 \eea
 and
 \bea
i \delta \Phi_9 &=& \bar\varepsilon \, \Gamma^9 \,  \Psi  = 
-i \overline{\varepsilon^{--}}_{\dot\alpha} (\bar\psi^{+-})^{\dot\alpha} -i    (\overline{\bar\varepsilon^{+-}})^{\alpha} \psi_\alpha^{--} - i (\overline{\bar \varepsilon^{-+}})^\alpha \psi^{++}_\alpha 
- i(\overline{\varepsilon^{++}})_{\dot\alpha} {(\bar\psi^{-+})}^{\dot\alpha} \\
&=&
-i (\bar\varepsilon^{++})_{\dot\alpha} (\bar\psi^{+-})^{\dot\alpha} -i   ({\varepsilon^{-+}})^{\alpha} \psi_\alpha^{--} -i  (\varepsilon^{+-})^\alpha \psi^{++}_\alpha - i
 (\bar\varepsilon^{--})_{\dot\alpha} (\bar\psi^{-+})^{\dot\alpha} 
 \eea
and thus 
\bea
i\,  \delta(\Phi_8 + i \Phi_9) &=&  2 \,\,  \left( \overline{\varepsilon^{--}}  \, \bar\psi^{+-} + {\overline{\bar\varepsilon^{+-}}} \, \psi^{--} \right) = 2 \,\,  \left( \bar\varepsilon^{++}  \, \bar\psi^{+-} + {\varepsilon^{-+}} \, \psi^{--} \right)  , \\
i \, \delta(\Phi_8 - i \Phi_9) &=&  -2 \, \left(\overline{\bar \varepsilon^{-+}} \, \psi^{++}   + \overline{\varepsilon^{++}} \, \bar\psi^{-+}  \right) = -2 \, \left({ \varepsilon^{+-}} \, \psi^{++}   + {\bar\epsilon^{--}} \, \bar\psi^{-+}  \right).
\eea

The variation of $A_m$ takes the form
\bea
i\, \delta A_m &=& (0,\overline{\epsilon^{--}})  \gamma_m 
\left(\begin{array}{c}
\psi^{++} \cr
0
\end{array}\right) +  (\overline{\bar\epsilon^{+-}},0) (-\gamma_m) 
\left(\begin{array}{c}
0 \cr
\bar\psi^{-+}
\end{array}\right)  \\
&& + 
( - \overline{\bar \epsilon^{-+}},0) (-\gamma_m)  \left(\begin{array}{c}
0 \cr
\bar\psi^{+-}
\end{array}\right)  + 
(0, - \overline{\epsilon^{++}}) \gamma_m 
\left(\begin{array}{c}
\psi^{--} \cr
0
\end{array}\right) .
\eea

With this we find
\bea
i \delta A_{\underline\beta}|_{\pm \pm} = 2 \, 
\left(\begin{array}{c}
 (\bar\epsilon^{\pm \pm})_{\dot1} \,  \psi_{1}^{\pm \pm}  +    (\bar\epsilon^{\pm \pm})_{\dot4} \,   \psi_4^{\pm \pm}                 \cr
 (\bar\epsilon^{\pm \pm})_{\dot2}  \,  \psi_{1}^{\pm \pm }   -   (\bar\epsilon^{\pm \pm })_{\dot4}  \,   \psi_3^{\pm \pm }         \cr
  (\bar\epsilon^{\pm \pm })_{\dot3} \,  \psi_{1}^{\pm \pm }   +  (\bar\epsilon^{\pm \pm })_{\dot4}  \,  \psi_2^{\pm \pm }                  \cr
\end{array}\right)
\eea
and
\bea
i \delta A^{\underline{\dot \beta}}|_{\pm \mp} = 2 \, 
\left(\begin{array}{c}
 (\epsilon^{ \pm \mp})^{1}   (\bar\psi^{\pm \mp})^{\dot 1}  +       (\epsilon^{\pm \mp})^{4}   (\bar\psi^{\pm \mp})^{\dot 4}      \cr
  (\epsilon^{ \pm \mp})^{1}  (\bar\psi^{ \pm \mp })^{\dot 2}    -     (\epsilon^{ \pm \mp})^{3}    (\bar\psi^{\pm \mp })^{\dot 4}         \cr
 (\epsilon^{ \pm \mp })^{1}   (\bar\psi^{\pm \mp })^{\dot 3}  +        (\epsilon^ {\pm \mp})^{2}  (\bar\psi^{ \pm \mp })^{\dot 4}         \cr
\end{array}\right).
\eea

After the twist only the terms involving $(\bar\epsilon^{++})_{\dot4}$ and  $(\bar \epsilon^{- +})^{1}$ survive as these correspond to singlets of $SU(3)$.
The  other variations  $i \delta A^{\underline{\dot\beta}}|_{\pm \pm}$ and $i \delta A_{\underline{\beta}}|_{\pm \mp}$ only contain combinations of  $\epsilon$ which do not survive the twist.


\subsection{Supersymmetry Variations for Twisted Theory}
\label{app:SUSYDetails}

To derive the supersymmetry variations in 2d, we  start with the 10d SYM  theory 
\be
\mathcal{L}_{10d}= -{1\over 4 g^2}\hbox{Tr} \left( F_{MN}F^{MN}\right) - {i \over 2 g^2} \hbox{Tr} \left(  \overline{\Psi} \Gamma^M D_M \Psi\right)\,,
\ee
which is invariant under the supersymmetry variations 
\be\ba
\delta A_M &=-i \bar\epsilon \Gamma_M \Psi \cr
\delta \Psi & = {1\over 2} F_{MN} \Gamma^{MN} \epsilon \,.
\ea\ee
Using the spinor and gamma-matrix decompositions in the last section, and noting that the supercharges that remain after the twist are 
\be
\epsilon_- = \epsilon^{--} \,,\qquad 
\bar\epsilon_{-} = \bar{\epsilon}^{-+} \,,
\ee
the variations of the gauge field $A_M$ reduce as follows
\be\ba
i \delta \Phi_8 &= 
\bar\epsilon \,  \Gamma^8 \,  \Psi =
{\epsilon^{--}}_{\dot 4} (\bar\psi^{+-})^{\dot4 } 
 -  {\bar \epsilon}^{-+\, 1} \psi^{++}_1 ,
\cr 
i \delta \Phi_9 &= \bar\epsilon \, \Gamma^9 \,  \Psi  = 
-i {\epsilon^{--}}_{\dot4} (\bar\psi^{+-})^{\dot 4} 
- i {\bar \epsilon}^{-+\, 1} \psi^{++}_1 \,,
\ea
\ee
and thus 
\be\ba
 \delta(\Phi_8 + i \Phi_9)  \equiv  \delta \varphi   &=  -2i \, {\epsilon^{--}}  \, \bar\psi^{+-}   \equiv  - \sqrt{2} \,   \epsilon_- \chi_+,\cr 
 \delta(\Phi_8 - i \Phi_9) \equiv  \delta \bar \varphi     &=  +2i \, {\bar \epsilon^{-+}} \, \psi^{++} \equiv  + \sqrt{2} \, \bar\epsilon_-  \bar\chi_+\,.
\ea\ee
The variation of the 6d gauge field $A_m$ takes the form
\be
i\, \delta A_m = (0,{\epsilon^{--}})  \gamma_m 
\left(\begin{array}{c}
\psi^{++} \cr
0
\end{array}\right) + 
( -{\bar \epsilon^{-+}},0) (-\gamma_m)  \left(\begin{array}{c}
0 \cr
\bar\psi^{+-}
\end{array}\right) \,.
\ee
Projecting onto the chiral and anti-chiral spinor components in ${\bf 3}$ and ${\bar{\bf 3}}$, respectively, yields
\be
\ba
\delta A_{\underline\alpha} \equiv \delta a &= 2 i \,  \epsilon^{--} \psi^{++}_{\underline\alpha}  \equiv  -\sqrt{2} \, \epsilon_- \psi_+, \cr 
\delta A_{\underline{\dot \alpha}}   \equiv \delta \bar{a} &= - 2 i \,  \bar\epsilon^{-+}  \bar{\psi}^{+-}_{\underline{\dot{\alpha}}} \equiv + \sqrt{ 2} \,  \bar\epsilon_- \bar\psi_+\,. 
\ea\ee
Furthermore, the variation of the 2d vector field components are 
\be
\ba
\delta v_0 = - \delta v_1 &=  i\epsilon_- \bar\eta_- -i\bar\epsilon_-  \eta_- \,.
\ea
\ee
Likewise the gaugino variation $\delta\Psi$ reduces to
\be
\ba
\delta\Psi^{  + + } &= \epsilon_{-} \left(\begin{array}{c} 
-F_{0,8}- F_{1,8}+i F_{0,9}+i F_{1,9}\cr 
  0_{31}
  \end{array}
   \right)
+
\bar\epsilon_{-} \left(\begin{array}{c} 
0\cr 
F_{0,6}+ F_{1,6} -i \left(F_{0,7} +F_{1,7}\right)\cr 
F_{0,4}+ F_{1,4}-i \left(F_{0,5}+F_{1,5}\right)\cr 
F_{0,2} + F_{1,2}-i \left(F_{0,3}+F_{1,3}\right)\cr 
  0_{28}
  \end{array}
   \right)
   \cr 
\delta\bar\Psi^{ + - } &=  \epsilon_{-} \left(\begin{array}{c} 0_{20} \cr 
F_{0,2}+ F_{1,2}+i( F_{0,3}+ F_{1,3})\cr 
-F_{0,4}- F_{1,4}-i (F_{0,5}+ F_{1,5})\cr 
F_{0,6}+F_{1,6} +i (F_{0,7}+ F_{1,7})\cr 
     0_{9}
  \end{array}
   \right) 
   +
    \bar\epsilon_{-} \left(\begin{array}{c} 0_{23} \cr 
F_{0,8}+ F_{1,8}+i (F_{0,9}+ F_{1,9} )\cr 
     0_{8}
  \end{array}
   \right)  \cr 
\delta\Psi^{  - - } &=\epsilon_{-} \left(\begin{array}{c} 0_{24} \cr 
-F_{0,1}+i \left(F_{2,3}+F_{4,5}+F_{6,7}-F_{8,9}\right) \cr 
F_{2,4}-  F_{3,5}+i \left(F_{2,5}+F_{3,4}\right)\cr 
 F_{3,7}-F_{2,6}-i \left(F_{2,7}+F_{3,6}\right)\cr 
F_{4,6}- F_{5,7}+i \left(F_{4,7}+F_{5,6}\right)\cr   0_{4}
  \end{array}
   \right)   
   + 
   \bar\epsilon_{-} \left(\begin{array}{c} 0_{25} \cr 
   F_{6,8}+F_{7,9}+i \left(F_{6,9}-F_{7,8}\right)\cr 
   F_{4,8}+F_{5,9}+i   \left(F_{4,9}-F_{5,8}\right)\cr 
   F_{2,8}+F_{3,9}+i \left(F_{2,9}-F_{3,8}\right)
\cr   0_{4}
  \end{array}
   \right)   
\cr 
\delta\bar\Psi^{  - + } &= \epsilon_{-} \left(\begin{array}{c} 0_{12} \cr 
-F_{2,8}+i F_{2,9}-i F_{3,8}-F_{3,9}\cr
F_{4,8}-i F_{4,9}+i F_{5,8}+F_{5,9}\cr 
-F_{6,8}+i F_{6,9}-i   F_{7,8}-F_{7,9}\cr 
0_{17}
     \end{array}
   \right) 
   + 
   \bar\epsilon_{-} \left(\begin{array}{c} 0_{12} \cr 
   F_{5,7} -F_{4,6}+i \left(F_{4,7}+F_{5,6}\right)\cr 
   F_{3,7} -F_{2,6}+i \left(F_{2,7}+F_{3,6}\right)\cr 
   F_{3,5}-F_{2,4}+i   \left(F_{2,5}+F_{3,4}\right)\cr 
   -F_{0,1}-i \left(F_{2,3}+F_{4,5}+F_{6,7}-F_{8,9}\right)   
  \cr
  0_{16}   \end{array}
   \right) 
   \, ,
   \ea
\ee
where the subscript of $\Psi$ indicates 2d chirality and R-symmetry charges, respectively, i.e. these are the projections of the 10d spinor onto the components with these 2d chiralities and R-symmetry. Furthermore, we projected onto either ${\bf 4}$ or ${\bar{\bf 4}}$. 
Rewriting this in terms of the component fields (\ref{table-matter1}) the variation of $\delta\Psi_{++}$ and $\delta \Psi_{--}$ gives rise to 
\be
\ba
\delta \psi^{++}_{1} &\equiv \frac{1}{i \sqrt{2}}   \, \delta \bar\chi_+   =  -\epsilon_- (D_0 + D_1) (\Phi_8 - i \Phi_9) \equiv  -\epsilon_- (D_0 + D_1) \bar{\varphi} \cr 
\delta \psi^{++}_{\underline{\alpha}} &\equiv -  \frac{1}{i \sqrt{2}}      \delta \psi_+ = - \bar\epsilon_- (D_0 + D_1) A_{\underline\alpha}  { + \bar\epsilon_- D_{\underline\alpha} (v_0 + v_1)} \equiv  -\bar\epsilon_- (D_0 + D_1) \, a   { + \bar\epsilon_- D_{\underline\alpha} v_+}    \cr 
\delta \psi^{--}_1 &\equiv -\delta \eta_-  = \epsilon_- (- F_{01} + i (F_{2,3} + F_{4,5} + F_{6,7} - F_{8,9}) )  \equiv \epsilon_- (- F_{01} - i \mathfrak{D})   \cr 
\delta\psi^{--}_{\underline{\alpha}} &\equiv \delta \rho_{-}= 
\epsilon_-   {\varepsilon_{\underline{\alpha}}}^{\dot{\underline{\beta}}\dot{\underline{\gamma}}} D_{\underline{\dot{\beta}}} A_{\dot{\underline{\gamma}}} 
{-} \bar\epsilon_- D_{\underline{\alpha}} \Phi_+    
\ea
\ee
\be
\ba
\delta \bar\psi^{+- }_{\dot{4}} &\equiv  \frac{1}{i \sqrt{2}}  \, \delta \chi_+   =  \bar\epsilon_- (D_0 + D_1) (\Phi_8 + i \Phi_9) \equiv \bar\epsilon_- (D_0 + D_1)\varphi \cr 
\delta \bar\psi^{+- }_{\underline{\dot{\alpha}}} &\equiv -  \frac{1}{i \sqrt{2}}   \, \delta \bar\psi_+ = \epsilon_- (D_0 + D_1) A_{\underline{\dot\alpha}} {- \epsilon_- D_{\underline{\dot\alpha}} (v_0 + v_1)}   \equiv  \epsilon_- (D_0 + D_1) \bar{a} {- \epsilon_- D_{\underline{\dot\alpha}} v_+}   \cr 
\delta \bar\psi^{-+}_{ \dot{4}} &\equiv  -\delta \bar\eta_-  = \bar\epsilon_- (- F_{01} - i (F_{2,3} + F_{4,5} + F_{6,7} - F_{8,9}) ) \equiv \bar\epsilon_- (- F_{01} + i \mathfrak{D} ) \cr 
\delta \bar\psi^{-+}_{ \underline{\dot\alpha}} &\equiv \delta \bar\rho_{-}=  \bar\epsilon_-   {\varepsilon_{\underline{\dot\alpha}}}^{{\underline{\beta}}{\underline{\gamma}}} D_{\underline{{\beta}}} A_{{\underline{\gamma}}} 
- \epsilon_- D_{\underline{\dot\alpha}} \Phi_- 
\,,
\ea
\ee
where $\varepsilon$ is the invariant tensor of $SU(3)$, satisfying ${\varepsilon_{\underline{\alpha}}}^{\dot{\underline{\beta}}\dot{\underline{\gamma}}} = - {\varepsilon_{\underline{\alpha}}}^{\dot{\underline{\gamma}}\dot{\underline{\beta}}}$, which enables the isomorphism between $\Lambda^2 {\bar{\bf 3}}$ and ${\bf 3}$. The variations of $\Psi_{-+}$ and $\Psi_{+-}$ yield the conjugates to these variations. 


\section{Examples: $SU(6)$ and $E_6$}
\label{app:Examples}

We collect various useful properties of elliptic fibrations, their singularity resolution, and intersection rings in the following. Whenever possible we refer back to the general analysis of resolutions in  \cite{Lawrie:2012gg}, which applies to four-folds, and only give details whenever necessary for the five-fold case. 

\subsection{$SU(6)$ Theories}
\label{app:SU6}

For illustration consider $SU(6)$. Again the general $k$ resolutions have appeared in \cite{Lawrie:2012gg}. 
The Tate form is 
\be
y^2+x y b_1 +y b_3 \zeta _0^3= x^3 +x^2 b_2 \zeta _0+x b_4 \zeta _0^3+b_6 \zeta _0^6 \,.
\ee
The classes of the coefficients are
\be
[b_1] =c_1\,,\quad 
[b_2] = 2 c_1-M_G\,,\quad 
[b_3] = 3 c_1-3 M_G\,,\quad 
[b_4]=4 c_1-3 M_G\,,\quad 
[b_6]= 6 c_1-6 M_G\,.
\ee
From the discriminant 
\be\ba
\Delta =& b_1^4 \left( b_4 \left(b_1 b_3+b_4\right)-b_1^2 b_6\right) \zeta _0^6\cr 
&+b_1^2 b_2 \left(8 b_1 b_3 b_4+8 b_4^2-b_1^2 \left(b_3^2+12 b_6\right)\right) \zeta _0^7\cr 
&-8 \left(b_2^2 \left(-2 b_1 b_3 b_4-2 b_4^2+b_1^2 \left(b_3^2+6 b_6\right)\right)\right) \zeta _0^8\cr 
&+\left(-16 b_2^3 \left(b_3^2+4
   b_6\right)+\left(b_1 b_3+2 b_4\right) \left(-32 b_1 b_3 b_4-32 b_4^2+b_1^2 \left(b_3^2+36 b_6\right)\right)\right) \zeta _0^9
   +O\left(\zeta _0\right)^{10} \,,
\ea
\ee
we identify the two codimension two loci $b_1=0$, which corresponds to matter in the $\Lambda^2 {\bf 6}= {\bf 16}$, and $P_{\bf 6} = b_1 b_3 b_4+b_4^2-b_1^2 b_6 =0$ associated with the fundamental ${\bf 6}$ representations. 

Consider the resolution sequence
\be
(x,y,\zeta _0;\zeta _1),\quad 
(x,y,\zeta _1;\zeta _2), \quad 
(x,y,\zeta _2;\zeta _3),\quad 
(y,\zeta _1;\zeta_4), \quad 
(y,\zeta _2;\zeta _5 )\,.
\ee
The exceptional sections correspond to the simple roots 
\be
(\alpha_0, \alpha_1, \alpha_2, \alpha_3, \alpha_4, \alpha_5) \leftrightarrow (\zeta _0,\zeta _1,\zeta _2,\zeta _3,\zeta _5,\zeta _4)\,.
\ee
Again the antisymmetric matter is localized along $b_1=0$, and the fundamental matter at $P_{\bf 6}=b_4^2+ b_1b_1 b_4 -b_1^2 b_6=0$. 
In codimension four, this model has a non-minimal locus $b_1 = b_2 =b_4=0$, where the Tate form vanishing orders are $(1,2,3,4,6)$. Thus, we need to remove this non-minimal locus 
\be
[b_1]\cdot [ b_2 ] \cdot [b_4]  = c_1 \cdot \left(4 c_1-3 M_G\right)\cdot  \left(2 c_1-M_G\right)   =0\,.
\ee
The fiber splittings were derived in general in \cite{Lawrie:2012gg} in codimension two and three.  As can be seen from the codimension three fibers therein, the codimension three enhancement to $I_m^*$, i.e. to a D-type singularity, is again monodromy reduced, as the fiber is characterized in terms of a quadratic equation. 
In the above resolution  $c_4(Y_5)$ is computed to be
\be\label{ASU637}
M_G \cdot_{Y_5}   c_4(Y_5) = M_G \cdot_{B_4} \left( 360 c_1^3-894 c_1^2 M_G+12 c_1 c_2+753 c_1 M_G^2-210 M_G^3\right)\,.
\ee
Anomaly cancellation can be checked with the following expressions for the chiralities of the matter, for trivial gauge bundle:
\be
\ba
\chi_{\rm bulk} & = \frac{1}{24} M_G \left(c_1-M_G\right) \left(-c_1 M_G+c_2+M_G^2\right) \cr 
\chi(b_1, {\bf 15}) &= \frac{1}{24} c_1 M_G \left(2  c_2+M_G^2\right) \cr 
\chi (P_{\bf 6}, {\bf 6}) &= \frac{1}{12} M_G \left(4 c_1-3 M_G\right) \left(-96 c_1 M_G+63 c_1^2+2 c_2+37 M_G^2\right) + \chi^{\rm sing}_{\bf 6} \,.
\ea
\ee
The  corrections due to the higher codimension singular loci take the form given in (\ref{ChiSUeven}), which accounts for the singular matter locus $P$ along $b_1=b_4=0$ and the additional contributions from the double curves when $\delta= b_3^2 + b_6=0$,
\be
\chi^{\rm sing}_{\bf 6}=-\frac{1}{4} c_1 M_G \left(7 c_1-6 M_G\right) \left(4 c_1-3 M_G\right) \,.
\ee
The anomaly contributions are, including the group theory factors,  
\be
\mathcal{A}_{\rm  surface} = {2} \chi ( b_1, {\bf 15} )+ {1\over 2} \chi (P_{\bf 6}, {\bf 6}) \,,\qquad 
\mathcal{A}_{\rm bulk} = - 6 \chi_{\rm bulk} \,,
\ee
and cancel the contribution from $\mathcal{A}_{3-7}$ detailed in  (\ref{ASU637}). 

The Chern-Simons terms are easily computed as well 
\be
\ba
c_4 (Y_5) \cdot D_1 & = -4 c_1 M_G^3+2 c_1^2 M_G^2+2 c_2 M_G^2-2 c_1 c_2 M_G+2 M_G^4 \cr 
c_4 (Y_5) \cdot D_2 & = -3 c_1 M_G^3+2 c_1^2 M_G^2+2 c_2 M_G^2+2 M_G^4\cr 
c_4 (Y_5) \cdot D_3 & = 760 c_1 M_G^3-874 c_1^2 M_G^2-10 c_2 M_G^2+336 c_1^3 M_G+14 c_1 c_2 M_G-220 M_G^4\cr 
c_4 (Y_5) \cdot D_4 & = -2 c_1 M_G^3+2 c_1^2 M_G^2+2 c_2 M_G^2+2 c_1 c_2 M_G+2 M_G^4\cr 
c_4 (Y_5) \cdot D_5 & =6 c_1 M_G^3-28 c_1^2 M_G^2+2 c_2 M_G^2+24 c_1^3 M_G+2 M_G^4 \,.
\ea
\ee
The box graphs (from which we determine $\sum_\lambda C_\lambda D_k$) for the even $SU(2k)$ groups have been determined in \cite{Hayashi:2014kca}, and confirm that the intersections of the Cartan divisors with $c_4(Y_5)$ can be written in terms of the chiralities as
\be
\ba
{1\over 24} c_4 (Y_5) \cdot D_1 &= - 2 \chi_{\rm bulk}    \cr 
{1\over 24} c_4 (Y_5) \cdot D_2 &= - 2 \chi_{\rm bulk} +  \chi ( b_1, {\bf 15} )\cr 
{1\over 24} c_4 (Y_5) \cdot D_3 &= - 2 \chi_{\rm bulk} +  \chi (P_{\bf 6}, {\bf 6})\cr 
{1\over 24} c_4 (Y_5) \cdot D_4 &= - 2 \chi_{\rm bulk} + 2 \chi ( b_1, {\bf 15} ) \cr 
{1\over 24} c_4 (Y_5) \cdot D_5 &= - 2 \chi_{\rm bulk} + 2 \chi ( b_1, {\bf 15} )  \,.
\ea
\ee
{Note that in the last equation, the constraint was used that the non-minimal locus $b_1= b_4=0$ does not contribute. }


\subsection{$E_6$ Theories}
\label{app:ExamplesE}

Finally, we discuss some properties of the exceptional gauge groups, which appear in the main text in section \ref{sec:E6Global}. 
The $E_6$ Tate form with vanishings $(1,2,2,3,5)$ is 
\be
y^2 +b_1 \zeta _0 x y+b_3 \zeta _0^2 y = x^3+ b_2 \zeta _0^2 x^2+ b_4 \zeta _0^3 x +b_6 \zeta _0^5 \,.
\ee
The only matter locus in codimension one above $\zeta_0=0$ is $b_3=0$, which gives rise to matter in the ${\bf 27}$. 
We resolve the model with the following chain of blowups
\be
\ba
\left(x,y,\zeta _0;\zeta _1\right)\,,\ 
\left(x,y,\zeta _1;\zeta _2\right)\,,\ 
\left(y,\zeta _1;\zeta _3\right)\,,\ 
\left(y,\zeta _2;\zeta_4\right)\,,\  
\left(\zeta _2,\zeta _3;\zeta _5\right)\,,\  
\left(\zeta _3,\zeta _4;\zeta _6\right)\,,\  
\left(\zeta _3,\zeta _5;\zeta _7\right) \,.
\ea
\ee
The simple roots are associated to the exceptional sections, and thus Cartan divisors, as follows\footnote{Note that in this resolution $\zeta_1=0$ implies $\zeta_3=0$, so these are not independent divisors.}
\be
 ( \alpha_1, \alpha_2, \alpha_3, \alpha_4, \alpha_5, \alpha_6, \alpha_0)\quad \leftrightarrow \quad 
 \left(\zeta _4,\zeta _6,\zeta _7,\zeta _5,\zeta _2,\zeta _1,\zeta _0\right)\,.
\ee
With this ordering, the intersections with $c_4(Y_5)$ are 
\be\label{c4DiE6}
\ba
c_4(Y_5) \cdot D_1 &=437 c_1 M_G^3-588 c_1^2 M_G^2-6 c_2 M_G^2+264 c_1^3 M_G+10 c_1 c_2 M_G-108 M_G^4\cr 
c_4(Y_5) \cdot D_2 &=35 c_1 M_G^3-50 c_1^2 M_G^2-2 c_2 M_G^2+24 c_1^3 M_G+4 c_1
   c_2 M_G-8 M_G^4\cr 
   c_4(Y_5) \cdot D_3 &=-4 c_1 M_G^3+2 c_1^2 M_G^2+2 c_2 M_G^2-2 c_1 c_2 M_G+2 M_G^4 \cr 
    c_4(Y_5) \cdot D_4 &= 35 c_1 M_G^3-50 c_1^2 M_G^2-2 c_2 M_G^2+24 c_1^3 M_G+4 c_1 c_2
   M_G-8 M_G^4 \cr 
   c_4(Y_5) \cdot D_5 &=-4 c_1 M_G^3+2 c_1^2 M_G^2+2 c_2 M_G^2-2 c_1 c_2 M_G+2 M_G^4\cr 
   c_4(Y_5) \cdot D_6 &=-4 c_1 M_G^3+2 c_1^2 M_G^2+2 c_2 M_G^2-2 c_1 c_2 M_G+2 M_G^4 \,.
\ea
%
\ee
These are matched with the chiralities in (\ref{E6CSCheck}).


\section{Type IIB Orientifolds on Four-folds with D7 and D3-branes } \label{app_IIB}

In this appendix we describe the weak coupling Type IIB orientifold limit of the 2-dimensional F-theory compactifications considered in the bulk of this work. 
In particular we will uncover a rich structure of Green-Schwarz-type couplings emanating from the Chern-Simons couplings of the branes.
Much of the discussion in this appendix parallels the analysis in \cite{Collinucci:2008pf,Blumenhagen:2008zz}  for 4-dimensional Type IIB compactifications on Calabi-Yau three-folds. We refer to this work and references therein for generalities on Type IIB orientifold compactifications with 7-branes.
  
 Consider therefore a Type IIB orientifold compactification on a Calabi-Yau four-fold $X_4$, endowed with a 
 holomorphic involution  $\sigma: X_4 \rightarrow X_4$. Its fix-point locus is given by an O7-plane wrapping a holomorphic divisor, i.e. a complex three-cycle, $D_{O7} \subset X_4$. For simplicity we assume the absence of O3-planes, which would wrap holomorphic curves on $X_4$; these are easily included into the framework.
A stack of  $n$ coincident D7-branes branes wrapping a divisor $D_i$ at generic position relative to the orientifold plane carries
a $U(n)$ gauge group. By generic we mean that $D_i \neq D_i'$ with $D_i' = \sigma(D_i)$ the orientifold image divisor.
Invariant branes give rise to gauge groups of type $Sp(n)$ or $SO(n)$. 
Divisors wrapped by single branes invariant under $\sigma$ as a whole, but not pointwise, are of Whitney umbrella type and exhibit a codimension one locus of double point singularities at the intersection with the O7-plane \cite{Collinucci:2008pf}. 

The singularity modifies the naive result for the Ramond-Ramond charges of such singular branes as detailed in \cite{Collinucci:2008pf} for divisors on Calabi-Yau three-folds,
 where the locus of double point singularities is a curve. This computation must be generalized to divisors on four-folds, where now the higher-dimensional nature of the singularities along a surface as opposed to a curve must be taken into account.
For simplicity we avoid this technical complication by focusing on non-invariant brane divisors $D_i \neq D'_{i}$, which are assumed to be smooth. 

The induced brane charges of this setup are computed by expanding the Chern-Simons action for the D7-branes and the O7-plane,
\be \label{CScouplingsFtheory}
\ba
S_{D7} &= 2 \pi \int_{D7} {\rm tr} \, e^{ \frac{1}{2\pi} \, \cal F}  \, \sum_{2p} C_{2p} \, \sqrt{  \frac{ \hat A({\rm T}D7)}{\hat A({\rm N}D7)}}, \cr
S_{O7} &= -  16 \pi \int_{O7} \sum_{2p} C_{2p} \, \sqrt{ \frac{L(\frac{1}{4} {\rm T}O7)}{L(\frac{1}{4} {\rm N}O7)}}.
\ea
\ee
Here ${\rm T}D7$ and ${\rm N}D7$ denote the tangent and normal space to the $D7$-brane (and similarly for the $O7$-plane) and ${\cal F} = F + \iota^*B_2$ in terms of the field strength $F$ of the D7-brane and the pullback of the B-field. We are working in conventions where $\ell_s = 1$. 
The relevant terms in the A-roof genus and the Hirzebruch L-genus are
\be
\ba
\hat A({\rm T}D) &= 1 - \frac{1}{24}\,  p_1({\rm T}D) + \ldots = 1 - \frac{1}{24} (c_1^2({\rm T}D) - 2 c_2({\rm T}D)) + \ldots, \cr
L({\rm T}D) &= 1 + \frac{1}{3} \, p_1({\rm T}D) + \ldots = 1 + \frac{1}{3}(c_1^2({\rm T}D) - 2 c_2({\rm T}D)) + \ldots , 
\ea
\ee
together with analogous terms for the normal bundles. 
As a result of the adjunction formula $c_1({\rm T}D) = - c_1({\rm N}D)$  for a holomorphic divisor $D$ on a Calabi-Yau four-fold $X_4$  and the fact that $c_2({\rm N}D) = 0$ the curvature terms follow as
\be
 \sqrt{  \frac{ \hat A({\rm T}D7)}{\hat A({\rm N}D7)}} = 1 + \frac{1}{24} c_2(D7),  \qquad \sqrt{ \frac{L(\frac{1}{4} {\rm T}O7)}{L(\frac{1}{4} {\rm N}O7)}} = 1 - \frac{1}{48} c_2(O7) \,,
 \ee
 where all omitted terms are forms of degree $8$ or higher.
 Under the orientifold action the field strength on each brane is mapped to its cousin on the orientifold image brane
 \bea
{\cal F}_i \rightarrow  -  {\cal F}'_i =  - \sigma^* {\cal F}_i \,,
\eea
where the minus sign is due to the worldsheet parity action. Furthermore, recall that $B_2$, $C_2$ and $C_6$ are orientifold odd, while $C_0$, $C_4$ and $C_8$ are orientifold even.
 
In general the compactification also includes a number of D3-branes filling $\mathbb R^{1,1}$ and wrapping  holomorphic curves $C_{\Xi}$ on $X_4$. 
The D3-brane action takes a similar form
\be
S_{D3} = - 2 \pi \int_{D3} {\rm tr} \, e^{ \frac{1}{2\pi} \cal F}  \, \sum_{2p} C_{2p} \, \sqrt{  \frac{ \hat A({\rm T}D3)}{\hat A({\rm N}D3)}}\,,
\ee
where the relative minus sign is crucial. For a D3-brane wrapping a complex curve on $X_4$ the geometric curvature terms  vanish for dimensional reasons.

\subsection{Tadpoles and Green-Schwarz terms} \label{app_TadpolesGS}

Let us now systematically reduce the Chern-Simons interactions to 2 dimensions. 
From the coupling of $C_8$ one deduces the standard condition for cancellation of the 
\be
{\rm D7-tadpole}: \qquad   \sum_i n_i (D_i + D_i') \stackrel{!}{=} 8 D_{O7} \,.
\ee
Next reduce the orientifold-odd 6-form $C_6$ in terms of a basis  $\{\omega^{{(4,-)}}_A\}$ of $H^4_-(X_4)$ and $\{\omega^{(6,-)}_M \}$ of $H^6_-(X_4)$ as
\bea
C_6 = c_2^A  \wedge  \omega^{{(4,-)}}_A + c_0^M  \, \omega^{(6,-)}_M\,,
\eea
where $c_2^A$ and  $c_0^M$ are the associated 2-forms and axionic scalar fields in the 2-dimensional field theory. 
Inserted into the Chern-Simons actions, this ansatz results in two types of terms, one of which is a 
 tadpole for $c_2^A$. In order for the compactification to describe a consistent vacuum we must require the cancellation of this 
\bea \label{app_D5tadpole}
{\rm D5-tadpole}: \quad  \sum_{i} \left( \int_{D_i }{\rm tr} {\cal F}_i \wedge  \omega^{{(4,-)}}_A -  \int_{D'_i }{\rm tr} {\cal F}'_i \wedge  \omega^{{(4,-)}}_A \right) \stackrel{!}{=} 0\,.
\eea
As in compactifications to 4 dimensions, the D5-tadpole constrains the choice of consistent gauge fluxes in Type IIB orientifolds, while it is automatically satisfied in the F/M-theory description of $G_4$-fluxes as elements of $H^4(Y_5)$.
The second type of terms couple the axions $c_0^M$ via Green-Schwarz-St\"uckelberg type interactions to the abelian part of the 7-brane gauge field strengths $F_i$ along $\mathbb R^{1,1}$,
\bea \label{S-Stgeom1}
S^{(1)}_{\rm GS} = \sum_{M, i}    \int_{\mathbb R^{1,1}} Q_{M i}   \, c_0^M  \, {\rm tr} {\cal} F_i, \qquad \quad Q_{M i} =  \int_{D_i}  \omega^{(6,-)}_M - \int_{D_i'}  \omega^{(6,-)}_M \,.
\eea
These terms are the 2-dimensional analogue of the geometric St\"uckelberg mass terms whose uplift to F-theory has been studied in detail in \cite{Grimm:2011tb} for compactifications to 4 spacetime dimensions. Note that these couplings can be non-zero only if $D_i \neq D'_i$ in homology.

A similar expansion of  the self-dual, orientifold even 4-form $C_4$ involves a basis $\{w_a^{(2,+)} \}$ of $H^2_+(X_4)$ and $\{w_k^{(4,+)} \}$ of $H^4_+(X_4)$,
\bea
C_4 = \sum_a c_2^{a} \wedge  w_a^{(2,+)} + \sum_k c_0^k \, w_k^{(4,+)}.
\eea
The tadpole for $c_2^{a}$ receives contributions from all D7-branes, D3-branes and the O7-plane.
 The total class $C = \sum_{\Xi} C_{\Xi}$ of all curves wrapped by the D3-branes is thus determined by requiring cancellation of this  
\be
\ba \label{app_D3Tadpole1}
{\rm D3-tadpole}: \quad C + C'& \stackrel{!}{=} \sum_i  \frac{n_i}{24}\left( D_i \wedge c_2(D_i)  +  D'_i \wedge c_2(D'_i) \right) + \frac{1}{6} D_{O7} \wedge c_2(D_{O7})  \cr
&+ \sum_i \frac{1}{8 \pi^2} \left( D_i \wedge {\rm tr} {\cal F}_i^2  + D'_i \wedge {\rm tr}{ {\cal F}'_i}^2 \right).
\ea
\ee
Furthermore we observe a flux-induced Green-Schwarz-St\"uckelberg term for the abelian part of the 7-brane field strengths $F_i$ 
 along ${\mathbb R}^{1,1}$ of the form
\be \label{app_SGS2}
\ba
S^{(2)}_{\rm GS} &= \sum_{k, i}  \int_{\mathbb R^{1,1}}  Q_{k i } \, c_0^k  \, {\rm tr F}_i , \qquad    
Q_{k i } =  \frac{1}{4 \pi} \left( \int_{D_i} {\rm tr}{\cal F}_i \wedge  w_k^{(4,+)} +  \int_{D'_i} {\rm tr} {\cal F}'_i \wedge  w_k^{(4,+)} \right).
\ea 
\ee 
Note that $H^4(X_4)$ contains both a $(3,1)$ and $(1,3)$ subspace and a $(2,2)$ subspace. Since BPS conditions exclude internal gauge fields of $(2,0)$ and $(0,2)$ Hodge type, only the terms associated with
\bea
w_k^{(4,+)} \in H^{2,2}_+(X_4)
\eea
contribute to (\ref{app_SGS2}) in a supersymmetric vacuum.

From expansion of $C_2$ in terms of a basis $\{\omega^{(2,-)}_p \}$ of $H^2_-(X_4)$,
\bea
C_2 = \sum_p c_0^p \, \omega^{2,-}_p ,
\eea
we receive first another contribution to the Green-Schwarz-St\"uckelberg coupling of the 7-brane $U(1)$ fields,
\be
\ba \label{S2Stflgeom}
S^{(3)}_{\rm GS} &= \sum_{p, i} \int_{\mathbb R^{1,1}} Q_{p i} \, c_0^p \, {\rm tr} F_i, \cr 
Q_{p i}  & = \int_{D_i} \left(\frac{1}{24\pi^2} {\rm tr} {\cal F}^2_i + \frac{1}{24} c_2(D_i)\right) \wedge \omega^{(2,-)}_p -  \int_{D'_i} \left(\frac{1}{24\pi^2} {\rm tr} {\cal F}'^2_i + \frac{1}{24} c_2(D'_i)\right) \wedge \omega^{(2,-)}_p 
\ea
\ee
The curvature induced terms are non-zero only if $D_i \neq D'_i$ in homology, in which case (\ref{S2Stflgeom}) contributes to the geometric St\"uckelberg mass terms for $F_i$ in addition to (\ref{S-Stgeom1}).
For the flux-induced part to be non-vanishing we need either $D_i \neq D'_i$ or ${\cal F}_i \neq {\cal F}'_i$ in homology.\footnote{Here we view the class ${\cal F}_i$ as a class on $X_4$ pulled back to $D_i$. This is justified because the part of ${\cal F}_i$ which is not in the image of the pullback map does not contribute to (\ref{S2Stflgeom}).} 
There is also a geometric Green-Schwarz-St\"uckelberg term for the $U(1)$ gauge fields originating from the D3-branes,
\bea \label{app_SGSD31}
S^{(1)}_{\rm GS, D3} = \sum_{\Xi, i} \int_{\mathbb R^{1,1}} Q_{\Xi p} \, {\rm tr F}_{\Xi}, \qquad Q_{\Xi p} = \frac{1}{2\pi} \left( \int_{C_{\Xi}} \omega^{(2,-)}_p - \int_{C'_{\Xi}} \omega^{(2,-)}_p \right) \,,
\eea
which is non-zero for $C_{\Xi} \neq C'_{\Xi}$ in homology.

It is worthwhile noting that there can be no F1 or D1-brane tadpole induced because $B_2$ and $C_2$ are orientifold-odd and thus their 2-form components along ${\mathbb R}^{1,1}$ are projected out.

Finally, the zero-form $C_0$  yields another contribution to the Green-Schwarz terms of the 7-brane gauge fields, 
 \be \label{app_SGS4}
  \ba
 S^{(4)}_{\rm GS} &= \sum_i  \int_{\mathbb R^{1,1}} Q_{0 i}  \, C_0 \, {\rm tr} F_i \cr
  Q_{0 i} &= \int_{D_i} \left( \frac{1}{24 (2 \pi)^3} {\rm tr} {\cal F}_i^3 + \frac{1}{48 (2 \pi)} {\rm tr} {\cal F}_i c_2(D_i) \right) + \int_{D'_i} \left( \frac{1}{24 (2 \pi)^3} {\rm tr} {\cal F'}_i^3 + \frac{1}{48 (2 \pi)} {\rm tr} {{\cal F}'}_i c_2(D'_i) \right)\,,
 \ea
 \ee 
and of the D3-brane $U(1)$ fields,
 \be \label{app_SGSD32}
S^{(2)}_{\rm GS,D3} = \sum_\Xi \int_{\mathbb R^{1,1}} Q_{0 \Xi}  \, C_0 \, {\rm tr} F_\Xi
 \qquad Q_{0 \Xi} =  - \int_{C_\Xi} \frac{1}{4\pi} {\rm tr} {\cal F}_\Xi -  \int_{C'_\Xi} \frac{1}{4\pi} {\rm tr} {\cal F}'_\Xi \,.
\ee 
However, a non-trivial gauge flux on the D3-brane necessarily induces a D-term. For vanishing VEVs of the localised charged matter states, this is not consistent with supersymmetry. More information on the D3-brane system will be provided in \cite{Lawrie:2016axq}.

\subsection{Anomaly Cancellation in a Prototypical Example}

In the remainder of this section we restrict ourselves to a simple example of a brane setup with $n$ 7-branes along a divisor $W$ and one extra D7-brane along the divisor $D$, each accompanied by their orientifold images. The 7-brane tadpole cancellation condition requires that 
\bea\label{7-tadpole}
n (W +W') + (D +D') = 8 \, D_{O7}.
\eea
We assume that all divisors can be chosen to be smooth, which must be verified in concrete examples. Modulo St\"uckelberg masses for the abelian gauge group factors, the gauge group from the 7-brane sector is now $U(n) \times U(1)$. The uplift of such models to F-theory contains either massless or St\"uckelberg massive $U(1)$  factors in addition to an $SU(n)$ gauge group \cite{Krause:2012yh}.

The total class of all wrapped spacetime-filling D3-branes is determined by (\ref{app_D3Tadpole1}), which becomes
 \be
 \ba
C + C' &=  Q_{\rm geom} + Q_{\rm gauge}, \cr
Q_{\rm geom} &= \frac{1}{24} \left( n \left(c_2(W) \wedge W + c_2(W') \wedge W'\right) + \left(c_2(D) \wedge D + c_2(D') \wedge D'\right)\right) \cr
&  + \frac{1}{6} c_2(D_{O7}) \wedge D_{O7}, \cr
Q_{\rm gauge} &= n \left({\rm ch}_2(L_W) \wedge W + {\rm ch}_2(L_W') \wedge W' \right)+ {\rm ch}_2(L_D) \wedge D +  {\rm ch}_2(L_D') \wedge D'.
\ea
\ee
Here we have introduced the line bundle $L_W$ with curvature $c_1(L_W) = \frac{1}{2\pi} {\rm tr} {\cal F}_i$ (and likewise for $L_D$).
For simplicity we are again  assuming vanishing closed string 3-form flux and absence of O3-planes.
It is convenient to organise the D3-brane curve class and its image as 
\be
\ba
 \label{CCprime}
C &= \frac{1}{24} \left( n \, c_2(W) \wedge W  + c_2(D) \wedge D   + 2 c_2(D_{O7}) \wedge D_{O7}  \right) +  {\rm ch}_2(L_W) \wedge W +{\rm ch}_2(L_D) \wedge D ,  \cr
C' &= \frac{1}{24} \left( n\,  c_2(W') \wedge W' +  c_2(D') \wedge D'  + 2 c_2(D_{O7}) \wedge D_{O7}  \right) + {\rm ch}_2(L_W') \wedge W' +{\rm ch}_2(L_D') \wedge D'. 
\ea
\ee

To compute the spectrum in the D7-D7-brane sector, we work in the upstairs geometry prior to orientifolding.
The analysis of sections \ref {sec:8dSYM} and \ref{sec_S2Matter}, especially the results (\ref{cohomologies-bulk}) and (\ref{table-matterspec}), carry over immediately. Alternatively at weak coupling an explicit analysis of open string vertex operators along the lines of \cite{Katz:2002gh} can be performed. The contributions of the bulk and surface matter to the $SU(n)$ gauge anomalies are:
\be \label{IIBWspectrum}
\begin{array}{c|c|c}
\hbox{Locus}  & \hbox{Representation}  &   \hbox{$SU(n)$ - anomaly contribution }\cr\hline
W & {\rm \bf Adj}_{0,0} & - n \,  {\chi}(W) \cr
W' &  {\rm \bf Adj}_{0,0} & - n \, {\chi} (W) \ \cr
W \cap D & \bar{\bf n}_{-1,1} & \frac{1}{2} \,  {\chi}(W \cap D) \cr
W \cap D' &  \bar{\bf n}_{-1,-1}  & \frac{1}{2} \, {\chi}(W \cap D') \cr
W' \cap D &  {\bf n}_{1,1}  & \frac{1}{2} \, {\chi}(W' \cap D)  \cr
W' \cap D' &  {\bf n}_{1,-1}  & \frac{1}{2} \, {\chi}(W' \cap D') \cr
W \cap W' & {\bf \Lambda^2 n}_{2,0} & 2 \times  \frac{n-2}{2}\,   {\chi}(W \cap W')
\end{array}
\ee
The subscripts denote the charge under $U(1) \subset U(n)$ and the $U(1)$ gauge group on $D$.
The first two lines denote the bulk spectrum on the $SU(n)$ D7-branes and their image, which we count as independent since we are working upstairs prior to taking the orientifold quotient.
Independent fundamental matter is localised at $W \cap D$ and $W \cap D'$. This matter is mapped to the matter at $W' \cap D'$ and  
$W' \cap D$ under the orientifold action. Furthermore we are assuming for simplicity that the intersection of $W$ with $W'$ is entirely contained inside the orientifold plane and therefore carries antisymmetric matter only.\footnote{More generally, $W \cap W' = W \cap O7 + {\cal C}_{\rm rest}$. The locus $ {\cal C}_{\rm rest}$, which is not contained inside the O7-plane, gives rise to matter in the symmetric and the antisymmetric representation of $U(n)$. This matter locus uplifts to a self-intersection of the $I_n$ discriminant locus in F-theory and is thus absent in generic Tate models.} 
In order to make this assumption we impose that $W \, W' - W \, O7 = 0 \in H^4(X_4)$. 
This constraint decomposes into two independent relations to be satisfied by the orientifold even and odd components $W_\pm \in H^2_{\pm}(X_4)$ of $W$.
Decomposing $W = W_+ + W_-$, $W' = W_+ - W_-$ and using that $W_- \wedge O7=0$ in homology since orientifold odd classes pull back trivially to the  O7-plane (see e.g. \cite{Blumenhagen:2008zz} in the present context), we arrive at the two constraints
 \be \label{homologicalrelation1}
 \ba
\frac{1}{4} (W + W') (W+W') - \frac{1}{2} (W + W') \, O7 &= 0 \in H^4(X_4), \cr
\frac{1}{4} (W - W') (W-W')  &= 0 \in H^4(X_4), 
\ea
\ee
to be imposed in all expressions that follow. Since we are working upstairs and counting the adjoint and the fundamental matter twice, we must do the same for the anti-symmetric matter. The group theoretic factors are the ones given in (\ref{SUnanomaly-factors}).
The relevant chiral indices are given in  (\ref{chibulk1}) and (\ref{chiS2}). For instance, for vanishing gauge flux  ${\cal F}=0$ the expressions reduce to
\be
\ba
\chi(W) &= \frac{1}{24} \int_W c_1(W) c_2(W), \cr
\chi (W \cap D) &= \int_{W \cap D} \frac{1}{12} (c_1^2(W \cap D)  + c_2(W \cap D)) + \frac{1}{2} c_1(W \cap D) c_1(K_{W\cap D}^{-1/2})  + \frac{1}{2} c_1^2(K_{W\cap D}^{-1/2}). 
\ea
\ee
These formulae assume are valid for smooth three-cycle $W$ and matter loci $W \cap D$, $W \cap D'$, where the standard form of the Hirzebruch-Riemann-Roch index theorem is valid. In the presence of singularities correction terms may become necessary.

The D3-branes wrapping the curve class $C$ and the image D3-branes intersect each of the 7-branes in a set of points. At each point a Fermi multiplet in the fundamental representation of the D7-brane gauge group is localised. In the upstairs geometry we thus  find the following charged matter in the D3-D7 sector, where the subscripts denote the charges under $U(1) \subset U(n)$ and under the $U(1)$ gauge group realized on the D3-branes (assuming that the latter come as single branes as opposed to stacks):
\be
\begin{array}{c|c|c}
\hbox{Locus}  & \hbox{Representation} &   \hbox{$SU(n)$ anomaly contribution }\cr\hline
W \cap C &  \bar{\bf n}_{-1,1} & - \frac{1}{2} \int_{X_4} C \wedge W   \cr
W \cap C' & \bar{\bf n}_{-1,-1}  & - \frac{1}{2} \int_{X_4} C' \wedge W \cr
W' \cap C & {\bf n}_{1,1}  & - \frac{1}{2}\int_{X_4} C \wedge W' \cr
W' \cap C' & {\bf n}_{1,-1}  & -\frac{1}{2} \int_{X_4} C' \wedge W'  \cr
\end{array}
\ee

The anomaly contributions from all sources of matter sum up to zero if we impose the D7-brane tadpole cancellation condition (\ref{7-tadpole}) as well as the two constraints (\ref{homologicalrelation1}) underlying the spectrum (\ref{IIBWspectrum}).

Likewise one systematically check anomaly cancellation in the presence of gauge flux.
In general, unless $W=W'$ and $D=D'$ in homology, the gauge flux is subject to the D5 -tadpole cancellation condition (\ref{app_D5tadpole}) and this constraint is crucial in order for the spectrum to be free of anomalies. Consider as the simplest example a setup where $W=W'$ and $D=D'$ in homology together with a line bundle $L_D$ on $D$ whose first Chern class is the pullback of some divisor class on $X_4$. 
The extra contribution due this gauge flux is first from the change of the chiral index counting matter localised on $W \cap D$ and $W \cap D'$ (plus images), see (\ref{chiS2}), and second due to the change in the D3-brane tadpole (\ref{CCprime}) and the resulting extra number of charged multiplets in the D7-D3-brane sector.
Both effects are found to precisely cancel,
\bea
\Delta {\cal A}& =& + \frac{1}{2} \,  {\rm ch}_2(L_D) \left( W D + W D' + W' D + W' D'\right)  + \\
&&+ \left(-\frac{1}{2}\right) \, {\rm ch}_2(L_D) \left(D W + D W' + D' W' + D' W\right) = 0 \,.
\eea
Generalisations to other flux configurations along these lines are immediate.

%

\providecommand{\href}[2]{#2}\begingroup\raggedright\endgroup


\end{document}